\documentclass[useAMS,usenatbib]{mn2e}

\usepackage{times}

\usepackage{graphicx}
\DeclareGraphicsExtensions{.jpg,.pdf,.mps,.png} 
\usepackage{subfigure}

\usepackage{amsmath, amsfonts, amssymb}
\usepackage{xspace}
\usepackage{hhline}
\usepackage{lscape}

\def\reff@jnl#1{{\rm#1\/}}
\def\aj{\reff@jnl{AJ}}                  
\def\araa{\reff@jnl{ARA\&A}}            
\def\apj{\reff@jnl{ApJ}}                
\def\apjl{\reff@jnl{ApJ}}               
\def\apjs{\reff@jnl{ApJS}}              
\def\ao{\reff@jnl{Appl.Optics}}         
\def\apss{\reff@jnl{Ap\&SS}}            
\def\aap{\reff@jnl{A\&A}}               
\def\aapr{\reff@jnl{A\&A~Rev.}}         
\def\aaps{\reff@jnl{A\&AS}}             
\def\baas{\reff@jnl{BAAS}}              
\def\jrasc{\reff@jnl{JRASC}}            
\def\memras{\reff@jnl{MmRAS}}           
\def\mnras{\reff@jnl{MNRAS}}            
\def\pra{\reff@jnl{Phys.Rev.A}}         
\def\prb{\reff@jnl{Phys.Rev.B}}         
\def\prc{\reff@jnl{Phys.Rev.C}}         
\def\prd{\reff@jnl{Phys.Rev.D}}         
\def\prl{\reff@jnl{Phys.Rev.Lett}}      
\def\pasp{\reff@jnl{PASP}}              
\def\pasj{\reff@jnl{PASJ}}              
\def\qjras{\reff@jnl{QJRAS}}            
\def\skytel{\reff@jnl{S\&T}}            
\def\solphys{\reff@jnl{Solar~Phys.}}    
\def\sovast{\reff@jnl{Soviet~Ast.}}     
\def\ssr{\reff@jnl{Space~Sci.Rev.}}     
\def\zap{\reff@jnl{ZAp}}                
\def\nat{\reff@jnl{Nature}}             

\renewcommand{\det}{\mathrm{det}~}

\newcommand{\fref}[1]{Fig.~\ref{#1}}

\newcommand{\sref}[1]{Section~\ref{#1}}
\newcommand{\eref}[1]{Equation~\ref{#1}}

\newcommand{\glamroc}{\textsc{glamroc}\xspace}
\newcommand{\slacs}{{SLACS}\xspace}
\newcommand{\sls}{{SL2S}\xspace}
\newcommand{\ie}{\textit{i.e.}\xspace}
\newcommand{\eg}{\textit{e.g.}\xspace}
\newcommand{\galfit}{\textsc{galfit}\xspace}
\newcommand{\JDEM}{{JDEM}\xspace}
\newcommand{\SNAP}{{SNAP}\xspace}
\newcommand{\LSST}{{LSST}\xspace}
\newcommand{\Euclid}{{Euclid}\xspace}
\newcommand{\hst}{{\it HST}\xspace}

\newcommand{\bd}{\begin{displaymath}}
\newcommand{\ed}{\end{displaymath}}
\newcommand{\be}{\begin{equation}}
\newcommand{\ee}{\end{equation}}
\newcommand{\beaa}{\begin{eqnarray*}}
\newcommand{\eeaa}{\end{eqnarray*}}
\newcommand{\bea}{\begin{eqnarray}}
\newcommand{\eea}{\end{eqnarray}}
\newcommand{\bs}[1]{\boldsymbol{#1}}

\newcommand{\zd}{z_{\rm d}}
\newcommand{\zs}{z_{\rm s}}
\newcommand{\Dd}{D_{\rm d}}
\newcommand{\Ds}{D_{\rm s}}
\newcommand{\Dds}{D_{\rm ds}}

\newcommand{\rc}{r_{\rm c}}
\newcommand{\Ie}{I_{\rm e}}
\newcommand{\re}{r_{\rm e}}

\title[Exotic Gravitational Lenses]
{An Atlas of Predicted Exotic Gravitational Lenses}

\author[]{%
Gilles Orban de Xivry,$^{1,2,3,4}$ 
Phil Marshall$^{1}$\\
$^{1}$ Physics Department, University of California, Santa Barbara, CA 93106, USA\\
$^{2}$ Observatoire Midi-Pyr\'en\'ees, 14 Avenue Edouard Belin, 31400 Toulouse, France \\
$^{3}$ Institut Sup\'erieur de l'A\'eronautique et de l'Espace, 10 Avenue Edouard Belin, 31055 Toulouse, France \\
$^{4}$ Facult\'e des Sciences Appliqu\'ees, Universit\'e de Li\`ege, Chemin des Chevreuils 1, Sart Tilman, 4000 Li\`ege, Belgique 
}

\begin{document}

\date{\today}

\pagerange{000--000} \pubyear{2008}

\maketitle

\label{firstpage}


\begin{abstract} 

Wide-field optical imaging surveys will contain tens of thousands of new
strong gravitational lenses. Some of these will have new and unusual image
configurations, and so will enable new applications: for example, systems with
high image multiplicity will allow more detailed study of galaxy and group
mass distributions, while high magnification is needed to super-resolve the
faintest objects in the high redshift universe.  
Inspired by a set of six unusual lens systems [including five selected from the
Sloan Lens ACS (SLACS) and Strong Lensing Legacy (SL2S) surveys, 
plus the cluster Abell 1703], we consider several
types of  multi-component, physically-motivated lens potentials, and use the
ray-tracing code \glamroc to predict exotic image configurations.  We also 
investigate the effects of galaxy source profile and size, and use realistic
sources to predict observable magnifications and estimate very approximate
relative cross-sections.
We find that lens galaxies with misaligned disks and bulges produce
swallowtail and butterfly catastrophes, observable as ``broken'' Einstein
rings.  Binary or merging galaxies show elliptic umbilic catastrophes, leading
to an unusual Y-shaped configuration of 4 merging images. While not the
maximum magnification configuration possible, it offers the possibility of
mapping the local small-scale mass distribution. We estimate 
the approximate abundance of
each of these exotic galaxy-scale lenses to be $\sim1$ per all-sky survey.
In higher mass systems, a wide range of caustic structures are expected, as
already seen in many cluster lens systems. We interpret the central ring and
its counter-image in Abell 1703 as a ``hyperbolic umbilic'' configuration,
with total magnification $\sim 100$ (depending on source size).
The abundance of such configurations is also estimated to
be $\sim 1$ per all-sky survey.
\vspace{2\baselineskip}
\end{abstract}



\section{Introduction}

Strong gravitational lenses, by producing multiple images and high
magnification, can teach us much about galaxies, how they form and how they
evolve. The numerous constraints allow us to determine the mass distribution
of galaxies, groups and clusters,  including invisible dark substructure.  On
the other hand, strong lens systems, by the high magnification they provide,
can act as cosmic telescopes,  leading to far more observed flux and thereby
opening unique windows into the early universe.

Until recently, apart from a very few exceptions~\citep[\eg][]{Rus++01}, the
lens configurations  observed have been Einstein ring, double and quadruple
image configurations. As the number of discovered lenses  increases, these
configurations will become commonplace and truly exotic ones should come into
view. Indeed, several recent strong lens surveys have proved  to be just big
enough to contain the first {\it samples} of  complex  galaxy-scale and
group-scale strong lenses capable of producing exotic high magnification image
patterns. One such survey is the Sloan Lens ACS survey  \citep[\slacs,
][]{Bol++06,Bol++08}, which uses the SDSS spectroscopic survey to identify
alignments of relatively nearby massive elliptical galaxies with star-forming
galaxies lying behind them, and then the Advanced Camera for Surveys (ACS) 
onboard the {\it Hubble Space Telescope} (\hst) to provide a high resolution image to
confirm that multiple-imaging is taking place. Another is the Strong Lens
Legacy Survey \citep[\sls, ][]{Cab++07}, which uses the 125-square degree,
multi-filter CFHT Legacy Survey images to find rings and arcs by their colors
and morphology. Together, these surveys have discovered more than 100 new
gravitational lenses. 

Moreover, strong lensing science is about to enter an exciting new phase: very
wide field imaging surveys are planned for the next decade that  will contain
thousands of new lenses for us to study and exploit. For example, the Large
Synoptic Survey Telescope \citep[\LSST, ][]{Ive++08} will provide deep
($\approx 27$th AB  magnitude), high cadence imaging of 20,000 square degrees
of the Southern sky in 6 optical filters, with median seeing $0.7$~arcseconds.
Similarly, the \JDEM and \Euclid \citep{Lau++08}  space observatories are
planned to image a similar amount of sky at higher resolution (0.1--0.2
arcseconds), extending the wavelength coverage into the near infra-red (the
\SNAP design, \citeauthor{Ald++04}~\citeyear{Ald++04}, 
with its weak lensing-optimised  small pixels
and large field of view, would be particularly effective at finding strong
lenses in very large numbers). With such panoramic surveys on the drawing
board, we may begin to consider objects that occur with number densities on
the sky of order $10^{-4}$ per square degree or greater.

In order to understand the  contents of these forthcoming surveys, it is 
useful to have an overview of the configurations, magnifications and
abundances we can expect from exotic gravitational lenses, namely those with 
complex mass distributions and fortuitously-placed sources. Several attempts
have already been made in  this direction,  generally focusing on individual
cases. This field of research is centred on the notion of critical points
(``higher-order catastrophes'' and ``calamities'') in the 3-dimensional space
behind a massive object. These points are notable for the transition in image
multiplicity that takes place there, and have been studied in a number of
different  situations by various authors in previous, rather theoretical
works~\citep[\eg][]{568B,KKB92,KSB94,KMW00,E+W01}. 

Here, we aim to build on this research to provide  an atlas to {\it illustrate
the most likely observable exotic lenses}. Anticipating the rarity of exotic
lenses, we focus on the most readily available sources, the faint blue
galaxies \citep[\eg][]{Ell97}, and investigate the effect of their light
profile and size on the images produced. Our intention is to improve our
intuition of the factors that lead to exotic image configurations. What are
the magnifications we can reach?  Can we make a first attempt at estimating
the abundances of lenses showing  higher-order catastrophes?  Throughout this
paper, we emphasize the critical points  associated with plausible physical
models, and try to remain in close contact with the observations. 

To this end, we select six complex lenses (three from the \slacs survey, two
from the \sls survey, and the galaxy cluster Abell 1703 from
\citeauthor{L08}~\citeyear{L08})  to motivate and inspire an atlas of exotic
gravitational lenses.  By making qualitative models of these systems we
identify points in space where,  if a source were placed there, an exotic lens
configuration would appear. We then extend our analysis to  explore the space
of lens model parameters: for those systems showing interesting caustic
structure, that is to say presenting higher-order catastrophes, we study the
magnifications we can reach, and make very rough  estimates of the
cross-sections and abundances of such lenses.

This paper is organised as follows. In \sref{sec:theory} we provide a review
of the basic theory of  gravitational lenses and their singularities, and then
describe in \sref{sec:magsources} our methodology for estimating the
source-dependent magnifications and cross-sections of gravitational lenses. 
We then present, in \sref{sec:targets}, our sample of unusual lenses,
explaining why we find them interesting. Suitably inspired, we then move on
the atlas itself,  focusing first on simple, \ie  galaxy-scale, lens models
(\sref{sec:simplemodels}), and then on more complex, \ie group-scale and
cluster-scale lenses (\sref{sec:complexmodels}). Finally, we use the
cross-sections calculated in Sections~\ref{sec:simplemodels} 
and~\ref{sec:complexmodels}  to make the first (albeit crude) estimates of the
abundance of exotic lenses for the particular cases inspired by our targets in
\sref{sec:abundance}.  We 
present our conclusions in~\sref{sec:conclude}. When calculating distances we
assume a general-relativistic Friedmann-Robertson-Walker (FRW) cosmology with
matter-density parameter $\Omega_m = 0.3$, vacuum energy-density parameter
$\Omega_{\Lambda}=0.7$, and Hubble parameter $H_0 = 70$ km s$^{-1}$
Mpc$^{-1}$.


\section{Gravitational lensing theory}\label{sec:theory}

This section is divided into two parts.  We first briefly review the
relevant basic equations in lensing theory, following the notation of
\citet{SKW06}.   We then introduce the qualitative elements of lens
singularity theory \citep{2001stgl.book.....P} relevant to this study.


\subsection{Basics}
  
\begin{figure}
\centering\includegraphics[width=5cm]{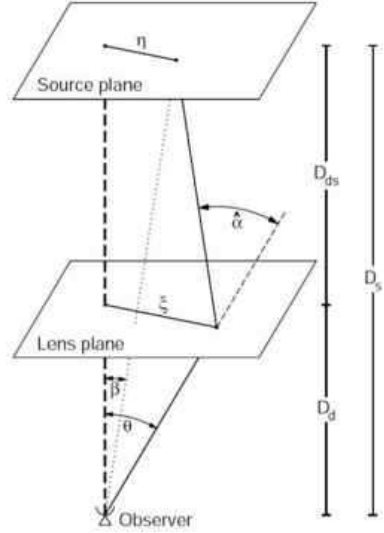}
%
\caption{Schema of a typical, single plane, gravitational lens system 
-- from \citet[][Figure~12, p.\ 20, reproduced with permission
\copyright ESO]{SKW06}.}
\label{fig:lens}
\end{figure}

We reproduce in \fref{fig:lens} a sketch, from \citet{SKW06}, of a typical
gravitational lens system.  In short, a lens system is a mass concentration at
redshift $\zd$ deflecting the light rays from a source at redshift $\zs$. The
source and lens planes are defined as planes perpendicular to the optical
axis.

The lens equation relates the position of the source  to its observed position
on the sky; in the notation of the sketch,  $\bs{\beta}  = \bs{\theta} -
\bs{\alpha}(\bs{\theta})$ where $\bs{\beta} = (\beta_1,\beta_2)$ denotes the
angular position of the source, $\bs{\theta} = (\theta_1,\theta_2)$ the
angular position of the image and $\boldsymbol{\alpha}(\boldsymbol{\theta}) =
\frac{\Dds}{\Ds} \boldsymbol{\hat{\alpha}}(\Dd \boldsymbol{\theta})$  is the
scaled deflection angle. $\Dd$, $\Ds$, and $\Dds$ are the angular diameter
distances to the lens, to the source, and between the lens and the source
respectively.

The dimensionless surface mass density or convergence is defined as 
\be
\kappa(\bs{\theta}) \equiv \dfrac{\Sigma(\Dd \bs{\theta})}{\Sigma_{\rm crit}} \quad
\text{with} \quad \Sigma_{\rm crit} \equiv \dfrac{c^2}{4 \pi G}\dfrac{\Ds}{\Dd D_{ds}}
\ee
where the critical surface mass density $\Sigma_{\rm crit}$ delimits the 
region producing multiple images (within which $\langle \kappa \rangle \geq
1$).  The critical surface mass density is, therefore, a characteristic value 
for the surface mass density which is the dividing line between ``weak'' and 
%
%
``strong'' lenses \citep[][p.\ 21]{SKW06}.

Gravitational lensing conserves surface brightness, but the shape
and size  of the images differs from those of the source. The distortion of
the images is described by the Jacobian matrix,
\begin{align}
{\bf J}(\bs{\theta})  &= \dfrac{\partial \bs{\beta}}{\partial \bs{\theta}} = \left( \delta_{ij} - \dfrac{\partial^2 \psi(\bs{\theta}}{\partial \theta_i \partial \theta_j} \right) \\
		&= \left( \begin{array}{cc}
				1-\kappa - \gamma_1   &  - \gamma_2 \\
				-\gamma_2	      &  1- \kappa + \gamma_1
			    \end{array}
		\right),
\end{align}
where we have introduced the components of the shear $\gamma \equiv \gamma_1 +
i \gamma_2$. Both the shear and the convergence can be written as combinations
of partial second derivatives of a lensing potential~$\psi$:
\bea
\kappa &=& \dfrac{1}{2} (\psi_{,11} + \psi_{,22}), \\
\gamma_1 &=& \dfrac{1}{2} (\psi_{,11} - \psi_{,22}), \quad \gamma_2 = \psi_{,12}.
\eea
The inverse of the Jacobian, ${\bf M}(\bs{\theta}) = {\bf J}^{-1}$, 
is called the magnification matrix,
and the magnification at a point $\bs{\theta_0}$ within 
the image is $|\mu (\bs{\theta_0})|$, where $\mu = \det {\bf M}$.

The total magnification of a point source at position $bs{\beta}$ is 
given by the sum of the magnifications
over all its images,
\be\label{eq:magtot}
\mu_{\rm p} (\bs{\beta}) = \sum_i |\mu(\bs{\theta_i})|,
\ee
The magnification of real sources with finite extent is given 
by the weighted mean of $\mu_{\rm p}$ over the
source area,
\be \label{eq:realmag}
\mu = \dfrac{\int d^2 \beta ~ I^{(s)} (\bs{\beta})\mu_{\rm p} (\bs{\beta})}{\left[ \int d^2 \beta ~I^{(s)} (\bs{\beta}) \right]}, 
\ee
where $I^{(s)} (\bs{\beta})$ is the surface brightness profile of the source.


\subsection{Singularity theory and caustic metamorphoses in gravitational
lensing} \label{ssec:ccc}

The critical curves are the smooth locii in the image plane on which the
Jacobian vanishes and the point magnification is formally infinite.  The
caustics are the corresponding curves, not necessarily smooth,  obtained
by mapping the critical curves into the source plane via the lens
equation. 

%
A typical caustic presents cusp points connected by fold lines,
which are the  generic singularities associated with strong lenses
(\citet{KKB92}). By this we mean that these two are stable in the source
plane,  \ie the fold and cusp are present in strong lenses for all $\zs
> \zd$.   By considering a continuous range of source-planes, the folds
can be thought of as surfaces in three-dimensional space extending
behind the lens, while cusps are ridgelines on these surfaces. 
%
Indeed, this is true of all caustics: they are best thought of as 
three-dimensional structures (surfaces)
lying behind the lens, which are sliced at
a given source redshift and hence renderable as lines on a 2-dimensional
plot. We will show representations of such three-dimensional caustics
later.
Other
singularities can exist but they are not stable, in the sense that they
form at a specific source redshift, or in a narrow range of source 
redshifts: they represent single points in three-dimensional space. The
unstable nature of these point-like singularities can be used to put
strong constraints on lens models  and can lead to high magnifications 
\citep[][]{Bag01} -- it is these exotic lenses that are the subject of
this paper. 

%
Excellent introductions to gravitational lenses and their critical
points can be found in the books by 
\citet[][where chapter~6 is particularly relevant]{SEF92} and
\citet{2001stgl.book.....P}; here we provide a brief overview.  
We will consider at various points the following critical points 
(known as ``calamities'' or ``catastrophes''):  lips, beak-to-beak, 
swallowtail, elliptic umbilic and hyperbolic umbilic. 
%
%
With these singularities, we associate metamorphoses: the beak-to-beak
and lips calamities, and the swallowtail catastrophe, mark the
transition (as the source redshift varies) from zero to two cusps, 
the elliptic umbilic
catastrophe from three to three cusps, and the hyperbolic umbilic
catastrophe involves the exchange of a cusp between two caustic lines
\citep[][section 6.3]{SEF92}.  
These are the five
types of caustic metamorphosis that arise generically from taking planar
slices of a caustic surface in three dimensions. 
As well as these, we will study the butterfly metamorphosis, which 
involves a
transition from one to three cusps. It can be constructed from
combinations of swallowtail metamorphoses; lenses that show butterfly
metamorphoses will have high
image multiplicities, hence our consideration of them  as ``exotic''
systems. \citet[][section 9.5]{2001stgl.book.....P} give an extensive
discussion of these caustic metamorphoses, and we encourage the reader
to take advantage of this.

\begin{figure}
\centering
\includegraphics[width=8cm]{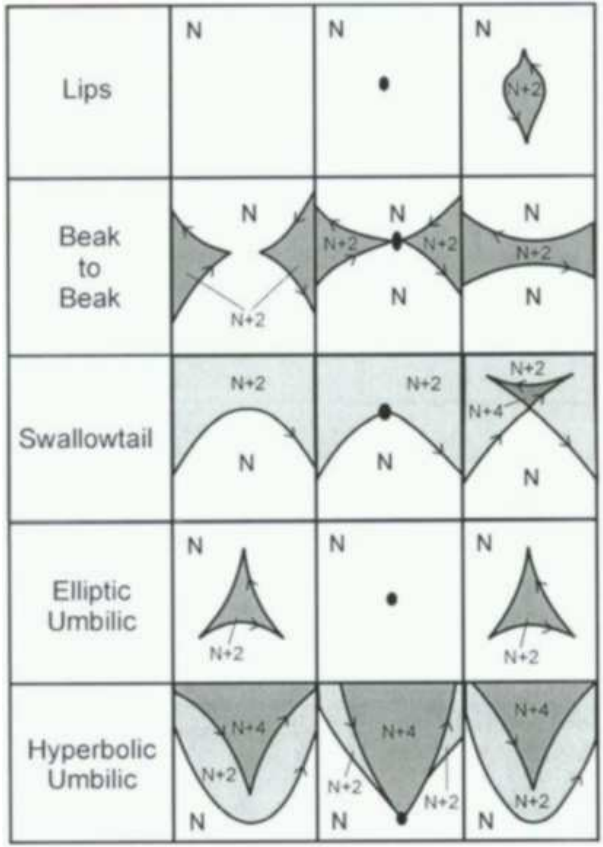}
%
%
\caption{Caustic metamorphoses and their lensed image multiplicities.
The points determine the critical points. The numbers $N$, $N+2$, $N+4$
indicate how many lensed images are produced of a light source lying in
the given region. The caustic curves are oriented so such that, locally,
light sources to the left generically have two more lensed images than
those to the right. Adapted from Figure~9.19 in
\citet[][p.\ 382]{2001stgl.book.....P}.
With kind permission of Springer Science$+$Business Media.}
\label{fig:critical-points}
\end{figure}

\begin{figure}
\centering
\includegraphics[width=8cm]{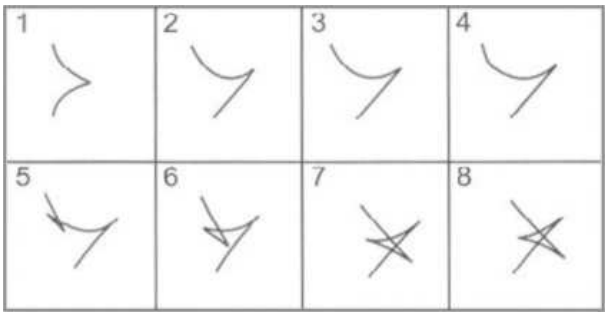}
\caption{A butterfly caustic metamorphosis, including 
a swallowtail as an
intermediate step. Adapted from Figure~9.17 in
\citet[][p.\ 379]{2001stgl.book.....P}.
With kind permission of Springer Science$+$Business Media.}
\label{fig:butterfly-formation}
\end{figure}

In \fref{fig:critical-points} and \fref{fig:butterfly-formation},
which are reproduced and extended from 
\citet{2001stgl.book.....P}, we sketch these critical points and their
related metamorphoses, showing the changes in image multiplicity~$N$ at
the transition. 
%
The features in the caustics after the swallowtail and
butterfly metamorphoses are also referred to as, respectively,
swallowtails and butterflies after their respective shapes.
A caustic shrinking to an
elliptic umbilic catastrophe and then appearing again is called a
``deltoid'' caustic \citep{2006MNRAS.366...39S}.  We note that varying
source redshift is not the only way to bring about a metamorphosis: the
independent variable  could also be the lens component separation in a
binary lens, the  isopotential ellipticity, the density profile slope
and so on. In this paper we aim to illustrate these metamorphoses in
plausible physical situations.


A number of theoretical studies of these critical points have already
been performed.
\citet{568B} undertook the first major study of complex gravitational lens 
caustics, and provided a classification of
gravitational lens images. 
Following this initial foray, 
\citet{KKB92} studied the lips and beak-to-beak ``calamities,'' 
applied to the long straight arc in Abell 2390.
\cite{KSB94} presented an analytical treatment of several 
isothermal elliptical gravitational lens models,
including the singular isothermal ellipsoid (SIE) and the
non-singular isothermal ellipsoid (NIE).
\citet{KMW00} focused on the most common strong lenses, namely 
isothermal elliptical density galaxies in
the presence of tidal perturbations.
\citet{E+W01} presented, via
the use of boxiness and skewness in the lens isopotentials, 
the formation of swallowtail and
butterfly catastrophes, while most recently \citet{S+E08} treated the case of binary galaxies
case, outlining in particular the formation of the 
elliptic umbilic catastrophe.

%

%
The mathematical definitions of the various catastrophes  
\citep[given in chapter 9 of][and
the above papers, along with some criteria for these to
occur]{2001stgl.book.....P}, 
involve various spatial derivatives of the
Jacobian, as well as the magnification itself. This is the significance of the
``higher-order'' nature of the catastrophes: they are points where the lens
potential and its derivatives are very specifically constrained. This fact
provides strong motivation for their study -- lenses showing image
configurations characteristic of higher-order catastrophes should allow
detailed mapping of the lens local mass distribution to the images. In
addition, the  large changes in image multiplicity indicate that these
catastrophe points are potentially associated with very high image
magnifications, perhaps making such lenses very useful as cosmic telescopes.


\section{Source sizes and observable magnification}
\label{sec:magsources}

We are interested in the most likely instances of exotic lensing by systems
close to catastrophe points. The most common strong lenses are galaxies lensed
by galaxies, and so it is on these systems that we focus our attention. The
observability of strong lensing depends quite strongly on the magnification
induced -- and this in turn depends on the source size, as indicated in
\eref{eq:realmag}. In this section we investigate this effect.

We compute maps of the total magnification in the source plane, $\mu_{\rm p}
(\bs{\beta})$ in \eref{eq:magtot}, using the ray-tracing code \glamroc
(described in the appendix). In order to study the magnification of realistic
extended sources, we convolve these magnification maps by images of plausible 
sources, described by elliptically symmetric surface brightness distributions
with Sersic -- Gaussian, exponential or de Vaucouleurs -- profiles (see  the
appendix for more details).  We then use these to quantify how likely a given
magnification is to appear by calculating: a) the fraction of the
multiply-imaged source plane area that has  magnification above some
threshold,  and conversely b) the magnification at a given fractional 
cross-sectional area. 

\begin{figure}
\centering
\includegraphics[width=0.9\linewidth]{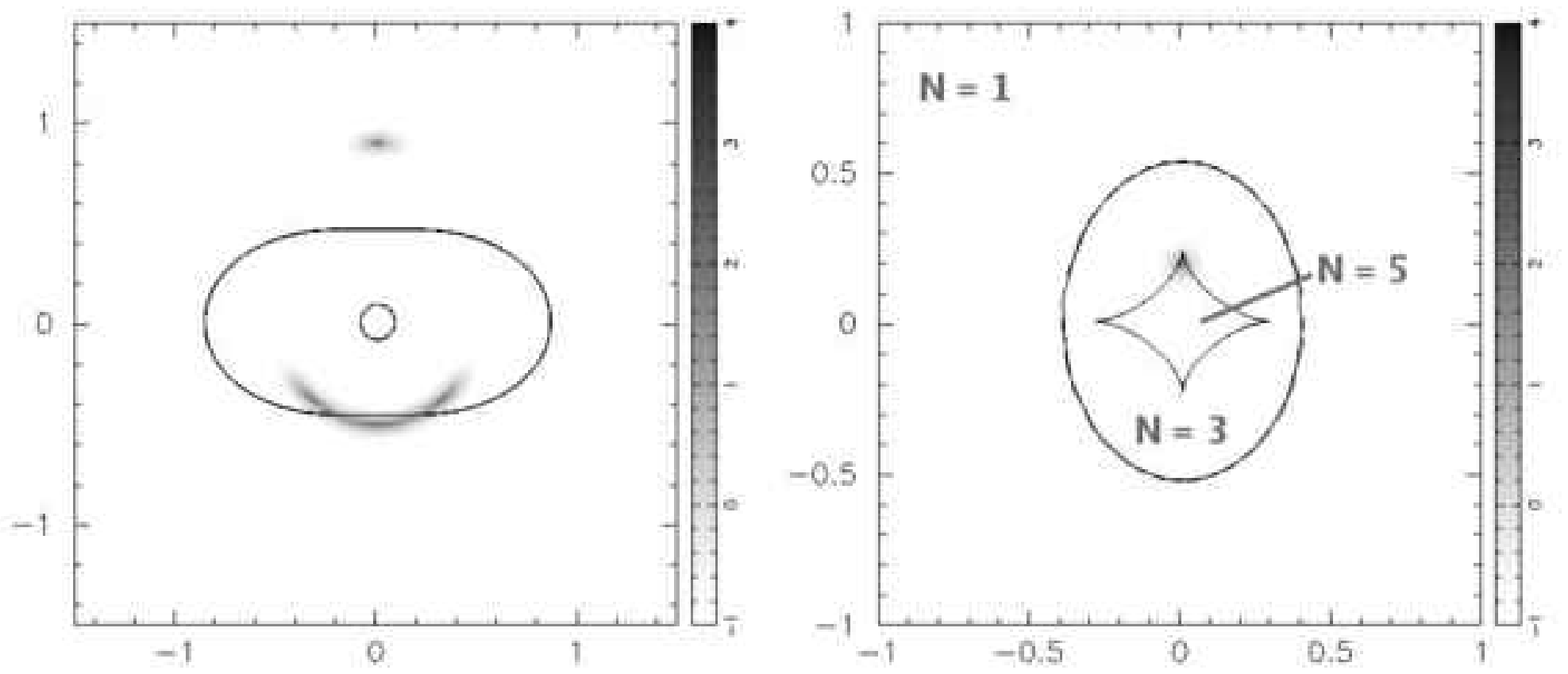}
\includegraphics[width=0.8\linewidth]{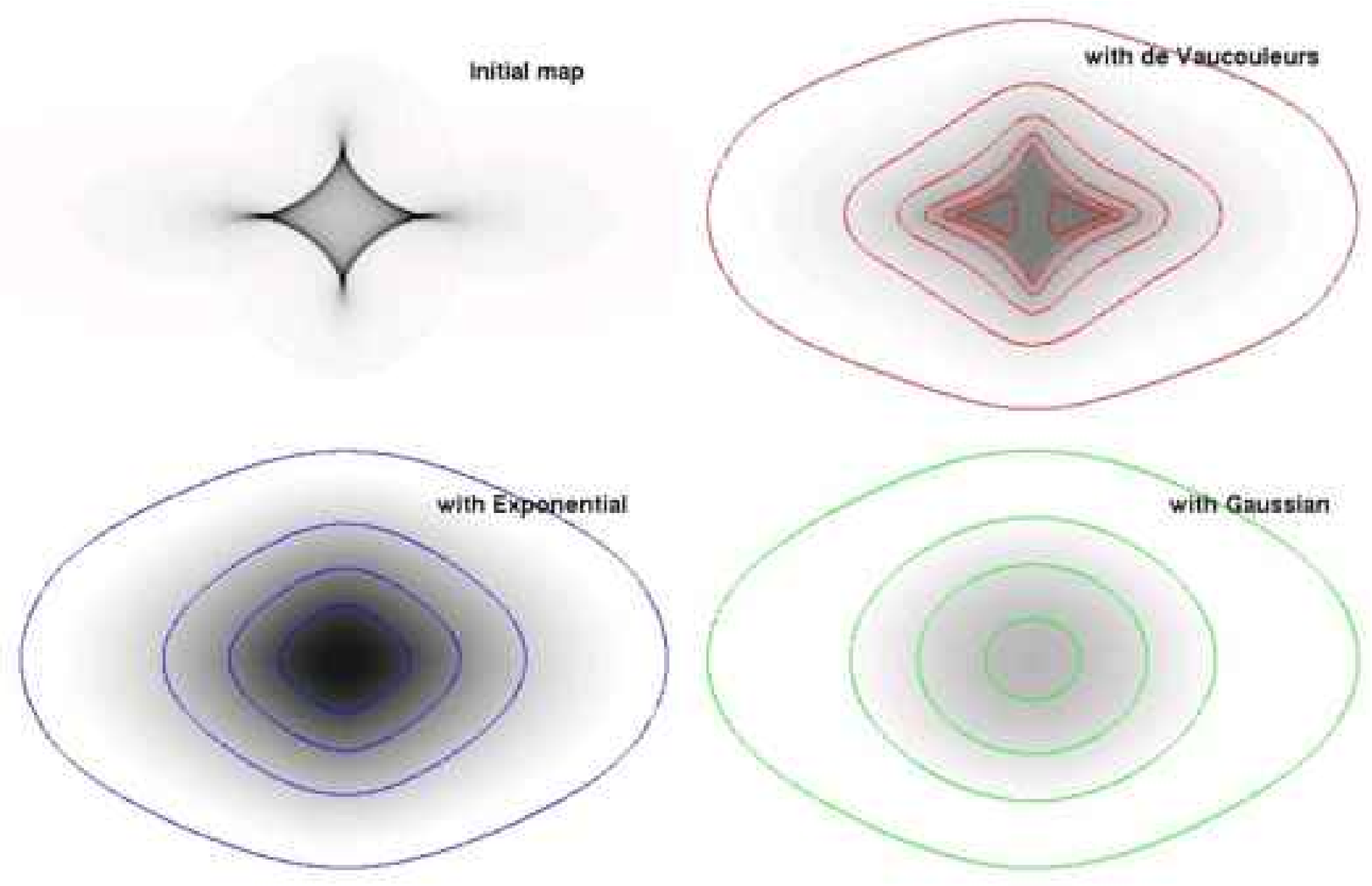}
\caption{
{\it Upper panel:} critical curves (left) and caustics (right) of an
elliptically-symmetric galaxy modelled by an NIE profile (appendix).
The source can be seen lying close to a cusp, giving three merging images, a
fainter counter-image and a de-magnified central image.
{\it Lower panel:} source-plane total
magnification maps convolved with four different source types:
point source, de Vaucouleurs bulge, exponential disk, Gaussian. 
Contours of equal magnification are shown on all 
three convolved maps, overlaid on the arbitrary
greyscale images: note the higher magnifications reached with the peakiest 
de Vaucouleurs profile.}
\label{fig:convol-effect}
\end{figure}

\begin{figure}
\centering
\includegraphics[width=7cm]{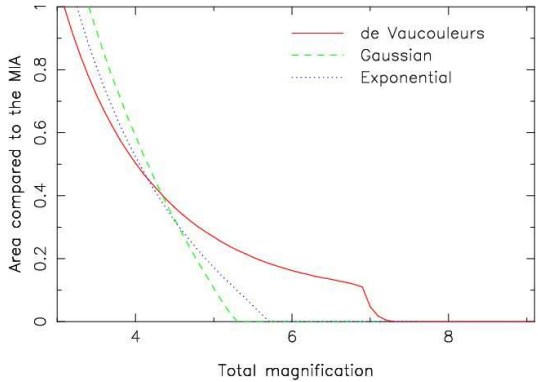}
%
\caption{The fractional cross-sectional area (relative to the 
multiple imaging area) as a function of threshold
total magnification, for a model NIE galaxy. 
The source half-light radius was fixed at 2.9kpc.}
\label{fig:cross-section-3profiles}
\end{figure}

We first use these tools to analyze the likely effect of source size and
profile on the observable magnification. We consider a standard lens model,
with a non-singular isothermal profile  (see appendix),  and a
fiducial  source at redshift $\zs = 2$ with half-light radius $2.9$kpc
\citep{Fer++04}.  The lens model and four magnification maps are presented in
\fref{fig:convol-effect}.
In \fref{fig:cross-section-3profiles} we then plot the corresponding
fractional cross-section area as a function of threshold total magnification.
We first remark that the area of multiple imaging corresponds approximately to
the area with magnification greater than 3, a value we will take as a point of
reference in later sections. The behaviour of the three source profiles is
fairly similar, with the peakier profiles giving rise to slightly higher
probabilities of achieving high magnifications. 
The two profiles we expect to be more representative of faint blue galaxies at
high redshift, the exponential and Gaussian, differ in their fractional
cross-sectional areas by less than~$10\%$. 

\begin{figure}
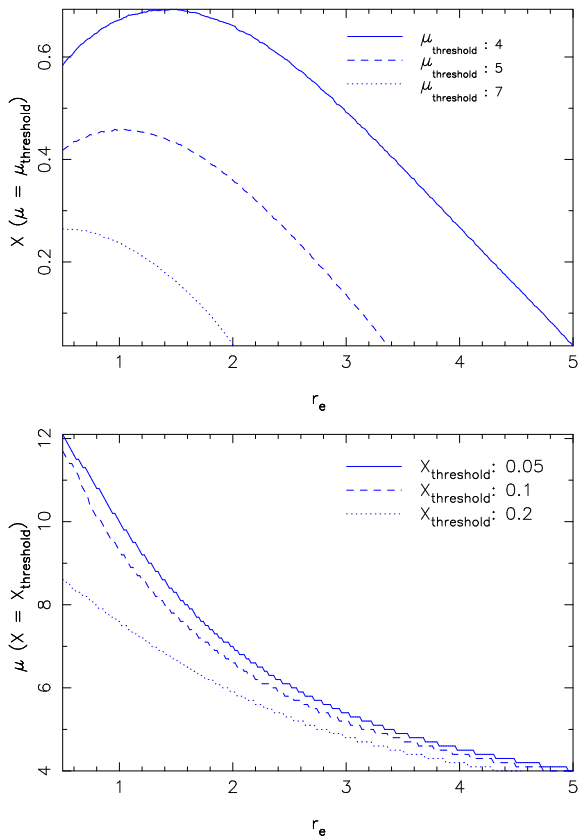

\centering
\subfigure{\includegraphics[width=0.9\linewidth]{figures/source-size-exp-fraction.eps}}
\subfigure{\includegraphics[width=0.9\linewidth]{figures/source-size-exp-mag.eps}}
\caption{{\it Upper panel:} for an exponential
profile source, we plot the fractional cross-sectional  area for
magnifications greater than various magnification thresholds, as a  function
of the half-light radius $\re$ (in kpc, $\zs=$). 
{\it Lower panel:} the corresponding magnification
for a given fractional  cross-sectional area threshold, also as a function of
$\re$.}
\label{fig:source-size}
\end{figure}

We now turn to influence of the half-light radius on the observable 
magnification, considering only the exponential profile and keeping the
fiducial source redshift of $\zs=2$.  We plot two graphs in
\fref{fig:source-size}: both as a function of half-light radius $\re$, we show
the fractional cross-sectional area greater than some magnification
threshold,  and also the magnification corresponding to a  given fractional
area threshold.  We study an inclusive range of half light-radii, from 0.5 kpc
to 5 kpc (\citet{Mar++07} give an example of a compact source of half-light
radius  $\simeq$0.6~kpc). 

We see that the magnification is, indeed, highly dependent on the
source size: the smaller the source, the
more likely a large total magnification is (lower panel).  
At small magnification, 
the fractional cross-sectional area reaches a maximum: larger sources smear
out the magnification map more, while 
smaller sources have already resolved the magnification map at that threshold
and cannot increase the interesting area.

Clearly the size of the source is more important than the form of the source
profile.  Henceforth we use only one profile, the exponential, and study three
different sizes: 0.5 kpc, 1.5 kpc and 2.5 kpc, keeping these constant in
redshift for simplicity in the analysis. At low redshift, these sizes are
somewhat smaller than the ones discussed in \citet{Fer++04}
(and the appendix). One might argue that such an arbitrary choice is not
unreasonable, given the uncertainty and scatter in galaxy sizes at high
redshift. The magnification bias noted here will  play some role in
determining the abundance of lensing events we observe; however, as we discuss
later, we leave this aspect to further work. In this paper we simply
illustrate the effects of source size on the observability of various lensing
effects, noting that 0.5--2.5~kpc represents a range of plausible source sizes
at $zs\gtrsim 1$.


\section{A sample of unusual lenses}\label{sec:targets}

The majority of the known lenses show double, quad and Einstein ring image
configurations well reproduced by simple SIE models. There are several cases
of more complex image configurations due to the presence of several sources;
there are also a few  more complex, multi-component lens systems, leading to
more complex caustic structures. One example of such a complex and interesting
lens is the B1359$+$154 system,  composed of 3 lens galaxies giving rise to 6
images \citet{Rus++01}. Another example is MG2016$+$112, whose quasar images
have flux and astrometric anomalies attributable to a small satellite
companion \citep[][and references therein]{Mor++09}.

We here show here six unusual lenses that have either been  recently
discovered or studied: three from the \slacs survey \citep{Bol++08}, two from
the \sls survey \citep{Cab++07}, and the rich cluster Abell 1703 \citep{L08}.
This set presents a wide range of complexity of lens structure. These lenses
are presented in this section but are not subject of further studies in the
following sections: their function is to illustrate the kinds of complex mass
structure that can arise in strong lens samples given a large enough imaging
survey. All six have extended galaxy sources, as expected for the 
majority of strong lenses detected in optical image survey data.

The three \slacs
targets are massive galaxies at low redshift ($\zd \sim 0.2$), 
that are not as regular as the majority of the SLACS sample \citep{Bol++08},
instead showing additional mass components such as a disk as well as a bulge, 
or a nearby satellite galaxy.
The two \sls targets are more complex still, being compact groups of galaxies.
The presence of several lensing objects makes them 
rather interesting to study and a good
starting points for our atlas.
The cluster Abell 1703 was first identified as a strong lens in the SDSS
imaging data by \citet{Hen++08}, and was recently studied with \hst by 
\citet{L08}. As we will see, elliptical clusters such as this
can give rise to a particular higher-order catastrophe, the hyperbolic
umbilic.


\begin{figure*}
\centering
\subfigure{\includegraphics[width=0.3\linewidth]{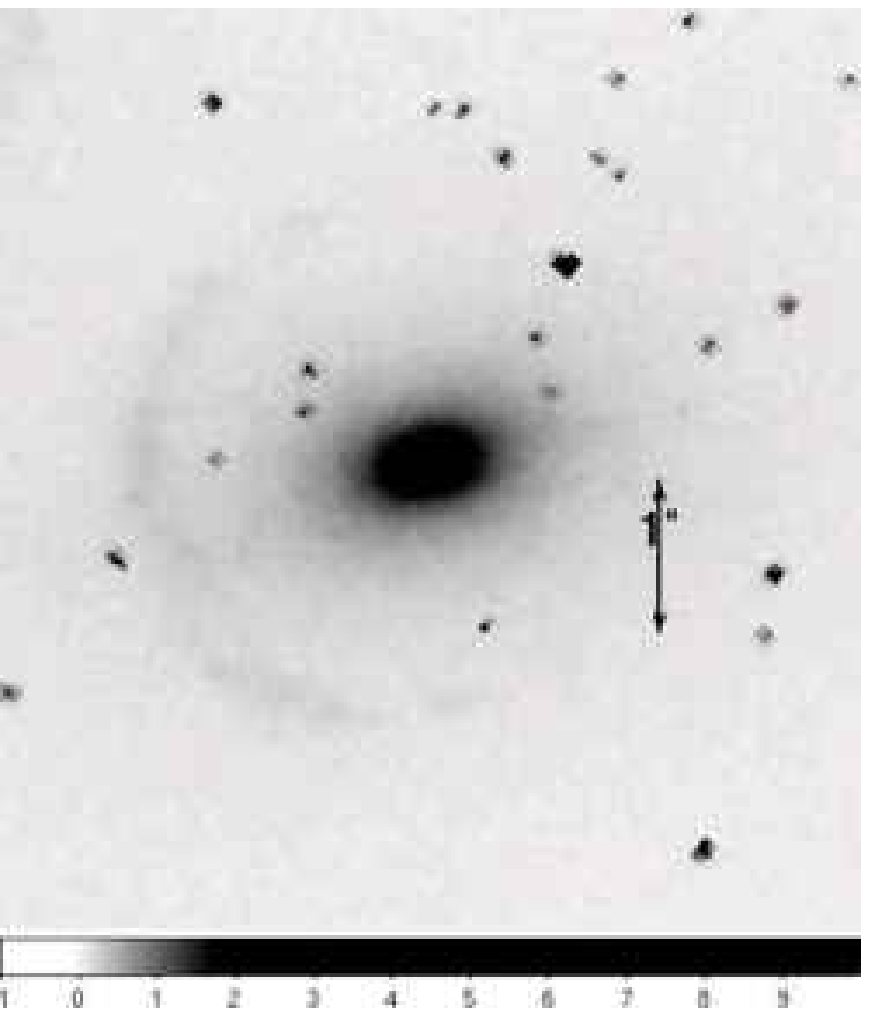}}
\subfigure{\includegraphics[width=0.315\linewidth]{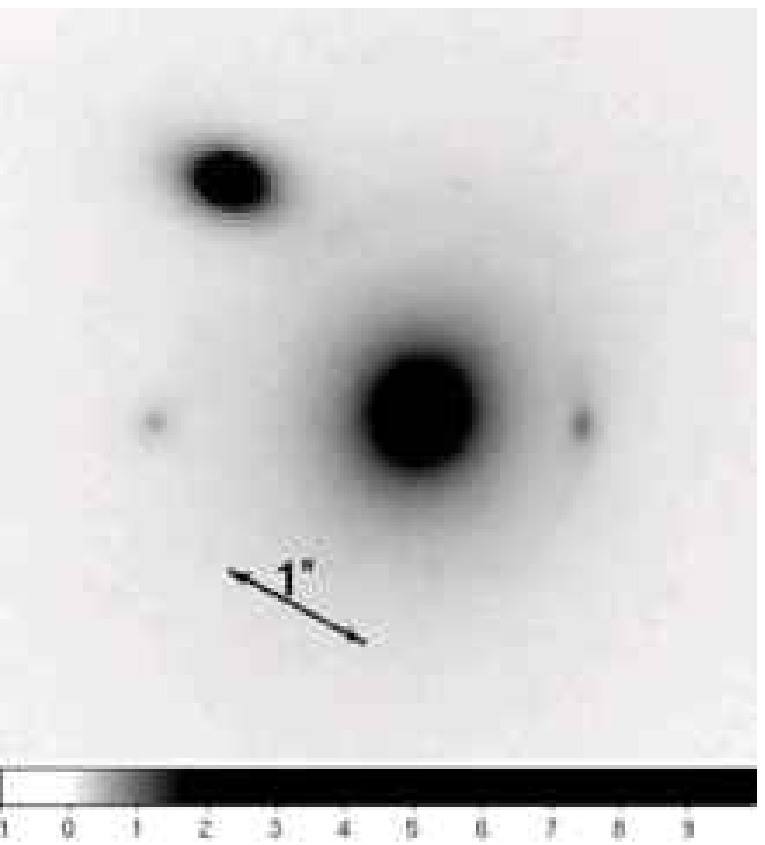}}
\subfigure{\includegraphics[width=0.3\linewidth]{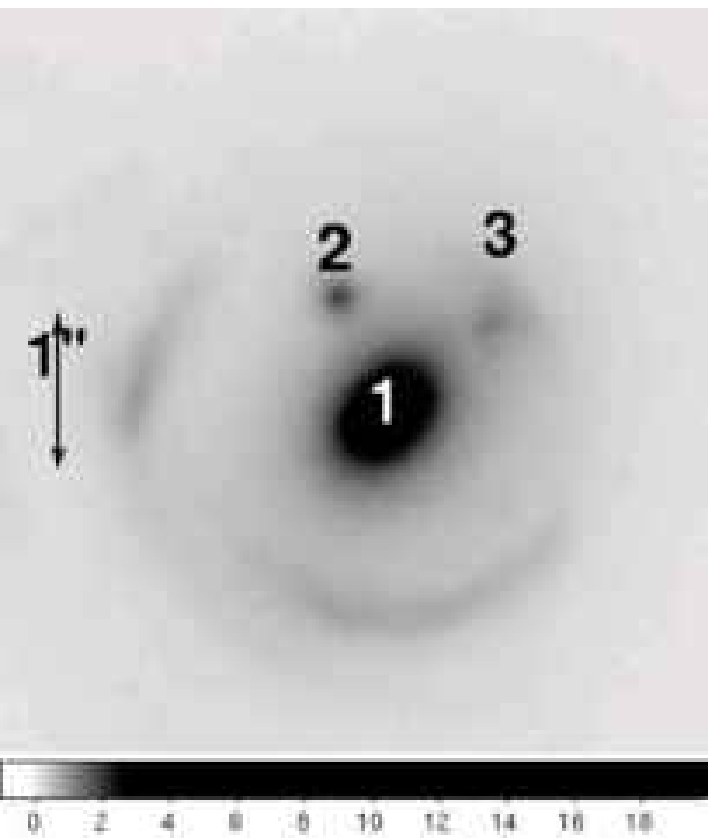}}
\hfill
\caption{Three unusual \slacs lenses. \hst/ACS images in the F814W filter,
approximately 6~arcsec on a side. {\it Left:} SDSSJ1100$+$5329; 
{\it centre:} SDSSJ0808$+$4706; {\it right:} SDSSJ0956$+$5100. 
In this final
image, we denote: 1) the main lens galaxy; 2) the satellite lens galaxy;
3) the counter-image to the extended arc system.}
\label{fig:SDSS}
\end{figure*}

\subsection{The three SLACS lenses}\label{ssec:slacs}

Three unusual \slacs 
lenses are presented in \fref{fig:SDSS}.
SDSSJ1100$+$5329 presents a large circular arc 
relatively far from the lens ($\sim1.5$''), broken at its centre. 
We were able to model this system with two SIE profiles 
of different ellipticities but with the same orientations, representing a
mass distribution comprising both a bulge and a disk. 
While a detailed modeling will be presented elsewhere, 
we take a cue from SDSSJ1100$+$5329 and study in 
\sref{sec:simplemodels} the catastrophe points 
present behind disk-plus-bulge 
lenses.

SDSSJ0808$+$4706 is a massive elliptical galaxy with a large satellite galaxy
lying just outside the Einstein ring and distorting its shape.
SDSSJ0956$+$5100 also has a satellite galaxy, this time 
inside the Einstein radius (the
arcs due to the lensing effect are at larger radius than the satellite). The
image component labeled 3 in \fref{fig:SDSS}, is considered 
to be due to the lensing effect: the system of images to which it belongs is
clearly distorted by the satellite. 
Inspired by these two systems, we study below the catastrophes lying 
behind binary lenses, a subject also investigated by \citet{S+E08a}.


\subsection{The two SL2S lenses}\label{ssec:sl2s}

\begin{figure}
\centering
\includegraphics[width=0.9\linewidth]{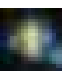}
\includegraphics[width=0.9\linewidth]{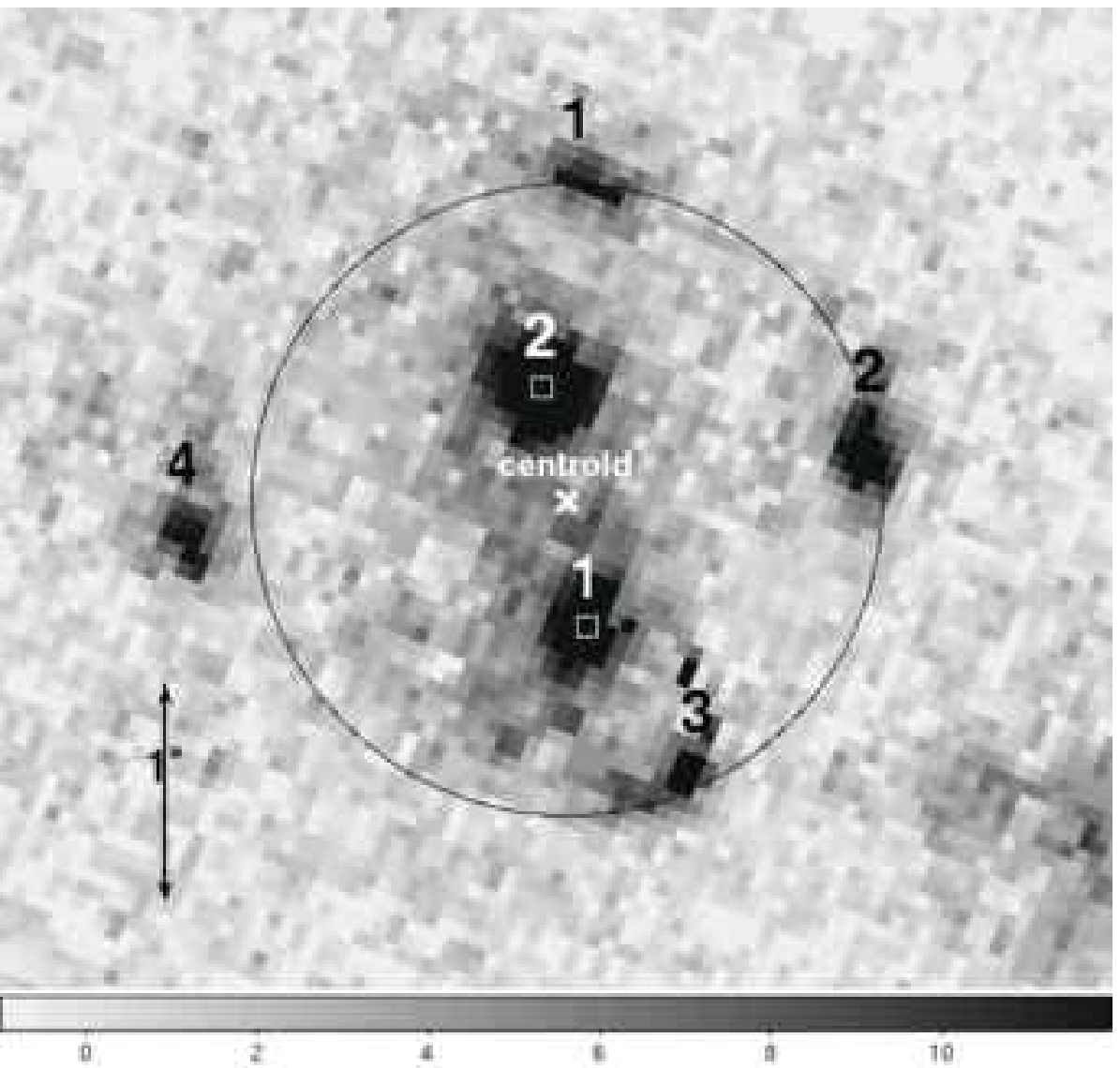}
\caption{SJ1405+5502, found in the CFHTLS data ({\it top}, 
g-r-i composite)
and followed-up with \hst/WFPC2 in Snapshot mode with the F606W filter
({\it bottom}).
Both images are approximately 6~arcsec on a side}
\label{fig:real140533}
\end{figure}

We consider two lenses from the \sls survey \citep{Cab++07}: SL2SJ0859$-$0345
and SL2SJ1405$+$5502. 
In SL2SJ1405$+$5502 we have an interesting binary system leading to an
asymmetric ``Einstein  Cross'' quad configuration (Figure~\ref{fig:real140533}). 
Binary systems, as we will
see in \sref{sec:complexmodels} below can develop exotic caustic structures
(as discussed by \citet{S+E08a}). We are able to  reproduce the image
configuration fairly well with a simple binary lens model,  and present its
critical curves and caustics in \sref{sec:complexmodels} below.

\begin{figure*}
\subfigure{\includegraphics[width=0.3\linewidth]{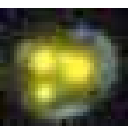}}
\subfigure{\includegraphics[width=0.3\linewidth]{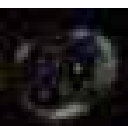}}
\subfigure{\includegraphics[width=0.3\linewidth]{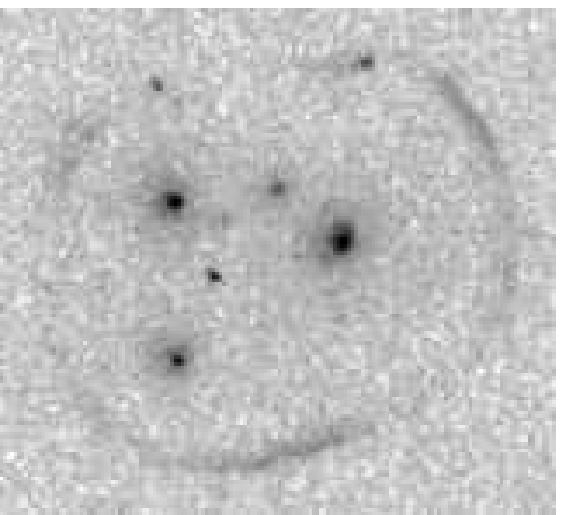}}
\caption{The complex lens system SL2SJ0859$-$0345. 
{\it Left:} g-r composite image from CFHTLS; {\it centre:} after subtraction of
the four bright lens components; {\it right:} at higher resolution, with
\hst/WFPC2 in the F606W filter. Images are approximately 15~arcsec on a side.}  
\label{fig:real085914}
\end{figure*}

SL2SJ0859$-$0345 is more complex, consisting of 4 or 5  lens components in a
compact group of galaxies \citep{Lim++08}. The image configuration is very
interesting: an oval-shaped system of arcs, with 6 surface brightness peaks
visible in the high resolution (low signal-to-noise) \hst/WFPC2 
(Wide-Field and Planetary Camera 2) image. It is in
this single-filter  image  that the fifth and faintest putative lens galaxy is
visible; it is not clear whether this is to be considered as a lens galaxy, or
a faint central image. In \fref{fig:real085914}  we show the WFPC2 and CFHTLS
images, and in the central panel a version of the latter with the four
brightest  lens components modelled with \galfit (in each filter
independently) and subtracted. The  central faint component remains undetected
following this process, due to its position between the sidelobes of the fit
residuals. We will improve this modelling elsewhere; here, in 
\sref{sec:complexmodels}, we simply investigate a qualitatively successful
4-lens galaxy model of the system, focusing again on the exotic critical curve
and caustic structures and the critical points therein.

\subsection{Abell 1703}\label{ssec:abell1703} 

\begin{figure}
\centering
\includegraphics[width=7cm]{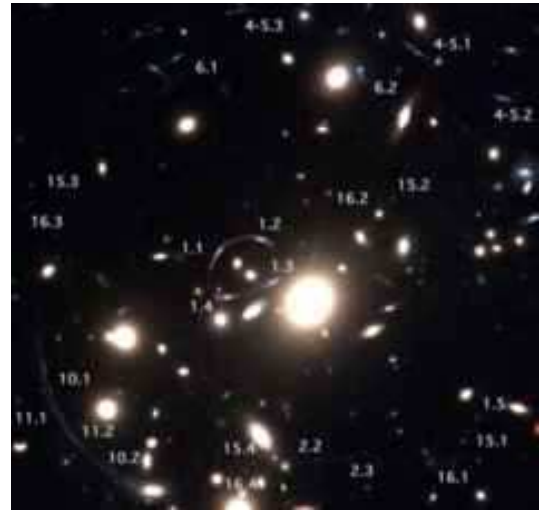}
\caption{Central part of the cluster Abell 1703, imaged with \hst/ACS and
reproduced from \citet{L08}.
We are interested in the bright central ring (images 1.1--1.4),  
and its counter-image (1.5).}
\label{fig:a1703}
\end{figure}

The strong lensing galaxy cluster Abell~1703 ($\zd=0.28$, shown in
\fref{fig:a1703}, reproduced from \citeauthor{L08}~\citeyear{L08}) exhibits an
interesting feature: a central ring with a counter-image. A naive
interpretation would link this central-quad feature to the two small galaxies
lying inside it; however, the large image separation across the quad and the
presence of the spectroscopically-confirmed  counter-image argue against this.
We instead interpret the image configuration as being due to a source lying
close to a hyperbolic umbilic catastrophe, a critical point anticipated for
elliptically-extended clusters such as this one~\citep{KKB92}.


\section{Simple lens models}\label{sec:simplemodels}

Motivated by our \slacs targets, we study in this section two different
``simple'' lens models: a single galaxy composed of both a bulge and a disk,
and a main galaxy with a companion satellite galaxy.  In this respect the
adjective ``simple'' could also be read as ``galaxy-scale.'' We will
investigate some ``complex,'' group-scale lens models in the next section.

Aa single isothermal elliptical galaxy yields at most four visible images;
higher multiplicities can be obtained when considering  deviations from such
profiles, leading to more than four images, as shown by \eg \citep{KMW00}.
Likewise, \citep{E+W01} showed how boxy or disky lens isodensity contours  can
give rise to six to seven image systems. Here we construct models from sums of
mass components to achieve the same effect,  associated with the swallowtail
and butterfly metamorphoses.


\subsection{Galaxies with both bulge and disk components}
\label{ssec:d+b}


\subsubsection{Model}

We model a galaxy with both a disc and a bulge using two concentric 
NIE models
with small core radii, different masses and different ellipticities. These two
NIE components are to be understood as a  bulge-plus-halo and a disk-plus-halo
components, as suggested by the ``conspiratorial'' results of the SLACS survey
\citep{Koo++06}. For brevity  we refer to them as simply the  bulge and disk
components.

We fix the total mass of the galaxy to be that of a fiducial massive
early-type galaxy, with overall velocity dispersion of $\sigma = 250$km
s$^{-1}$, and impose the lens Einstein radius to correspond to this value of
$\sigma$. We then divide the mass  as $\sigma_{\rm bulge} = 200 $ and
$\sigma_{\rm disk} = 150$km s$^{-1}$ to give a bulge-dominated galaxy with
prominent disk, such as that seen in SDSSJ1100$+$5329. Integrating the
convergence, we calculate the ratio of the mass of the bulge component to the
total mass, and find $f = \frac{M_{\rm E, bulge}}{M_{\rm E, total}} = 0.64$,
where $M_{\rm E}$ indicates the mass enclosed within  the Einstein radius.
This ratio is somewhat low for a lens galaxy, but not extremely so:
\citet{2003MNRAS.345....1M} studied a theoretical population of disky lenses,
and found that around one quarter of their sample had bulge mass fraction less
than this.

\begin{figure*}
\raggedright
\begin{minipage}[t]{0.95\linewidth}
  \begin{minipage}[t]{0.32\linewidth}
    \vspace{-0.94\linewidth}
    \raggedleft\includegraphics[width=0.78\linewidth]{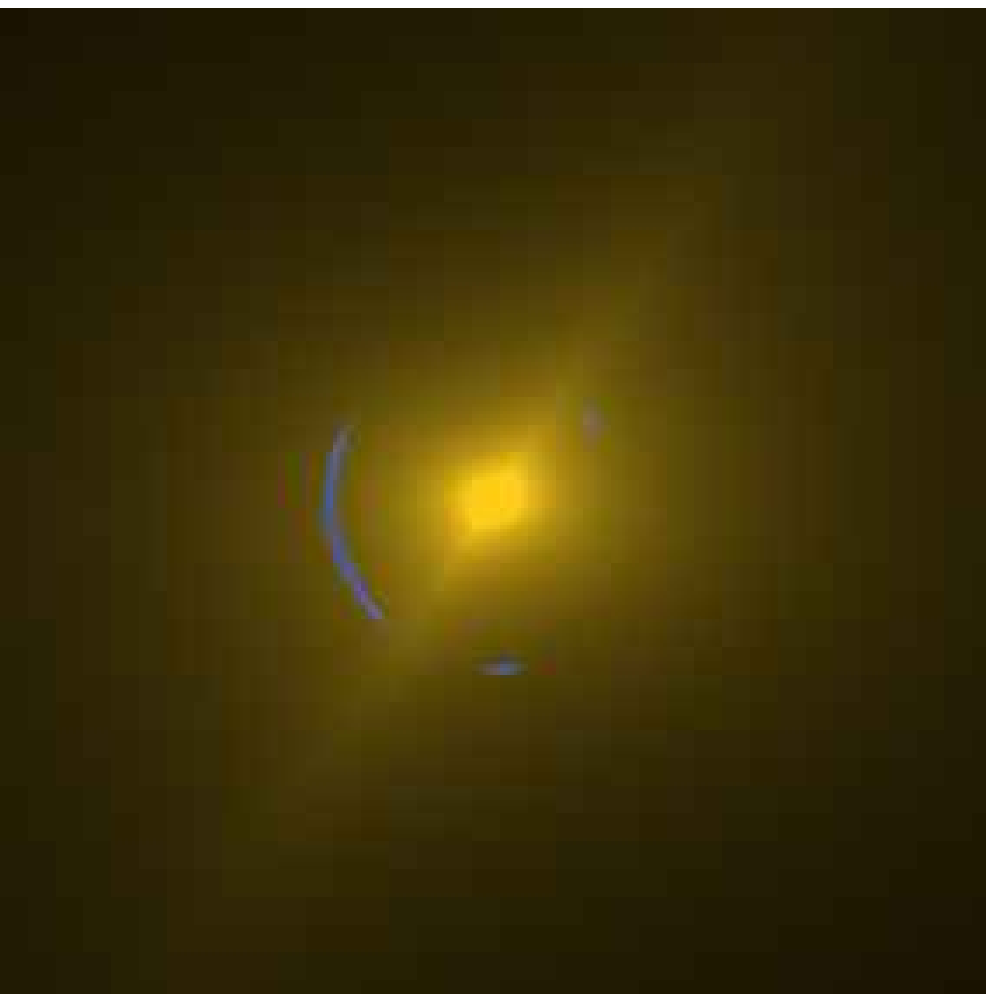}
  \end{minipage}\hfill
  \begin{minipage}[t]{0.67\linewidth}
    \centering\includegraphics[width=\linewidth]{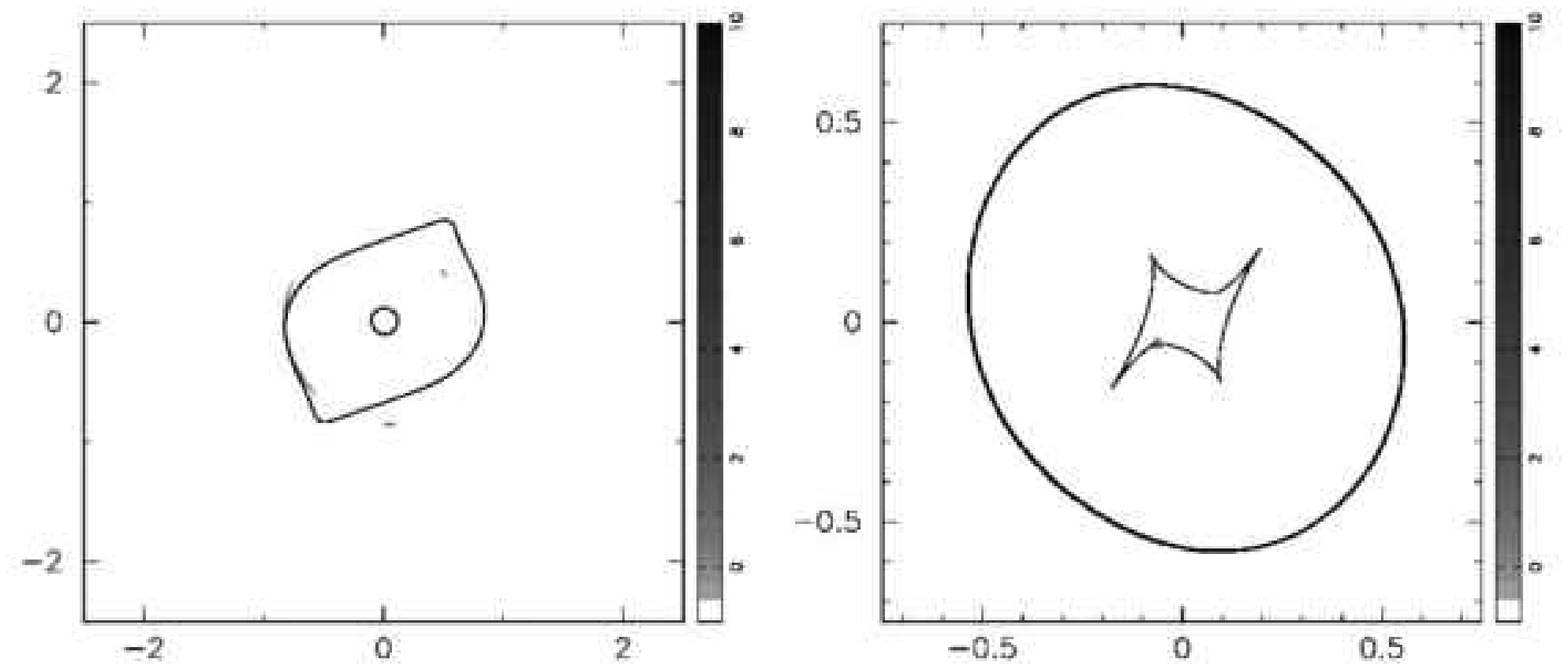}
  \end{minipage}
\end{minipage}
\begin{minipage}[t]{0.95\linewidth}
  \begin{minipage}[t]{0.32\linewidth}
    \vspace{-0.94\linewidth}
    \raggedleft\includegraphics[width=0.78\linewidth]{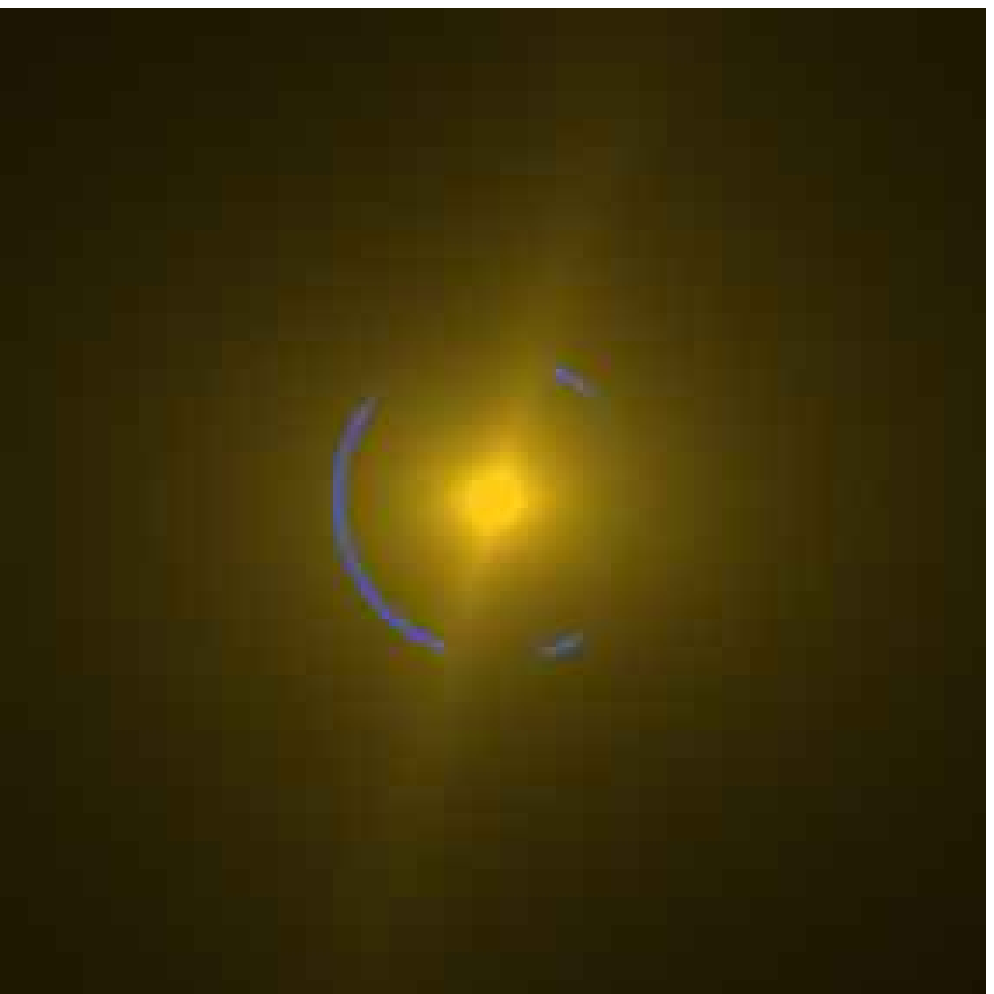}
  \end{minipage}\hfill
  \begin{minipage}[t]{0.67\linewidth}
    \centering\includegraphics[width=\linewidth]{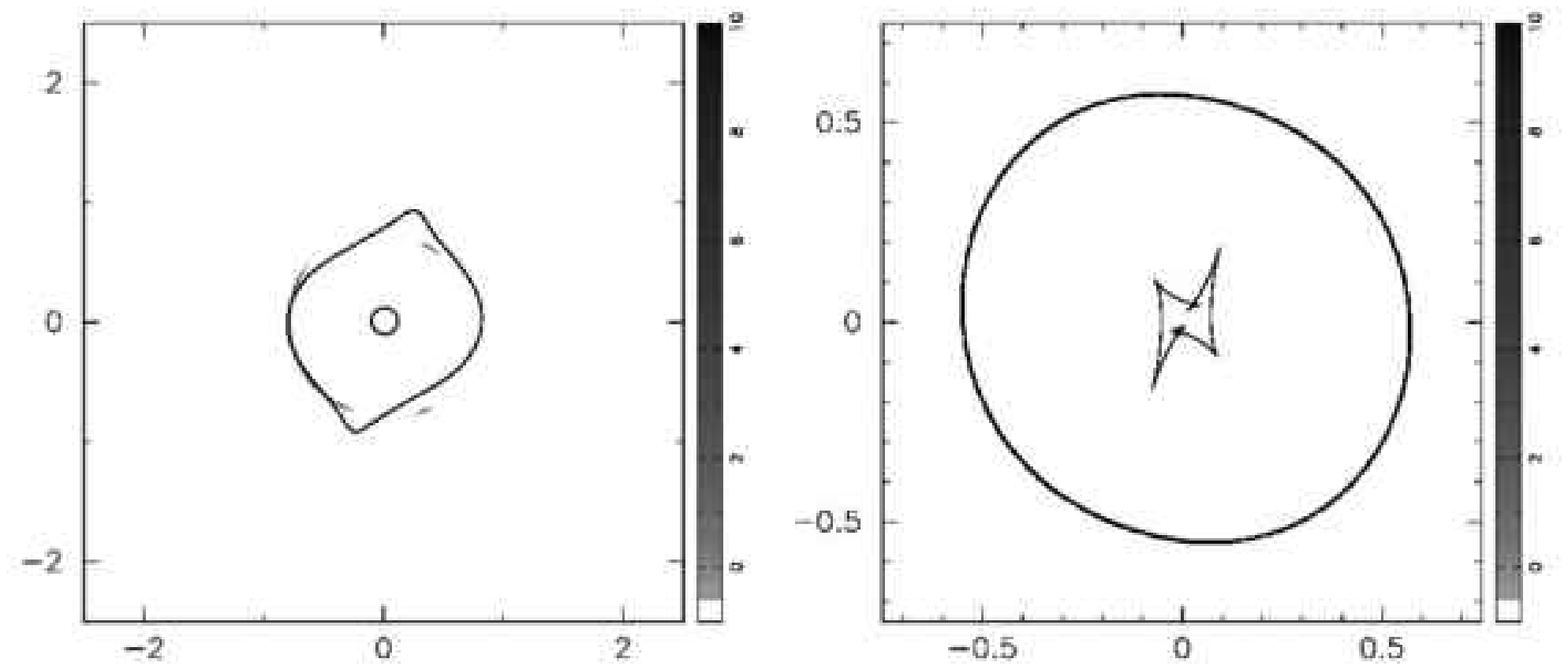}
  \end{minipage}
\end{minipage}
\begin{minipage}[t]{0.95\linewidth}
  \begin{minipage}[t]{0.32\linewidth}
    \vspace{-0.94\linewidth}
    \raggedleft\includegraphics[width=0.78\linewidth]{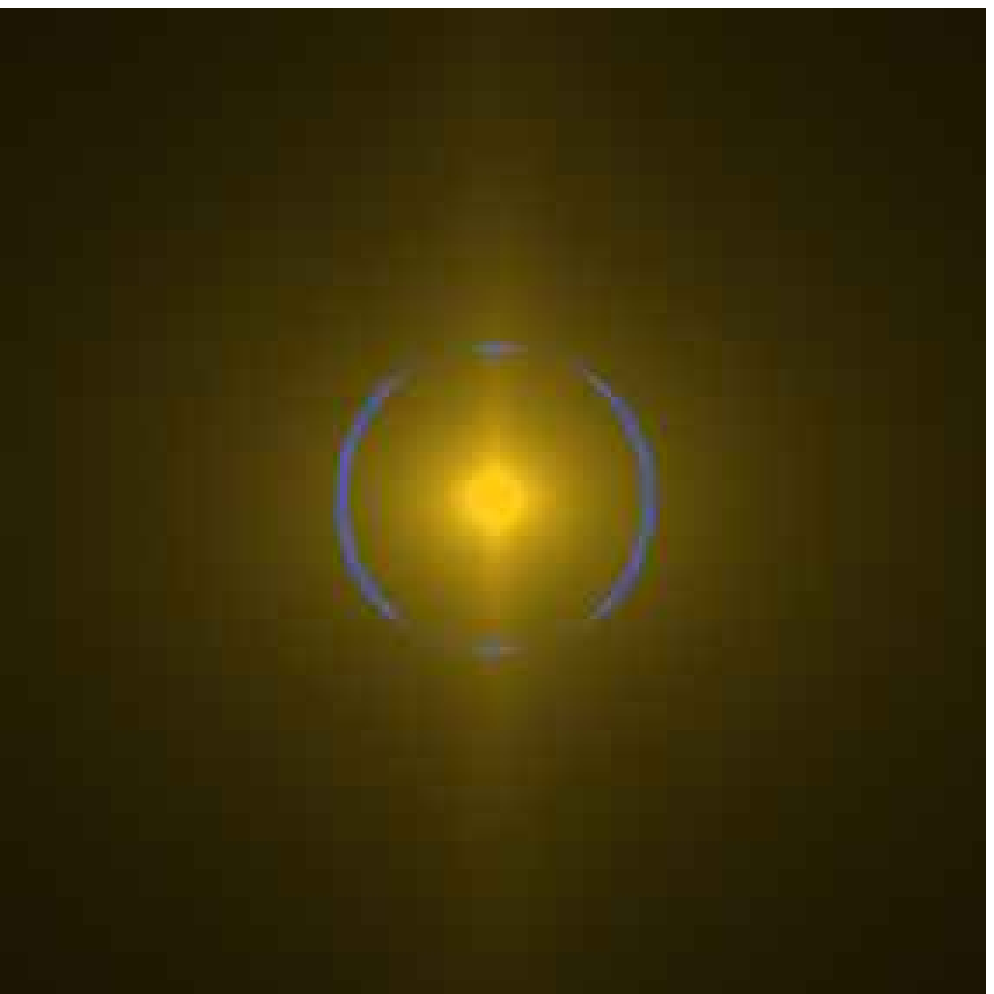}
  \end{minipage}\hfill
  \begin{minipage}[t]{0.67\linewidth}
    \centering\includegraphics[width=\linewidth]{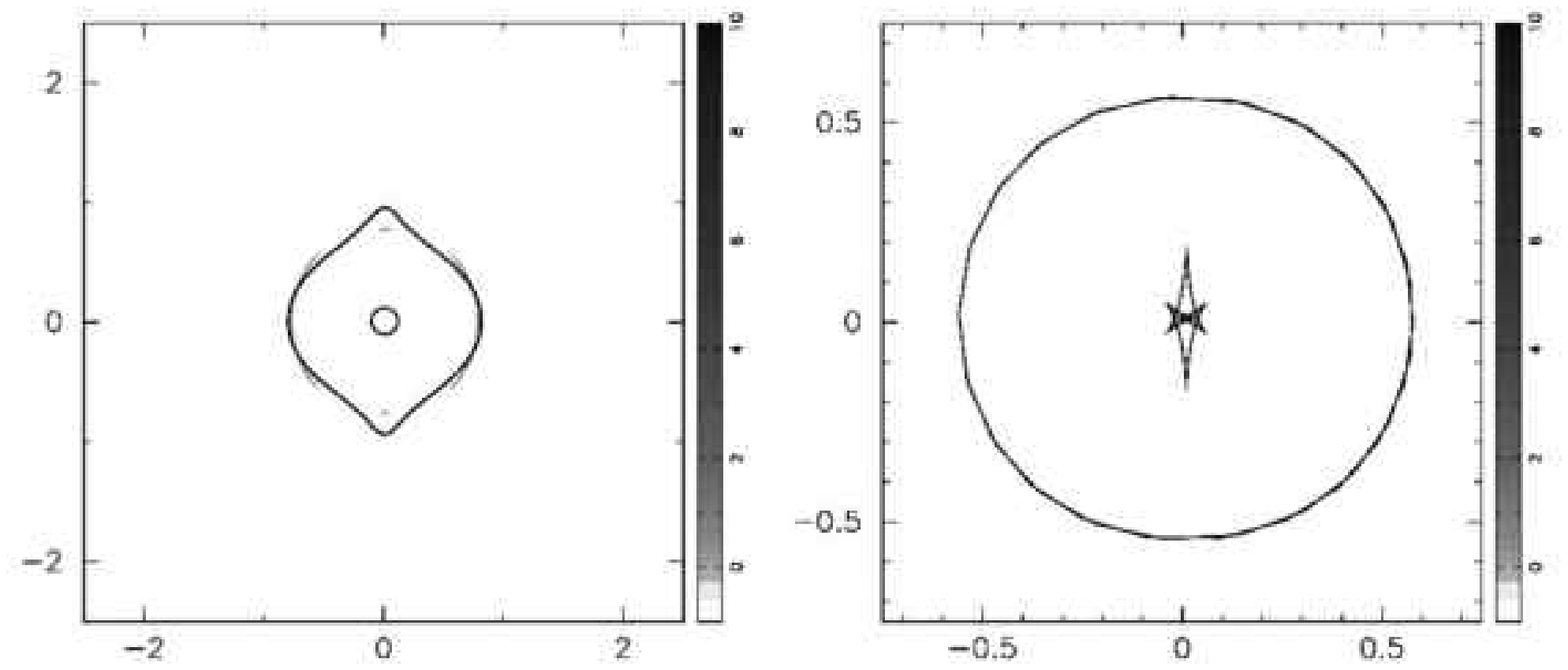}
  \end{minipage}
\end{minipage}
\caption{Gravitational lensing due to a model bulge-dominated, disky
galaxy. Three different bulge-disk misalignment angles~$\phi$ are
shown:  from top to bottom, $\phi = 1.0,~1.3,~1.57$~radians. {\it Left
to right:} a predicted high-resolution  optical image; the critical
curves and image positions (note the small inner critical curve arising
from the presence of the small core radius in the NIE profile);  the
source position relative to the caustics.   For the anti-aligned case,
$\phi = 1.57$, two butterflies overlap in the centre of the source-plane
giving rise to an Einstein ring image system composed of 8~merging
images, visible here for the case of an almost point-like source.}
\label{fig:diskbulge}
\end{figure*}

The parameters of our double NIE model are summarised as follows:
\begin{center}
	\begin{tabular}{lrrrrrr}
	\hline
	Type &     $\sigma$ & $\epsilon'$ & $\rc$ & $\phi$\\
	\hline\hline
	Bulge    &      200 & 0.543       & 0.1   & 0.0\\
	Disk     &      150 & 0.8         & 0.1   & $\phi$\\
	\hline
\end{tabular}
\end{center}
Here the variable 
$\phi$ represents the relative orientation of the two components' 
major axes, and
$\sigma$ is measured in
km s$^{-1}$, $\rc$ in kpc, $\theta$ in arcseconds and $\phi$ in radians.
The ellipticity~$\epsilon'$ refers to the mass distribution (see
the appendix for more details).
Predicted lens system appearances, and critical and caustic curves, are shown
in \fref{fig:diskbulge}, for three values of the misalignment angle~$\phi$.

\subsubsection{Critical curves, caustics and image configurations}

The mapping of the outer critical curve to the source plane (right panels of
\fref{fig:diskbulge}) gives quite complex caustic curves, depending of the
orientation. For $\phi = 0$, the profile is quasi-elliptical, and the inner
caustic is a familiar astroid shape. As $\phi$ increases, the astroid caustic
becomes skewed. At $\phi \simeq 1$, two of the folds ``break'' in to  two
swallowtail catastrophes (at opposing points).  This critical value of $\phi$
varies with disk-to-bulge ratio, and the ellipticities of the two components:
larger ellipticities or disk-to-bulge ratios give more pronounced asymmetry,
and a smaller $\phi$ is required to break the folds. For  $\phi > \phi_{\rm
crit}$, we observe two swallowtails presenting two additional cusps each.   As
$\phi$ approaches $\pi / 2$, each swallowtail migrates to a cusp, and
produces  a butterfly (we can also obtain two butterflies with $\phi < \pi/2$,
if we increase the asymmetry). If the asymmetry is sufficiently large, the two
butterflies overlap,  producing a nine-imaging region (eight visible images
and a de-magnified central image). In the same way, two swallowtails can
overlap, producing also a seven-imaging region  (six images in practice).
\citet{E+W01} and \citet{KMW00}  give analytical solutions to produce such
regions by using, respectively, deviations of isophotes (boxiness and skew)
and by adding external shear corresponding to external perturbations. All we
show here is that the combination of realistic disk and bulge mass components
is a physical way to obtain such distorted isodensity contours; the
disk-to-bulge ratio and their misalignment can be  straightforwardly estimated
using standard galaxy morphology tools.

\begin{figure*}
\centering
\subfigure{\includegraphics[width=0.3\linewidth]{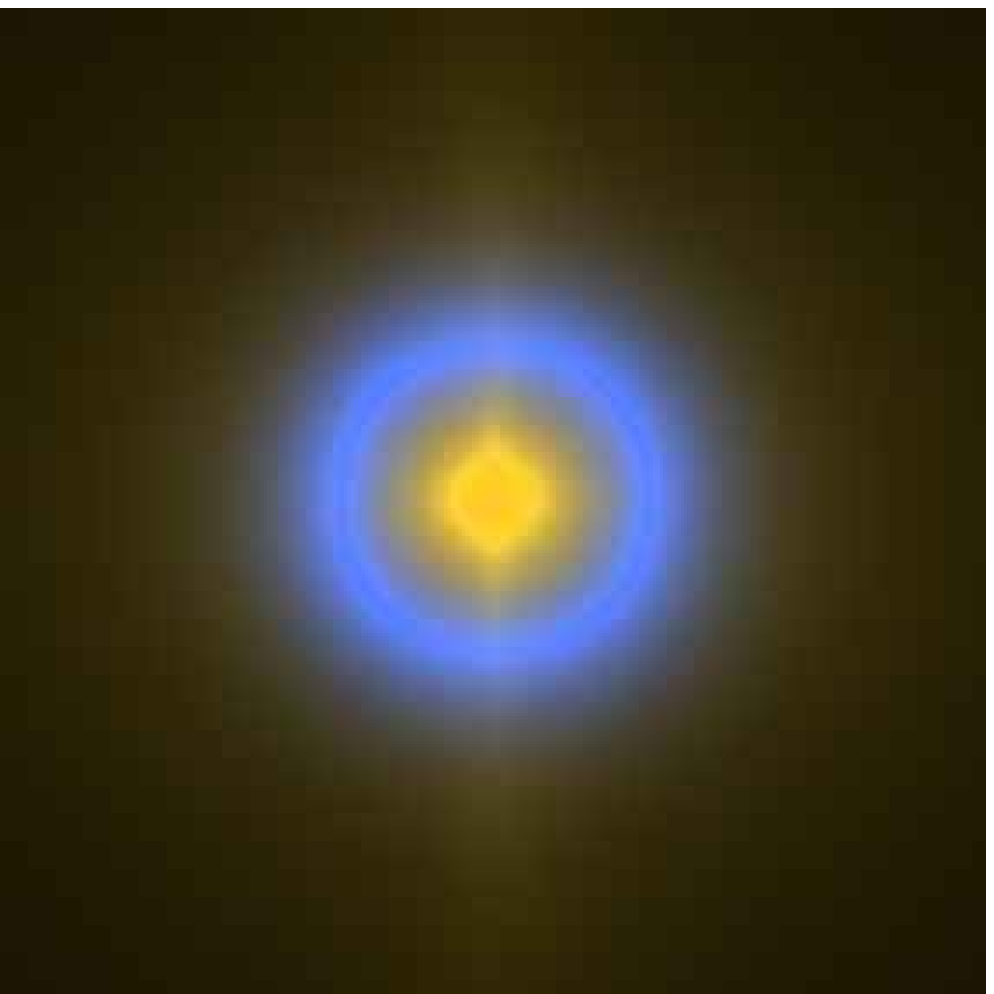}}
\subfigure{\includegraphics[width=0.3\linewidth]{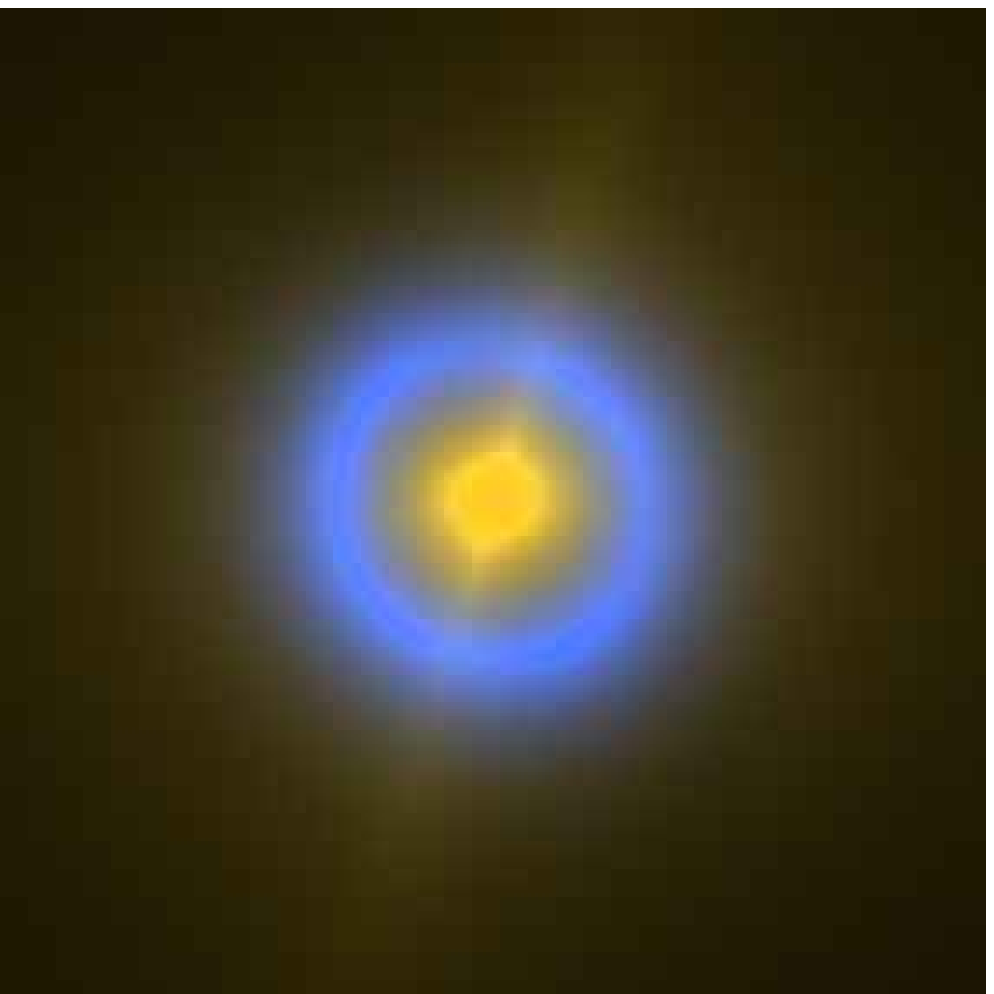}}
\subfigure{\includegraphics[width=0.3\linewidth]{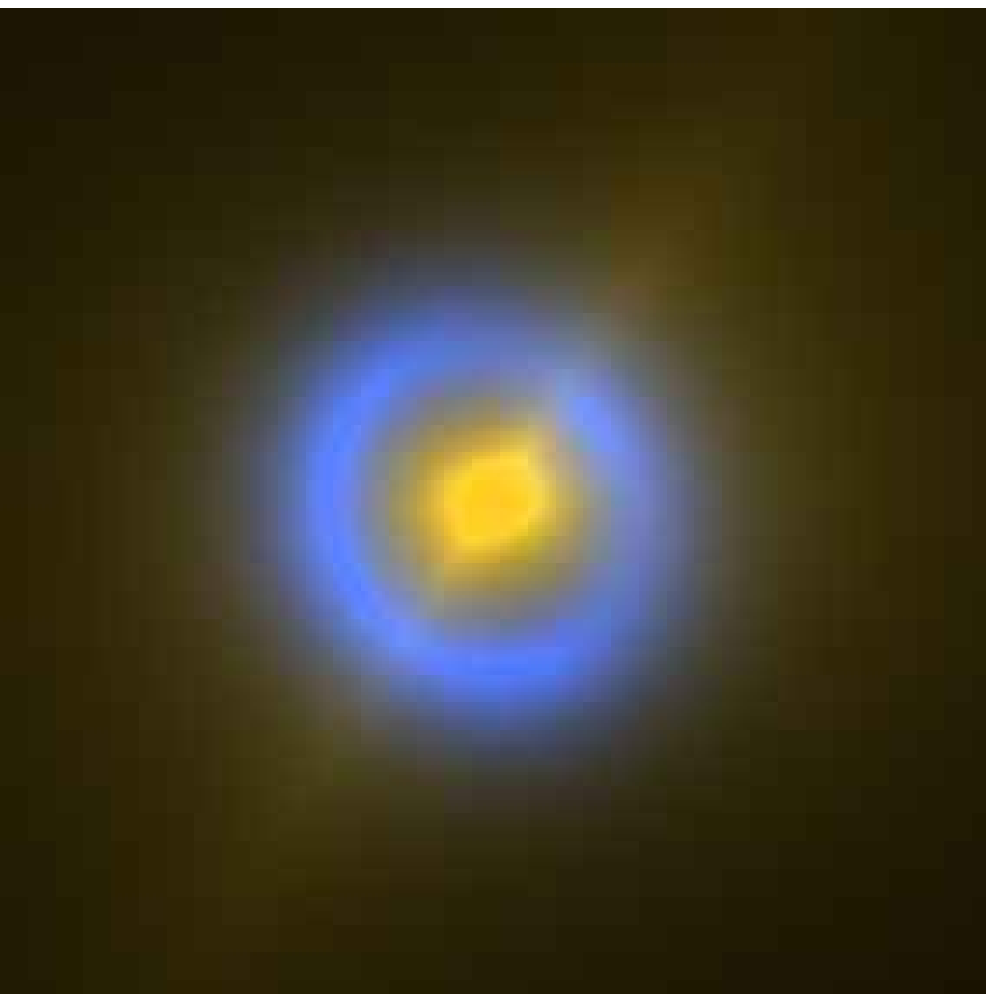}}
\caption{Gravitational lensing due to a model
bulge-dominated, disky galaxy, and a
realistic faint blue galaxy source. The same three bulge-disk misalignment
angles~$\phi$ as in  \fref{fig:diskbulge} are shown:  $\phi =
1.0,~1.3,~1.57$~radians, from left to right.  
The source is at $z_s = 1.2$ and has a half-light radius of 2~kpc.}
\label{fig:realdiskbulge}
\end{figure*}

Observationally speaking, it is important to note that the various critical
areas (the swallowtail and butterfly, and the regions of overlap between them)
are very small, and close to the optical axis of the lens. A  small source
lying in one of these regions will appear either as a very large arc  (four
images merging in the case of a swallowtail, five images merging in the case
of a butterfly), or as a ``broken'' Einstein ring (in the nine-imaging region
cases). However, in practice, with realistic source sizes, these image
configurations will appear as Einstein rings, around which the surface
brightness varies: this is illustrated in \fref{fig:realdiskbulge}, where we
predict optical images for the same lens models as in \fref{fig:diskbulge} but
with a source at $\zs=1.2$ with $\re=2$~kpc.   These fluctuations can tell us
something about the detailed structure of the large-scale potential, and
should be borne in mind when modelling such rings \citep{Koo05,V+K09}.

\subsubsection{Magnification}

\begin{figure}
\centering
\includegraphics[width=0.9\linewidth]{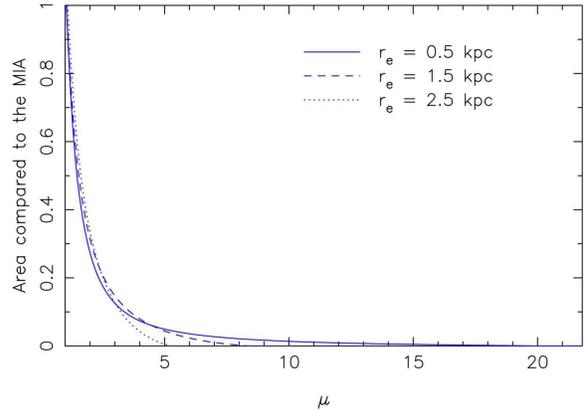}
\caption{Fractional cross-section as a 
function of total magnification produced for our model
disk-plus-bulge galaxy, with relative orientation $\phi = 1.55$.}
\label{fig:xmag}
\end{figure}

In order to study the magnification of this model system, we first plot in
\fref{fig:xmag}the fractional cross-sectional area as a  function of the
magnification for the almost maximally-misaligned case, $\phi = 1.55$.  We
show three different source sizes (as in \sref{sec:magsources}). For a typical
extended source ($\re =$ 2.5 kpc), the total magnification only reaches a
maximum of around $\mu = 5$, no larger than we might expect for a simple NIE
lens. At fixed source size, Einstein rings have similar  magnification
regardless of whether they have catastrophe-induced fluctuations.
We can see from \fref{fig:diskbulge} that when we increase the misalignment
$\phi$, the astroid caustic area decreases. As the source is typically more
extended than this astroid caustic, the regions of high magnification have a
small contribution to the fractional cross-section. When we decrease the
size of the source, the magnification
shifts to higher values (10--15), but still comparable to those obtainable
with a simple NIE model.

\begin{figure*}
\centering
\subfigure{\includegraphics[width=0.3\linewidth]{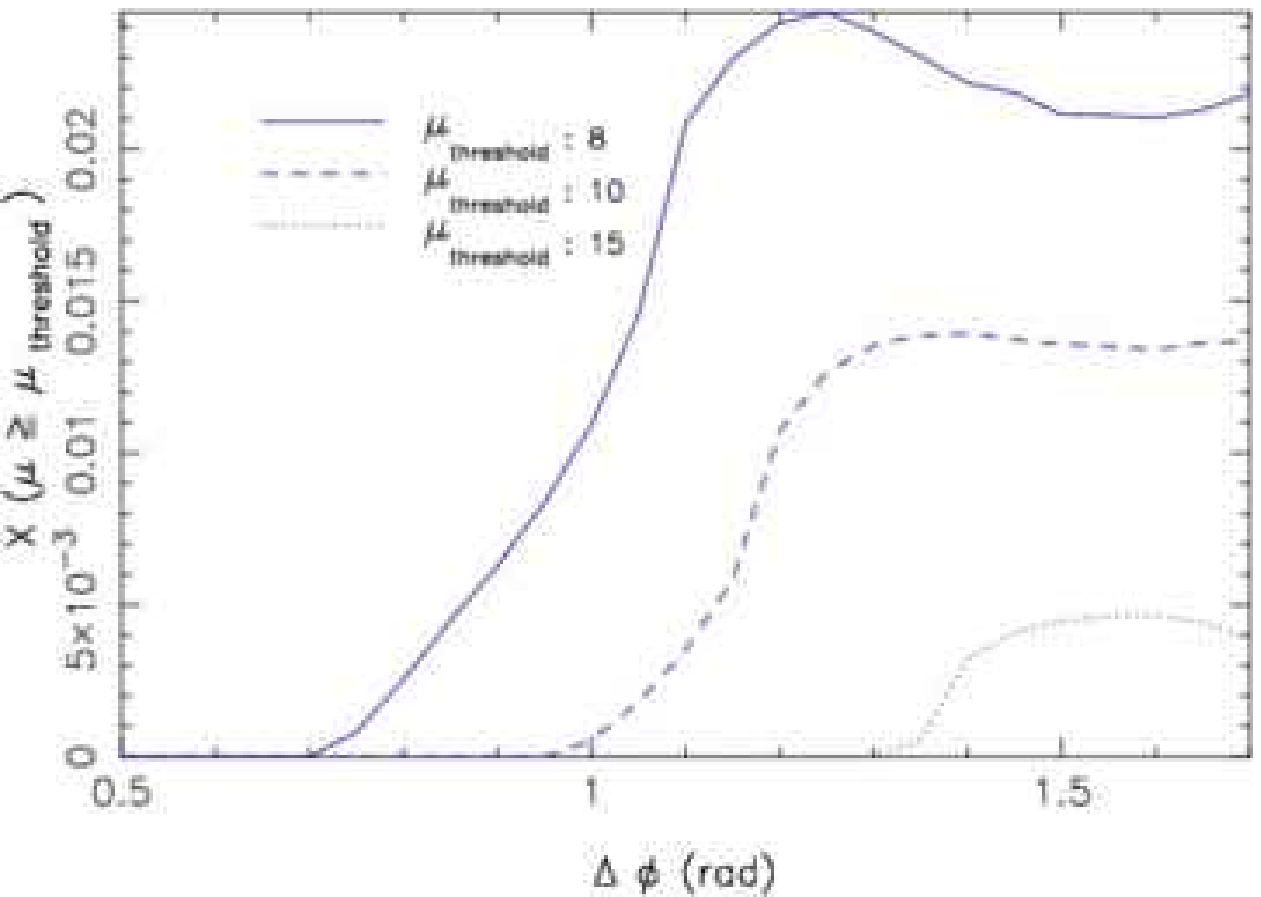}}
\subfigure{\includegraphics[width=0.3\linewidth]{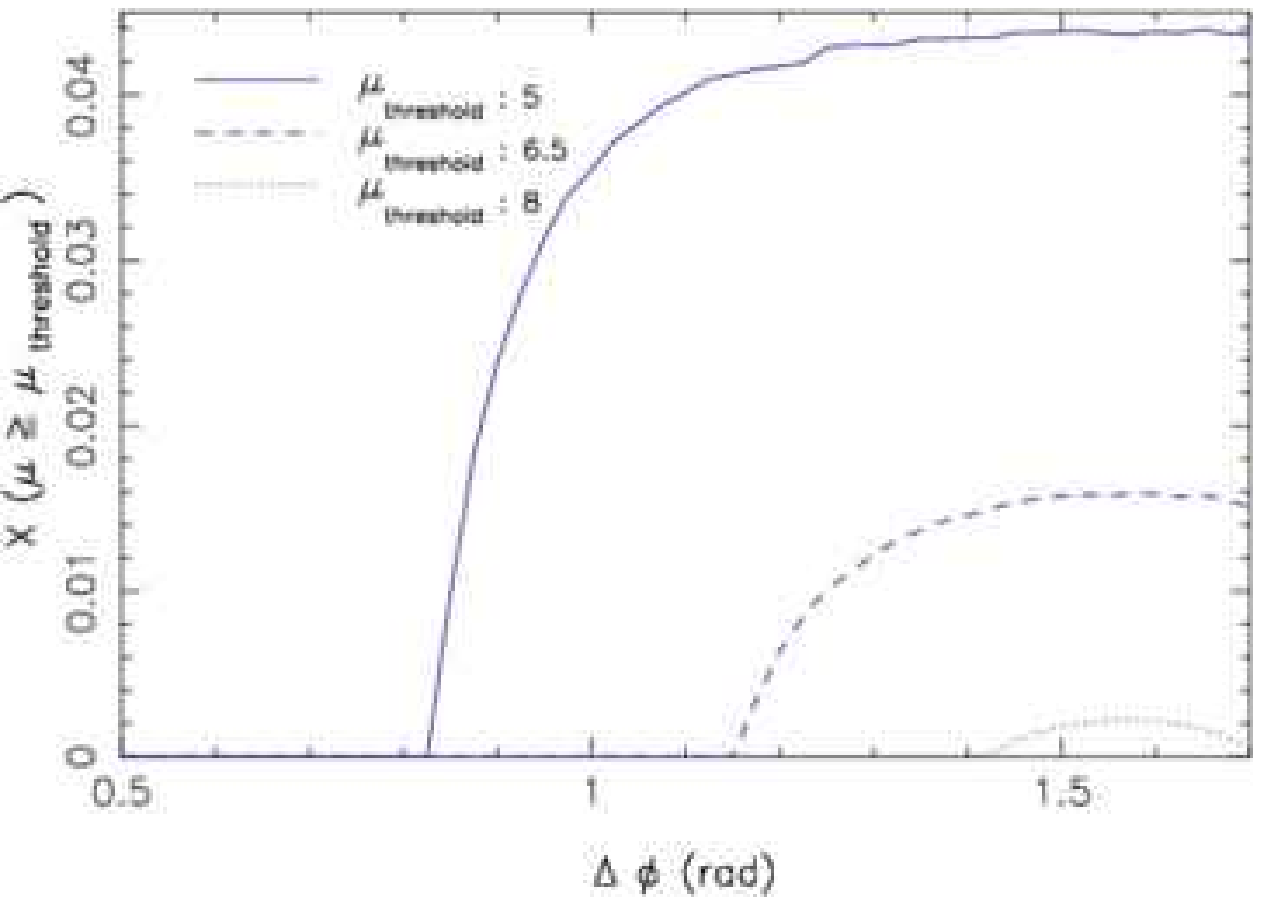}}
\subfigure{\includegraphics[width=0.3\linewidth]{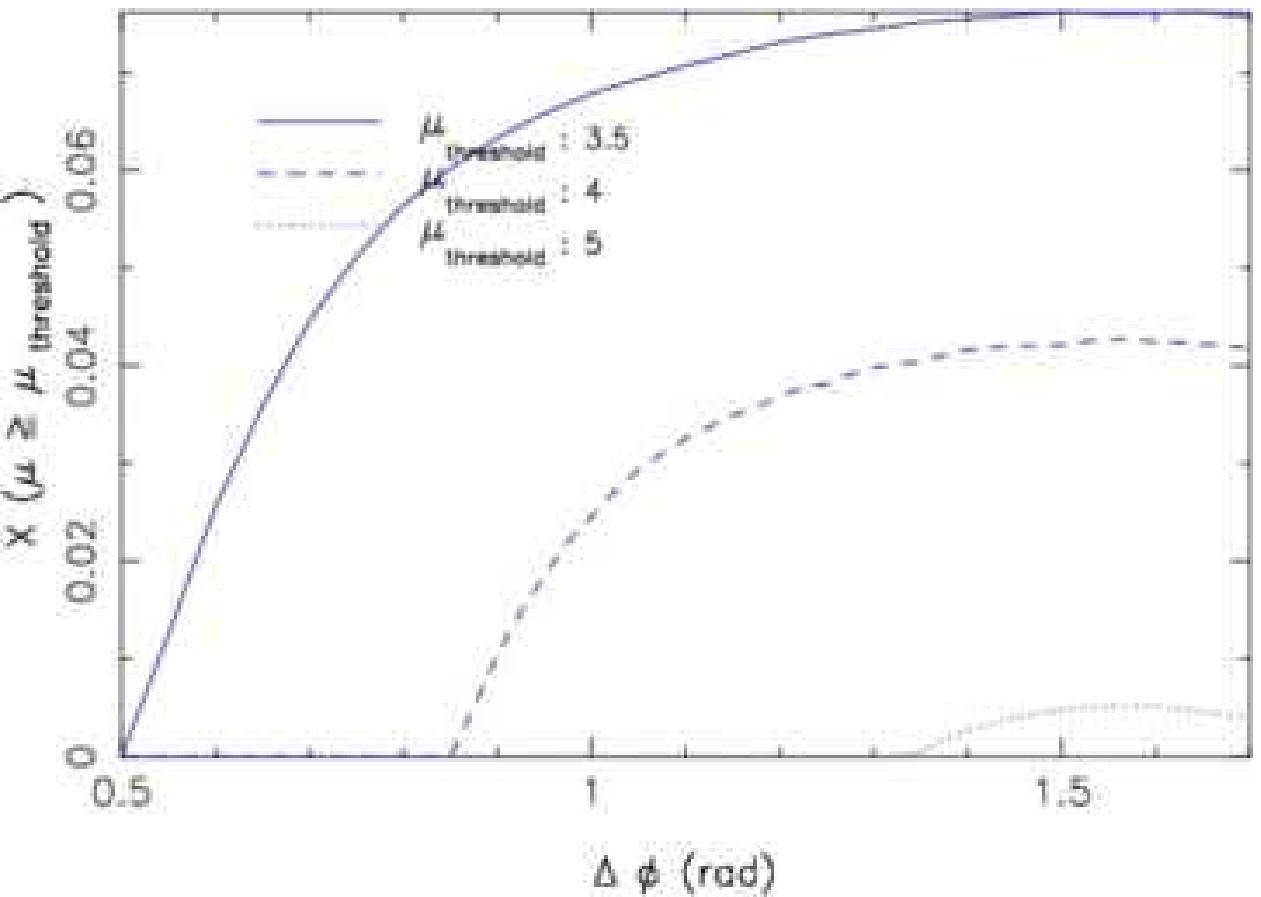}}
\caption{Fractional cross-section area giving total magnification greater than
a given threshold, as a function of disk/bulge misalignment angle.  We
consider our 3 fiducial sources, from left to right: $\re = 0.5, 1.5,
2.5$~kpc}
\label{fig:disk-bulge-fraction}
\end{figure*}

We then compute the fractional cross-section of the lens having a
magnification higher or equal than a given threshold,
\fref{fig:disk-bulge-fraction}, as explained in \sref{sec:magsources}. With a
source size of $\re = 0.5$ kpc, we see that for a low magnification threshold 
($\mu>8$), the maximum is at $\phi = 1.25$ and not at  $\phi = 1.57$ as we 
expected. Increasing the threshold magnification,  the maximum does
shift to $\phi = 1.57$.   Increasing the misalignment $\phi$, regions of
higher magnification but  smaller fractional area appear, while at the same
time the astroid caustic  cross-section decreases. By considering higher
magnification thresholds, we focus on more the higher magnification regions.
For example, at $\phi = 1.25$, the caustic presents two relatively large
swallowtails, which are regions of medium-high magnification.  For higher
magnification thresholds (\eg $\mu>15$), the swallowtail regions (and indeed
the four cusp regions) no longer provide significant cross-section, and the
maximum shifts to the butterfly regions (and the region where they overlap). 
However, this region is correspondingly smaller, and the cross-section 
decreases to just $\sim 5\times10^{-3}$ of the multiple imaging area.

These considerations can be used to try and estimate the  fraction of
butterfly configurations we can observe. The problem for the swallowtails is
to find a  magnification threshold which isolates these regions (and separates
them from the effects of the 4 cusps). The butterfly regions are more cleanly
identified in this way:  only the place where the two butterflies overlap  can
give rise to such a small fractional area yet high magnification. When
estimating butterfly abundance in \sref{sec:abundance},  we will make use of
the  approximate fractional cross-sectional area of  $\sim 5\times10^{-3}$.
For the swallowtails, we note that  these form over a wider range of
misalignment angles, perhaps $\pm 0.5$~rad from anti-alignment compared to
$\pm 0.05$~rad for butterfly formation:  we predict very roughly that we would
find ten times more swallowtails than butterflies just from their appearing
for a ten times wider range of disk/bulge misalignments.

\subsection{Galaxies with a satellite}\label{ssec:sat}

As illustrated in section~\ref{ssec:slacs} with the system SDSSJ0808$+$4706
and SDSSJ0956$+$5100, main galaxies can have small satellites, relatively
close. We expect these satellite to perturb the caustic structures, and 
perhaps give rise to higher order catastrophes \citep[\eg][]{C+K04}.


\subsubsection{Model}

We model a system comparable to SDSSJ0808$+$4706, \ie a main galaxy with a
small satellite, by two NIE profiles (with very small cores). The
model parameters were set as follows:
\begin{center}
 	\begin{tabular}{rrrrrrr}
       \hline
	  Type &    $\sigma$ & $\epsilon_{\rm p}'$ & $\rc$  &   $\theta_1$  & $\theta_2$ & $\phi$\\
	 \hline\hline
	  NIE  & 236.  & 0.05   & 0.05   &    0     &  0   & 1.3     \\
        NIE  & 80.   & 0.15   & 0.01   &  -0.735 & 1.81 & -0.3  \\
       \hline
	\end{tabular}
\end{center}
where $\sigma$ is in km s$^{-1}$, $\rc$ in kpc, $\theta$ in arcseconds and
$\phi$ in radians.

The redshifts of the lens and source planes were set at  $\zd =0.2195$ and
$\zs = 1.0215$ (to match those of SDSSJ0808$+$4706).  We arbitrarily chose the
velocity dispersion of the satellite to be one third that of the main
galaxy, and used a double-peaked source to better reproduce the arcs seen in 
SDSSJ0808$+$4706.

\begin{figure}
\centering
\includegraphics[width=9cm]{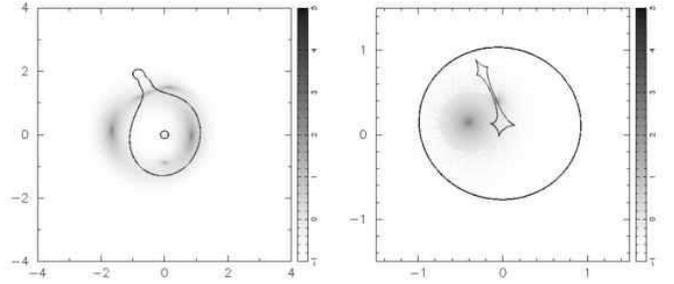}
\caption{Critical curves and caustics for our model of
SDSSJ0808$+$4706, with predicted images overlaid.}
\label{fig:0808b}
\end{figure}


\subsubsection{Critical curves, caustics and image configurations} 

The critical curves and caustics for this model system are shown in 
\fref{fig:0808b}. The two astroid caustics merge to give
an astroid caustic with six cusps instead of four, a 
beak-to-beak metamorphosis (see section \ref{ssec:ccc}). 

\begin{figure}
\centering
\includegraphics[width=9cm]{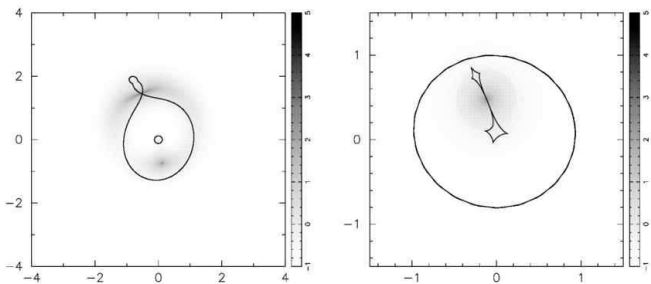}
\caption{Critical curves and caustics for a model main galaxy plus
satellite lens, presenting a 
beak-to-beak calamity, and
a resulting image configuration.}
\label{fig:beak-to-beak}
\end{figure}

In \fref{fig:beak-to-beak}, we present the beak-to-beak calamity  image
configuration  by slightly adjusting the velocity dispersion parameter of the
satellite from $\sigma = 80$ km s$^{-1}$ to $\sigma = 67.7$ km s$^{-1}$.   The
image configuration of the beak-to-beak configuration is the merging of three
images in a straight arc (we refer also to
\citeauthor{KKB92}~\citeyear{KKB92}, who discuss the beak-to-beak in order to
explain the straight arc in Abell 2390). The beak-to-beak calamity does not
lead to higher magnification  (compared to the cusp catastrophes), as it is
still the merging of just  three images. 

As the lens potential of the satellite increases, the caustics become more
complex; such lenses are better classified as ``binary'' and are studied below
in \sref{sec:complexmodels}. On the other hand, small (and perhaps dark) 
satellite galaxies are expected to modify the lensing flux ratio
\citep[\eg][]{S+E08}: such  substructure can easily create very local
catastrophes such as swallowtails \citep{Bra++04}, which may be
observable in high resolution ring images, as discussed in
the previous section \citep{Koo05,V+K09}.


\section{Complex lenses}\label{sec:complexmodels}

We expect group and cluster-scale lenses to produce higher multiplicities,
higher magnifications, and different catastrophes. Motivated by two \sls
targets, SL2SJ1405$+$5502 and SL2SJ0859$-$0345, and the cluster Abell~1703, we
first study a binary system, and the evolution of its caustics with lens
component separation and redshift, and the apparition of elliptic umbilic
catastrophes. Then we illustrated a four-lens component system based on the
exotic SL2SJ0859$-$0345 lens. Finally, we study  the cluster Abell 1703, which
presents an image system characteristic of a hyperbolic umbilic catastrophe.


\subsection{Binary lenses}\label{ssec:binary}


\subsubsection{Model}

\begin{figure}
\centering
\subfigure{\includegraphics[width=5.5cm]{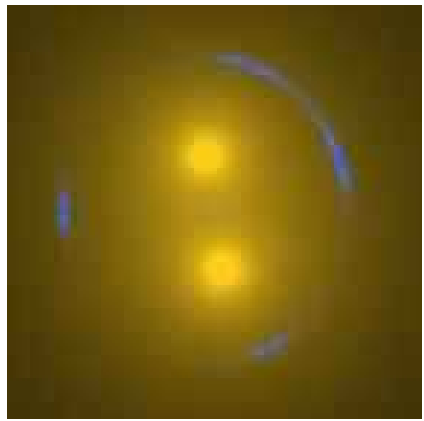}}\hfill
\subfigure{\includegraphics[width=9cm]{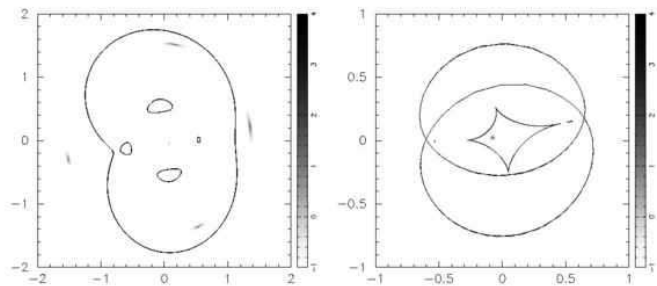}}
\caption{{\it Top:} mass distribution of our model of SL2SJ1405$+$5502,
and surface brightness of the 
predicted images. {\it Bottom:} critical curves and caustics}
\label{fig:140533}
\end{figure}

Given the asymmetry of the pair of lens galaxies visible in SL2SJ1405$+$5502
(\sref{ssec:sl2s}), we expect the critical curves and caustics to be quite
complex. We first make a qualitative model of the lens, using the following
mass components:

\begin{center}
\begin{tabular}{rrrrrrr}
\hline
type &    $\sigma$ & $\epsilon_{\rm p}'$ & $\rc$   &   $\theta_1$  & $\theta_2$ & $\phi$\\
\hline\hline
NIE    &      250 & 0.06 & 0.1    & 0.0774 &-0.55&1.5\\
NIE    &      260 & 0.04 & 0.1    &-0.0774 & 0.55& -0.6\\
\hline
\end{tabular}
\end{center}
where $\sigma$ is in km s$^{-1}$, $\rc$ in kpc, $\theta$ in arcseconds  and
$\phi$ in radians. The redshifts chosen are $\zd = 0.6$ and $\zs = 1.2$,
reasonable estimates for the SL2S objects.
As \fref{fig:140533} shows, the model-predicted image configuration matches
the observed images fairly well. We note the presence of a predicted faint
central image, which may be responsible for some of the brightness in between
the lens galaxies in \fref{fig:real140533}. 

We now explore more extensively this type of binary system. We note that the
galaxies don't need to have any ellipticity or relative mis-alignment in order
to have an asymmetric potential. We define a simple binary lens model
consisting of two identical non-singular isothermal spheres of $\sigma = 250$
km s$^{-1}$ and core radii of 100~pc, $\zd = 0.6$ and $\zs = 1.2$. 

\subsubsection{Critical curves, caustics and image configurations}

\begin{figure}
\centering
\includegraphics[width=8cm]{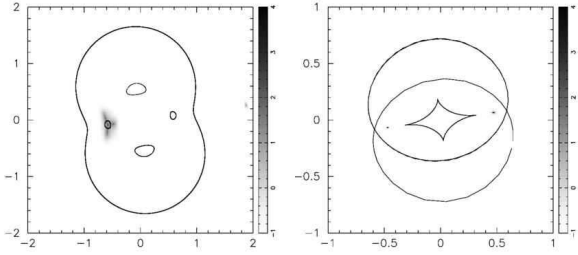}
\caption{An ``elliptic umbilic image configuration.''}
\label{fig:binarydeltoids}
\end{figure}

In a recent paper, \citet{S+E08a} studied strong lensing by binary galaxies,
noting the appearance of several catastrophes.  Even with isothermal spheres
instead of the ellipsoids used in modelling SL2SJ1405$+$5502 above,  the
binary system can produce an inner astroid caustic with two ``deltoid''
(three-cusped) caustics \citep{2006MNRAS.366...39S}. The latter characterize
an elliptic umbilic metamorphosis, and can be seen in
\fref{fig:binarydeltoids} as the small source-plane features in the right-hand
panel, lying approximately along the $\theta_1$-axis. The general caustic
structure is the result of the merging of the two astroid caustics belonging
to the two galaxies. 

\begin{figure*}
\centering
\includegraphics[width=17cm]{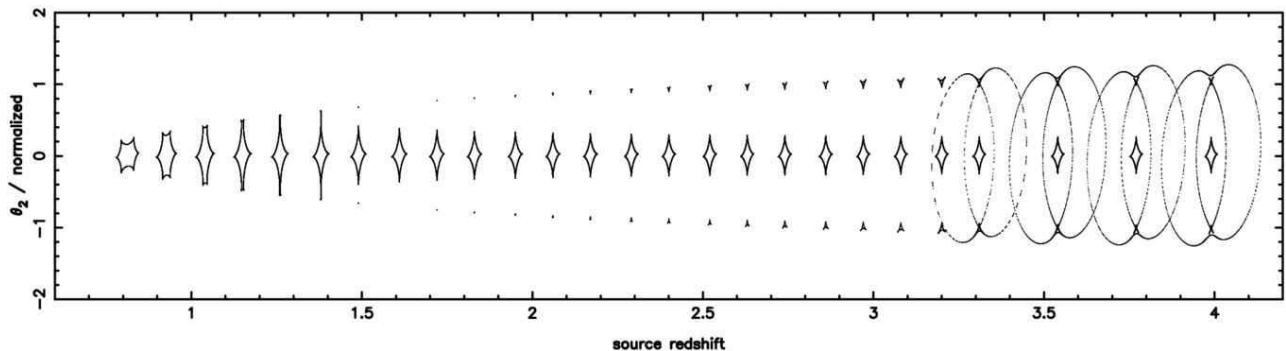}
\caption{Evolution of the caustics with source redshift
behind a fixed-separation binary lens system.}
\label{fig:binary-z-evolution}
\end{figure*}

\citet{S+E08a} studied the different metamorphoses undergone as the two mass
components are brought closer together. In order to give a further
illustration of the caustic structure, we show in
\fref{fig:binary-z-evolution} the evolution with source redshift  of the
caustics behind our  binary model system for a given separation $\Delta \theta
= 1.5''$. 
Plots like this emphasise the three-dimensional nature of the caustic
structure, slices of which are referred to as ``caustics'' in the
gravitational lensing literature.
This is equivalent to the evolution with  separation, since the
important parameters are the ratio of the  Einstein radii to the separation
$\frac{\theta_{E;1,2}}{\Delta \theta}$ and the Einstein radius increases with
the redshift. The condition to obtain an elliptic
umbilic catastrophe with two isothermal spheres is given by \citet{S+E08a}, 
\be\label{eq:ellumb}
\Delta \theta =  \sqrt{\theta_{E,1}^2 + \theta_{E,2}^2}.
\ee

At low source redshift, we observe a six-cusp astroid caustic (the aftermath
of a beak-to-beak calamity at the merging of the two four-cusp caustics). As
the source redshift increases, we observe another two beak-to-beak calamity
(the system is symmetric) leading to two deltoid caustics, with the
beak-to-beak calamity again marking a transition involving the gain of two
cusps. These deltoid caustics then shrink two points, the two elliptic umbilic
catastrophes visible at around $z_s = 1.6$. Beyond this source redshift we
then observe two different 
deltoid caustics  (with the elliptic umbilic
catastrophe marking the transition between one deltoid and the next).  For
clarity, we have not represented the outer caustics until the deltoid caustics
meet these curves to form a single outer caustic and a lips with two
butterflies.
In \fref{fig:binarydeltoids} we present an ``elliptic umbilic image
configuration,'' where four images merge in a Y-shaped feature -- yet to be
observed in a lens system. 

The abundance of such image configurations can be estimated very roughly from
their fractional cross-section.  We estimate geometrically the fractional
cross-section of the deltoid caustics in our model system, compared to the
multiple imaging area, as being around $10^{-3}$. 


\subsubsection{Magnification}

In order to study the magnification provided by such binary lenses, we again
plot the fractional cross-section area with magnification above some
threshold,  against source redshift. The reference cross-section was taken to
be that   where the magnification is $\geq$ 3, which corresponds roughly to 
the multiple imaging area. We make this plot for our three fiducial sources.
We first note that the cross-section for  a magnification of 10 is quite big
($\sim$0.2): here is a high probability of attaining a high total
magnification with such lenses. This probability persists even with large
source sizes, although the cross-section for even higher magnification does
decrease quite quickly.

\begin{figure*}
\centering
\subfigure[$\re$ = 0.5kpc]{\includegraphics[width=5.8cm]{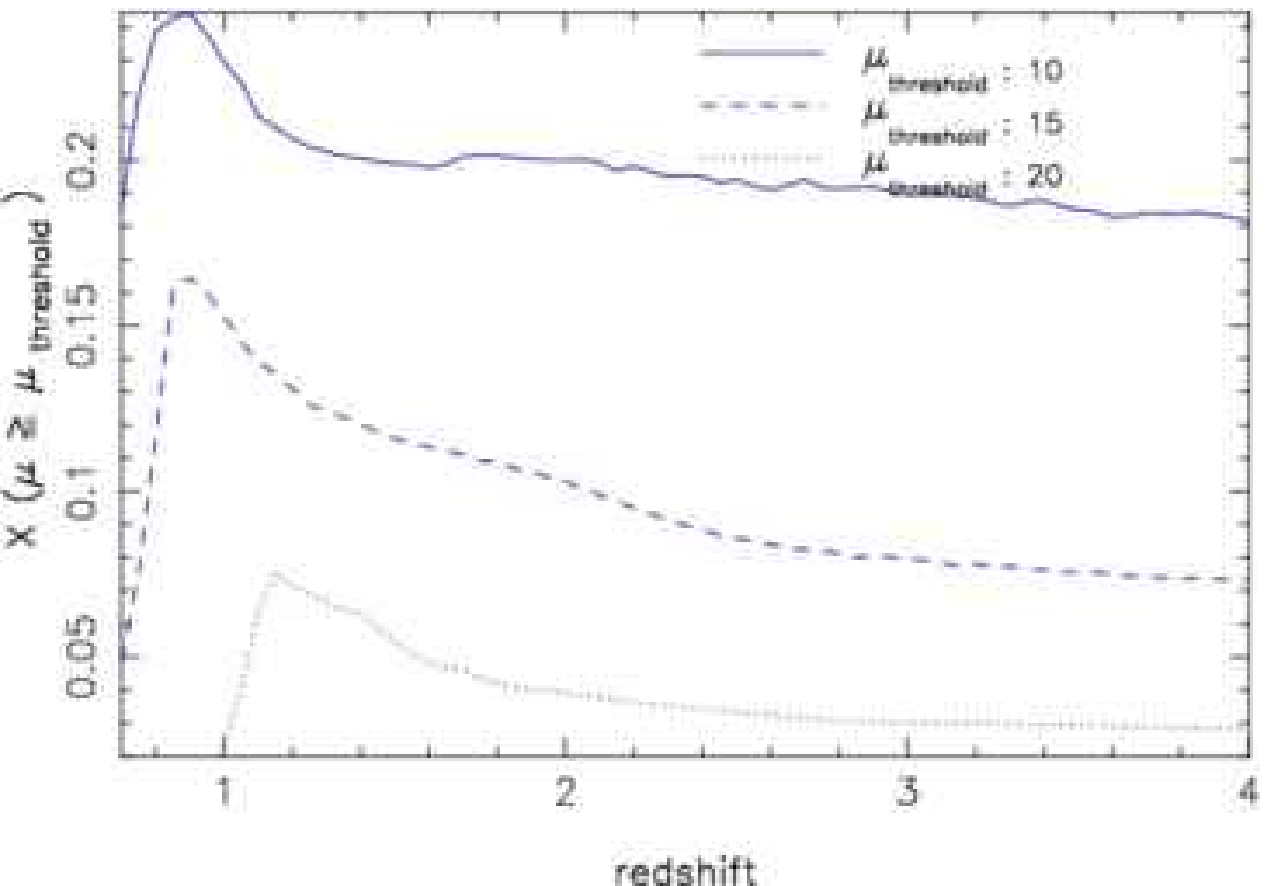}}
\subfigure[$\re$ = 1.5kpc]{\includegraphics[width=5.8cm]{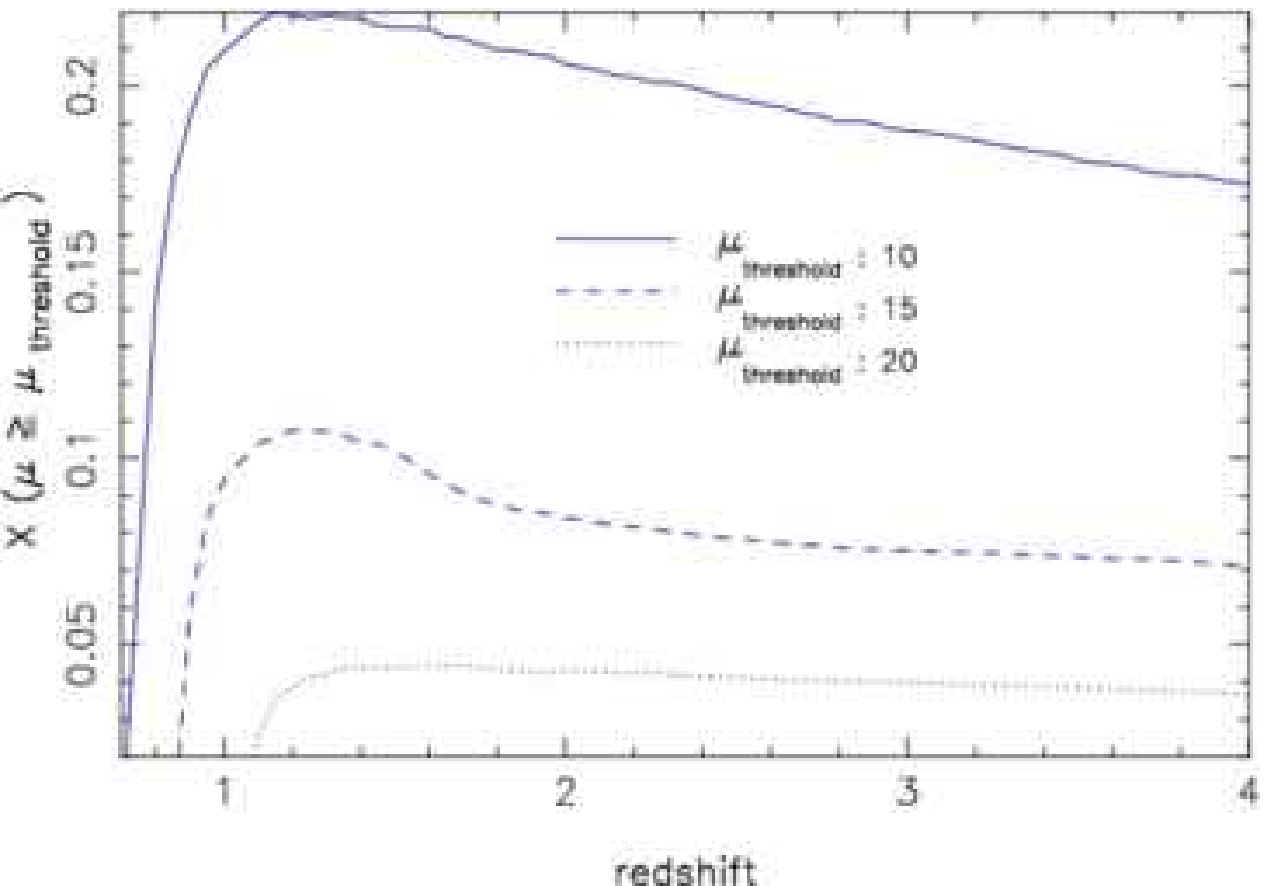}}
\subfigure[$\re$ = 2.5kpc]{\includegraphics[width=5.8cm]{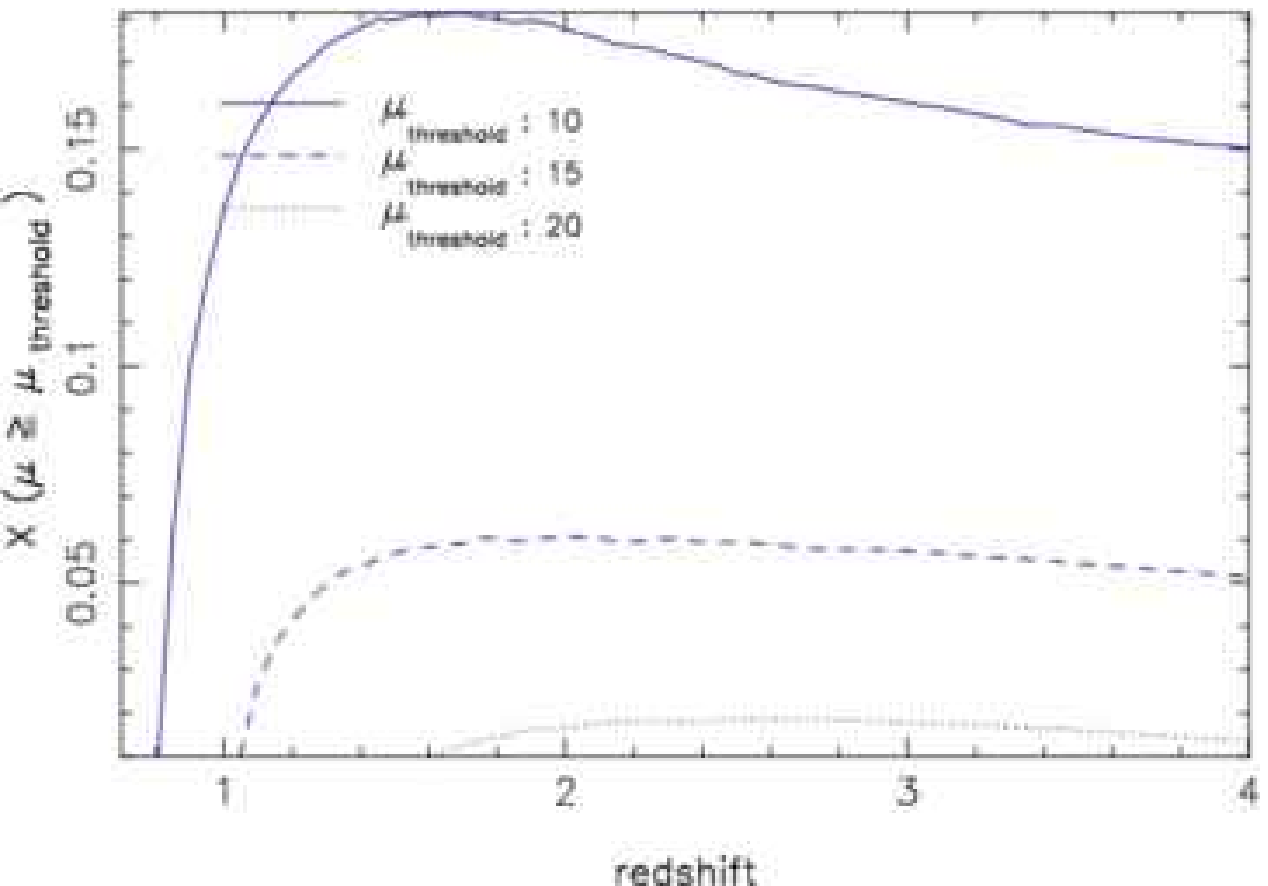}}
\caption{Fractional cross-section for magnification higher than a given
threshold (compared to area with $\mu \geq 3$) for the binary lens model. 
We consider 3 fiducial source sizes, from left to right: 
$\re = 0.5, 1.5, 2.5$ kpc.}
\label{fig:red}
\end{figure*}

\begin{figure}
\centering
\includegraphics[width=8cm]{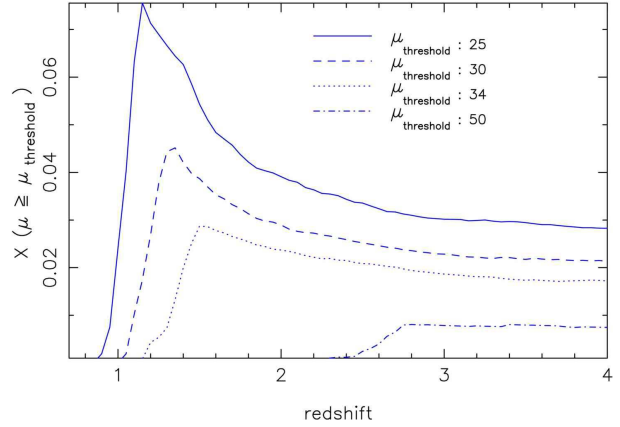}
\caption{As \fref{fig:red}, but showing higher magnification thresholds.}
\label{fig:binary-high-mag}
\end{figure}

For small sources, $\re = 0.5$ kpc, the pair of two cusps meeting in a
beak-to-beak  metamorphosis dominates the magnification $\simeq10$
cross-section, with the maximum cross-sectional area occurring  when the two
cups are close but before the beak-to-beak calamity itself. Re-plotting  the
same small source cross-section with higher magnification thresholds (25, 30,
34, 40, Figs.~\ref{fig:red} and~\ref{fig:binary-high-mag}), 
we see that the magnification
$\simeq30-40$ cross-section is dominated by the deltoids and their
neighbouring cusps. To reach even higher magnifications requires a very high
redshift (and hence small in solid angle) source placed at the center of the
main astroid.

\subsubsection{Two merging galaxies}

As an extreme case of the binary system illustrated above,  we briefly
consider two merging galaxies.  The model used is similar except that we use
elliptical profiles with $\epsilon_{\rm p}' = 0.2$ instead of spheres.

\begin{figure}
\raggedright
\begin{minipage}[t]{0.95\linewidth}
  \begin{minipage}[t]{0.45\linewidth}
    \raggedleft\includegraphics[width=\linewidth]{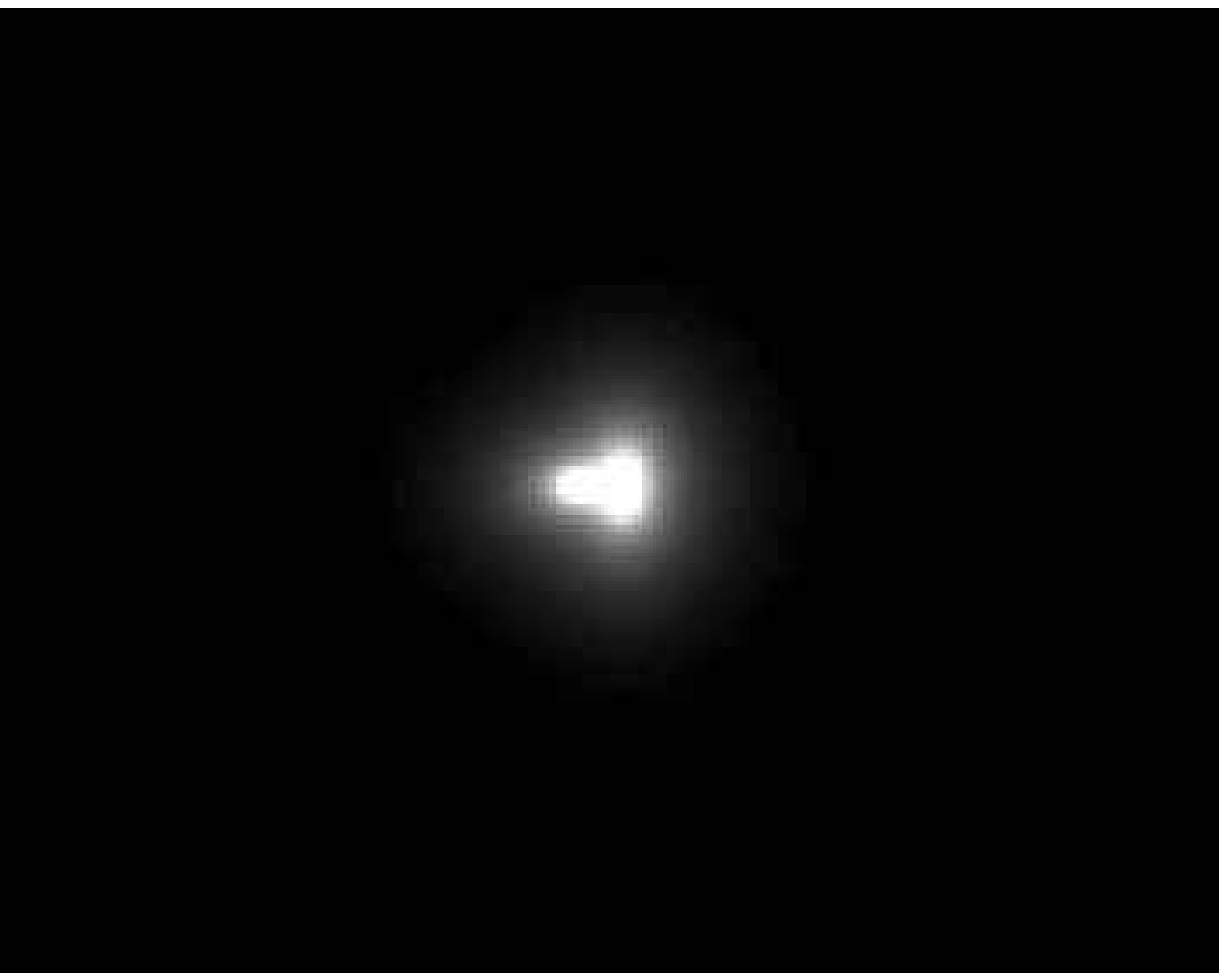}
  \end{minipage}\hfill
  \begin{minipage}[t]{0.535\linewidth}
    \centering\includegraphics[width=0.8\linewidth]{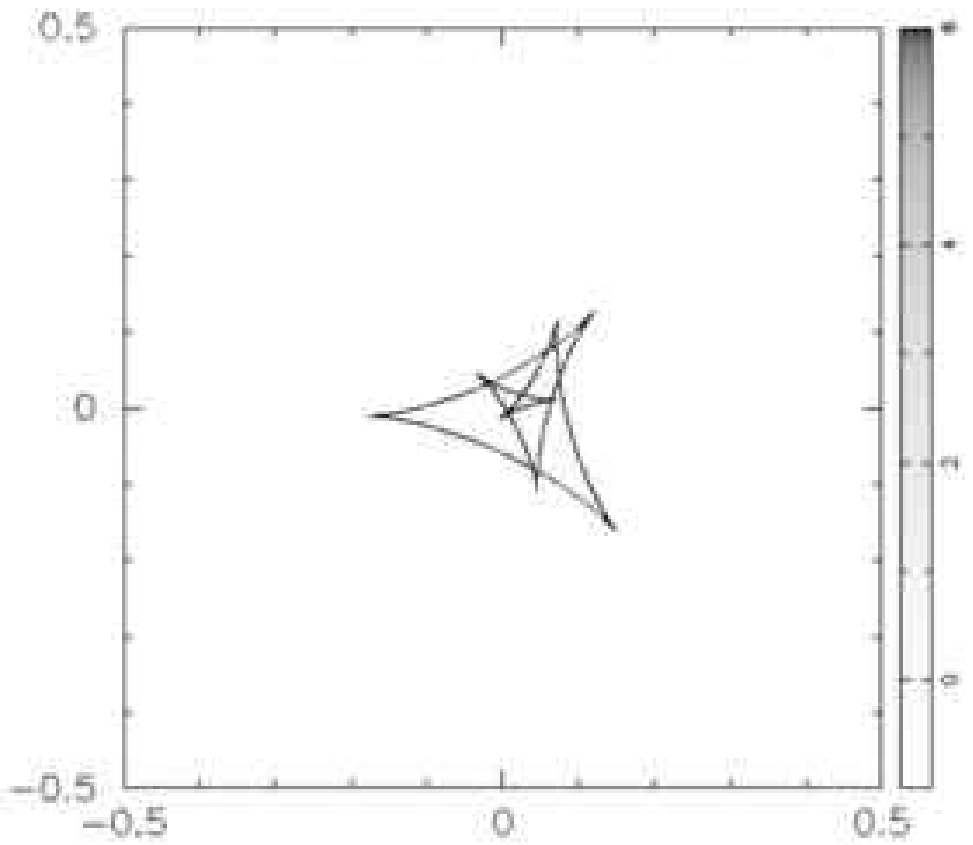}
  \end{minipage}
\end{minipage}
\caption{Two elliptical galaxies with a fixed separation 
$\Delta \theta =0.4$ arcseconds and $\phi = 1.55$ radians.
Lens convergence map {\it left}, and 
a zoomed-in view of the astroid caustic in the source plane 
{\it right}.}
\label{fig:sbo}
\end{figure}

\begin{figure*}
\raggedright
\begin{minipage}[t]{0.95\linewidth}
  \begin{minipage}[t]{0.32\linewidth}
    \vspace{-0.94\linewidth}
    \raggedleft\includegraphics[width=0.78\linewidth]{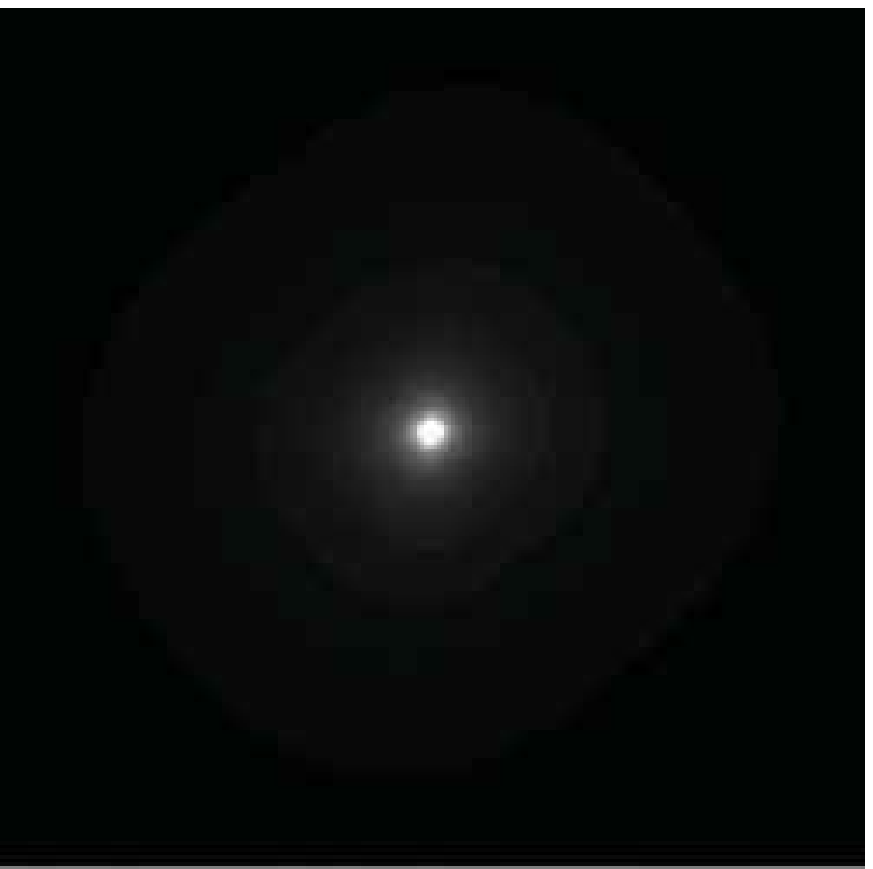}
  \end{minipage}\hfill
  \begin{minipage}[t]{0.67\linewidth}
    \centering\includegraphics[width=\linewidth]{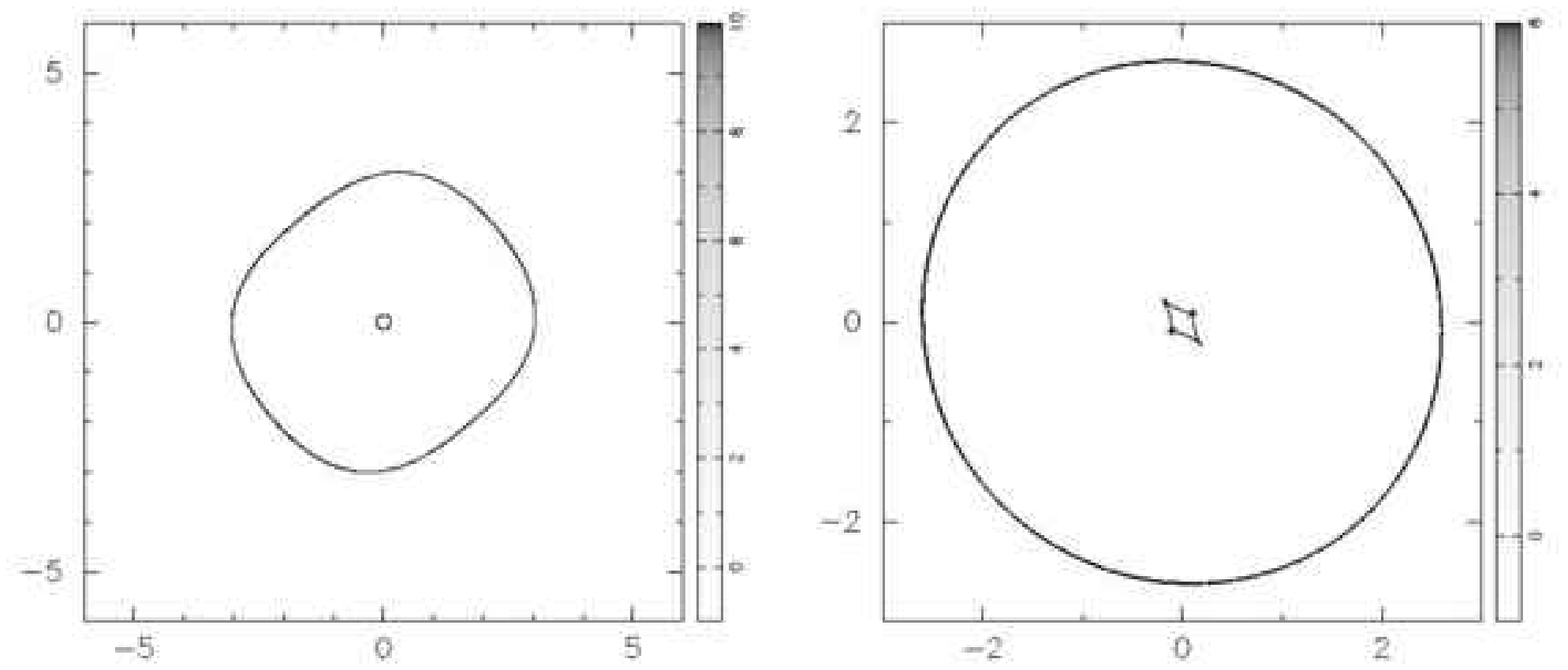}
  \end{minipage}
\end{minipage}
\caption{Two elliptical galaxies with a relative orientation of 1.2
radians, separation $\Delta \theta = 0$~arcsec.}
\label{fig:sbb}
\end{figure*}

In \fref{fig:sbo}, we plot the convergence and an enlarged view of the 
caustics for a mass component separation of $\Delta \theta = 0.4$ arcseconds 
and a relative orientation $\phi = 1.55$. When we vary the orientation, the
fold breaks leading to a swallowtail, then a butterfly and then a region of
nine-images multiplicity (as presented in the \fref{fig:sbo}), similar to that
seen in  the disk and bulge model \sref{ssec:d+b}. In \fref{fig:sbb}, we then
plot the convergence, the critical curves and the caustics for zero separation
and a relative orientation of $\phi = 1.2$, as might be seen in a
line-of-sight merger or projection. This model has boxy isodensity contours
(as also seen in the critical curves), while the caustics include two small
butterflies. This model is therefore a a physical way to obtain the boxiness
needed to generate such catastrophes \citep[\eg][]{E+W01,S+E08a}.

\subsection{A four-component lens system: SL2SJ0859$-$0345}


\subsubsection{Model} 

\begin{figure}
\centering
\includegraphics[width=6cm]{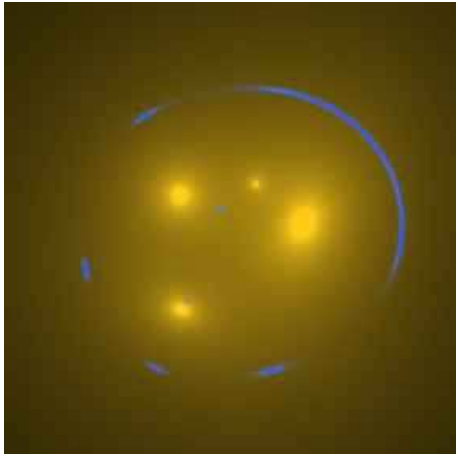}
\caption{Convergence and predicted images for a qualitative model of 
SL2SJ0859$-$0345, showing the 4 lens components and complex oval system of
arcs.}
\label{fig:085914}
\end{figure}

SL2SJ0859$-$0345 is an example of a lens with still more complex critical
curves and caustics; we represent it qualitatively with the following
four-component model:

\begin{center}
  \begin{tabular}{rrrrrrr}
  \hline
  type &    $\sigma$ & $\epsilon_{\rm p}'$ & $ \rc$   &   $\theta_1$  & $\theta_2$ & $\phi$\\
  \hline\hline
  NIS & 250 & 0.0& 0.1&  -1.3172 & 0.87 & 0   \\
  NIE & 330 & 0.1& 0.1&   1.9458 & 0.12 & 1.2 \\
  NIE & 200 & 0.1& 0.1&  -1.2872 &-2.17 &-0.3\\
  NIS & 120 & 0.0& 0.1&   0.6885  &1.17 & 0 \\
  \hline
  \end{tabular}
\end{center}
where once again
$\sigma$ is in km s$^{-1}$, $\rc$ in kpc, $\theta$ in arcseconds and 
$\phi$ in radians.
The redshifts chosen were $\zd = 0.6$ and $\zs = 2$.
The velocity dispersions and the ellipticities were chosen to reflect 
the observed surface brightness.
In \fref{fig:085914} we show our best predicted images,  for comparison with
the real system in \fref{fig:real085914}. 

\subsubsection{Critical curves, caustics and image configurations}

\begin{figure*}
\begin{minipage}{0.48\linewidth}
\includegraphics[width=8cm]{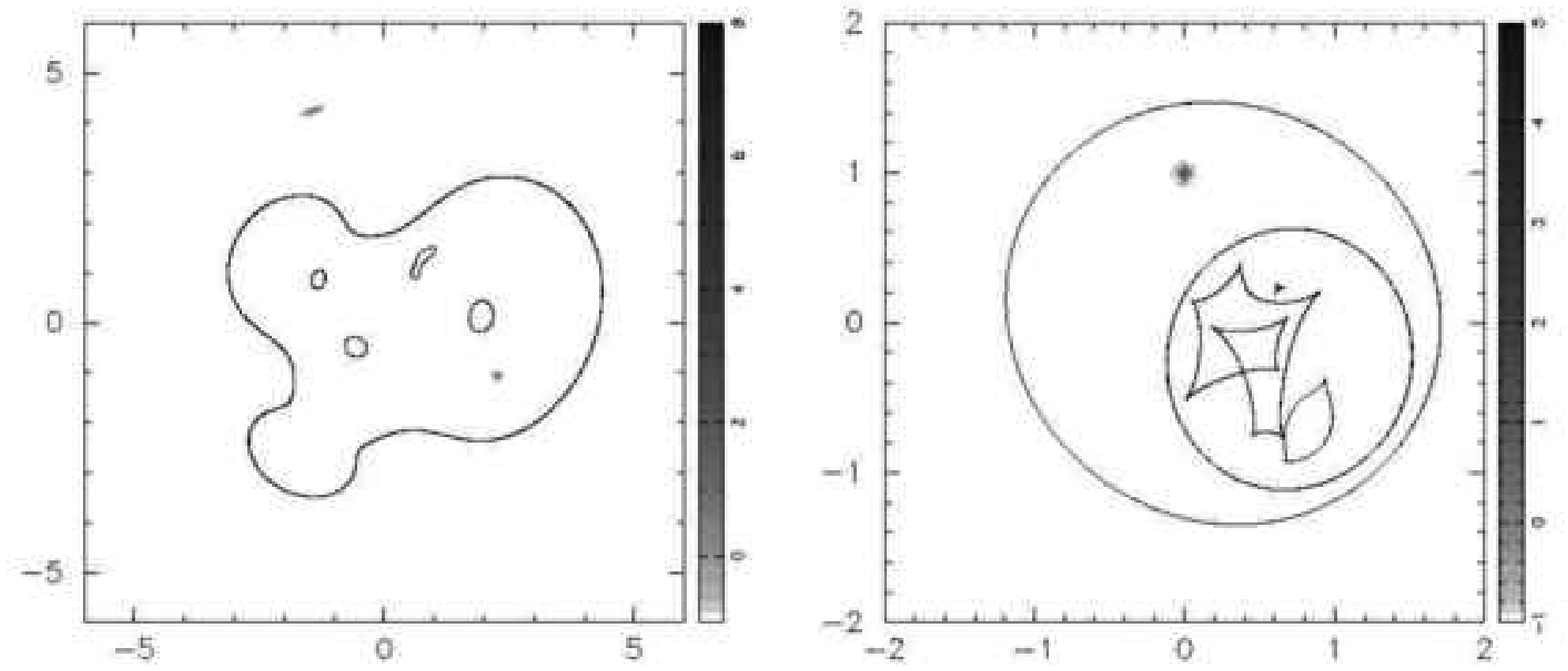}\hfill
\includegraphics[width=8cm]{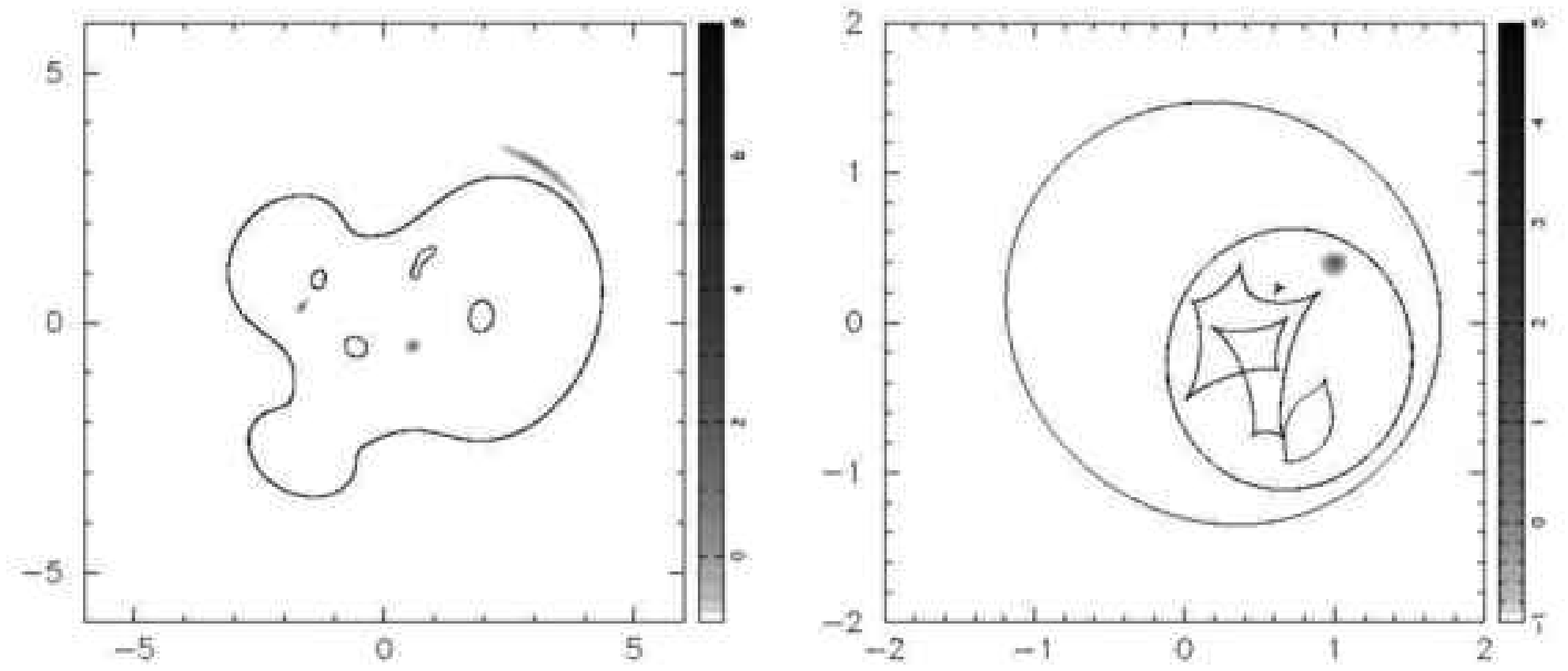}\hfill
\includegraphics[width=8cm]{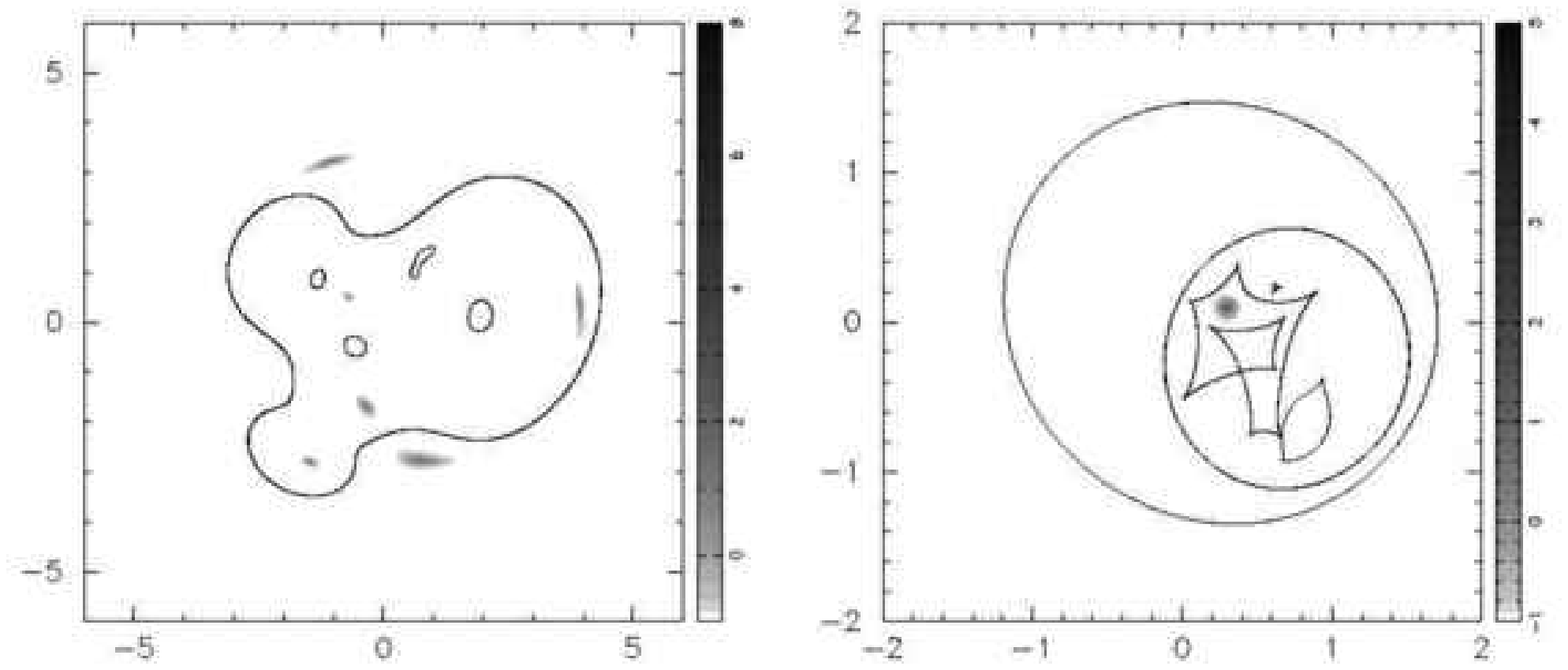}\hfill
\includegraphics[width=8cm]{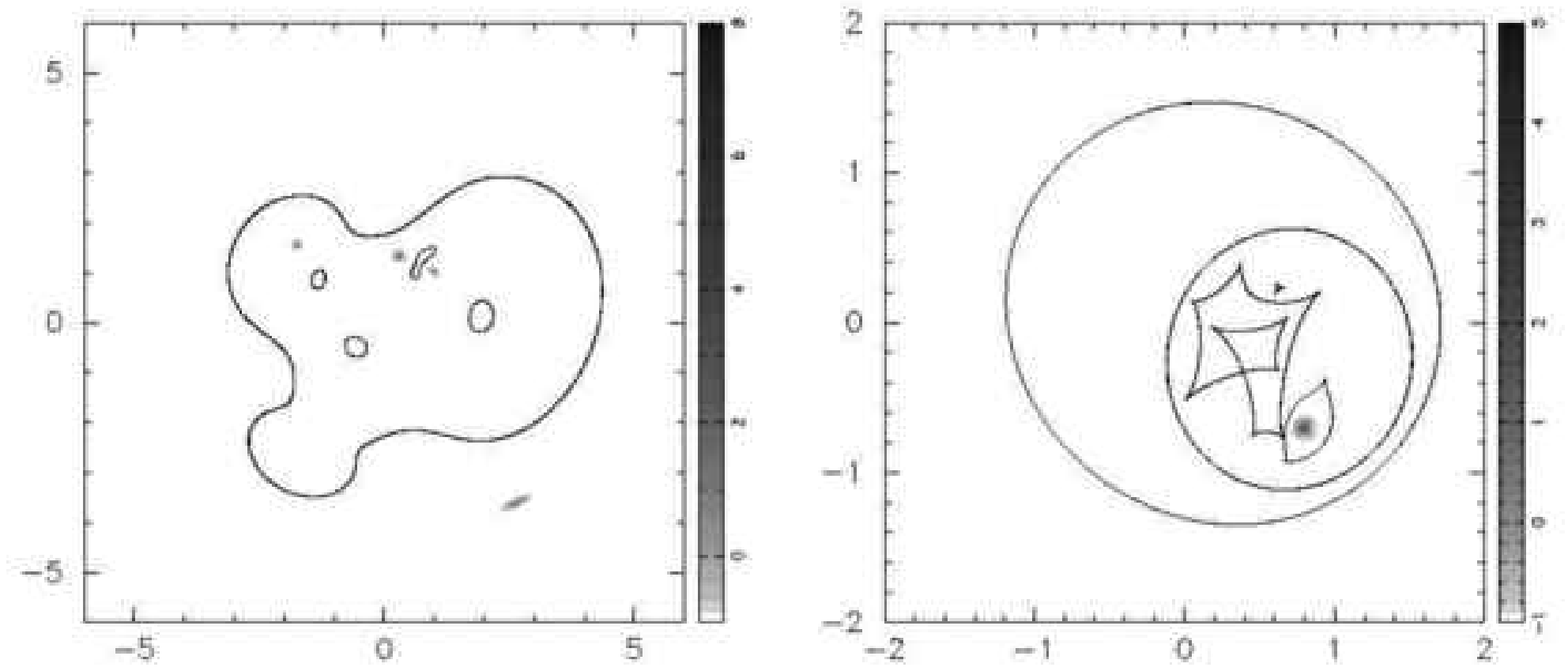}\hfill
\includegraphics[width=8cm]{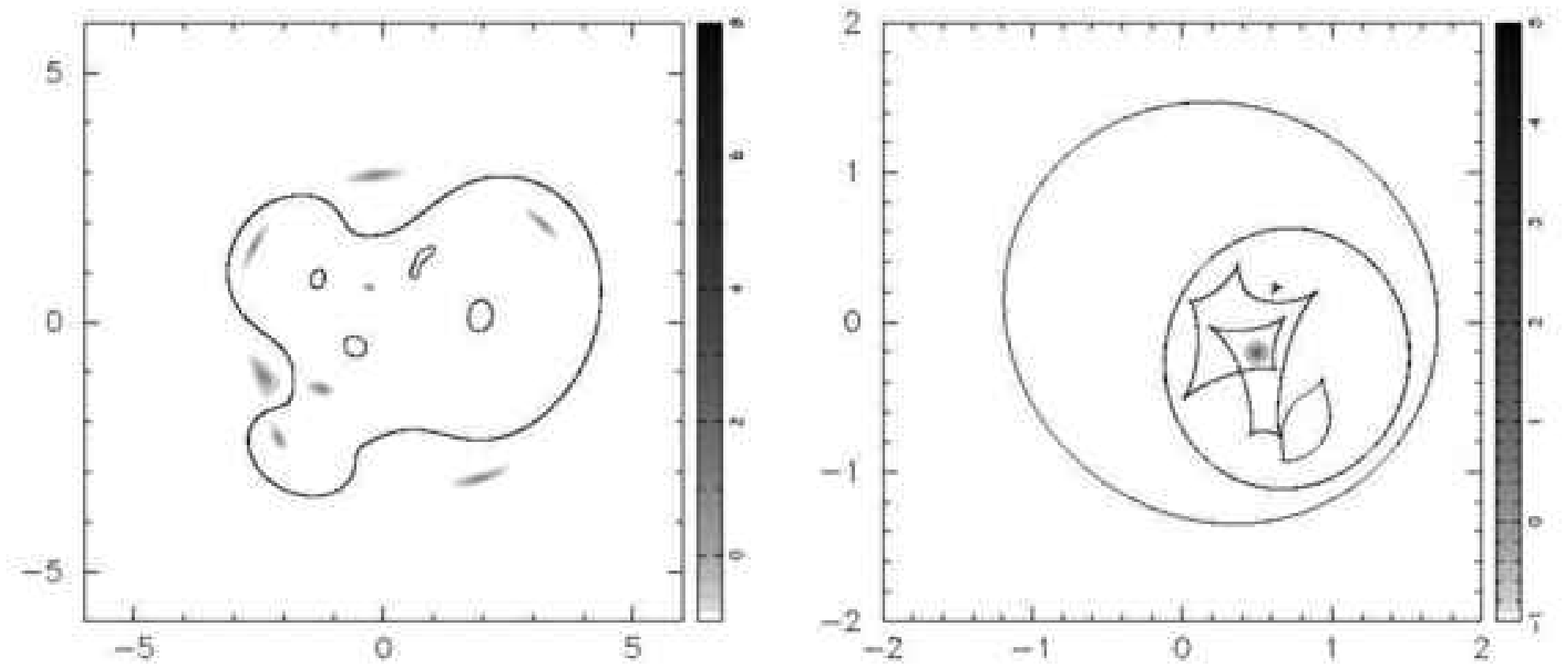}\hfill
\end{minipage}\hfill
\begin{minipage}{0.48\linewidth}
\includegraphics[width=8cm]{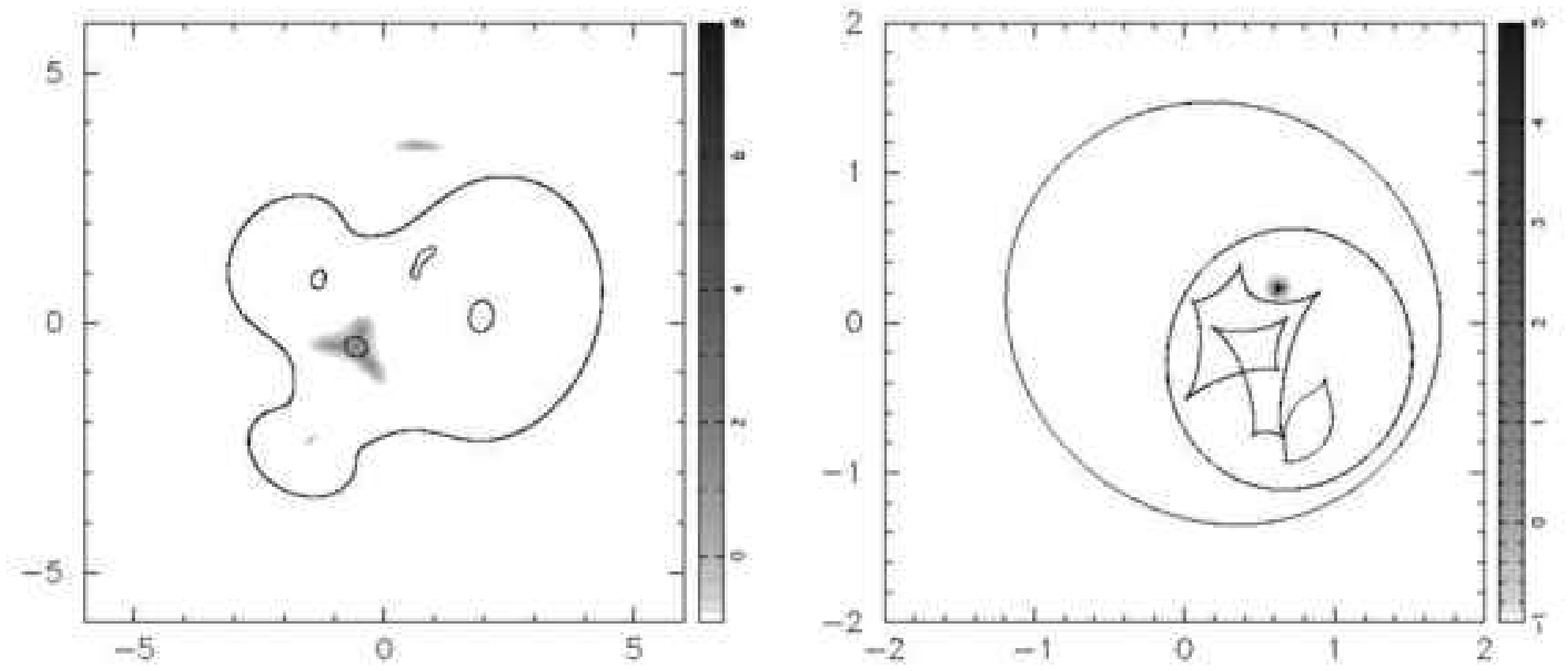}\hfill
\includegraphics[width=8cm]{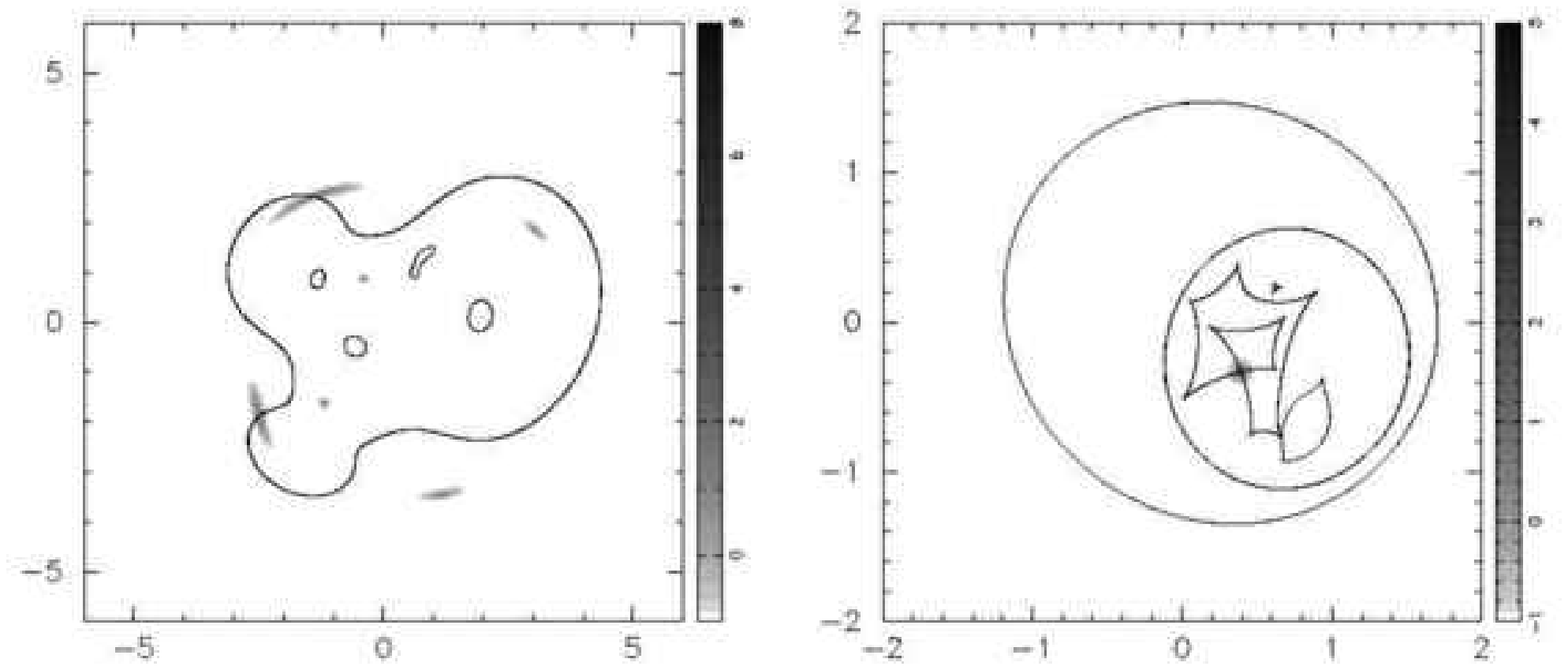}\hfill
\includegraphics[width=8cm]{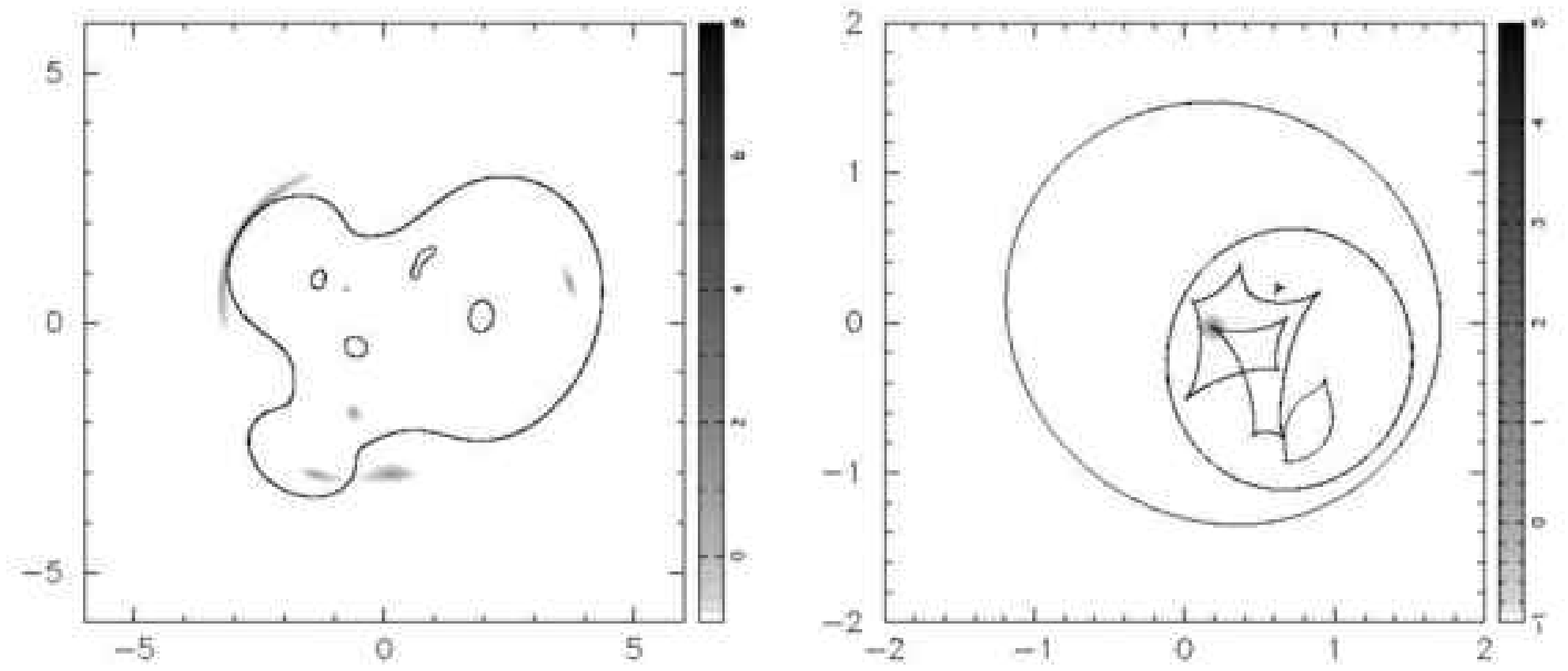}\hfill
\includegraphics[width=8cm]{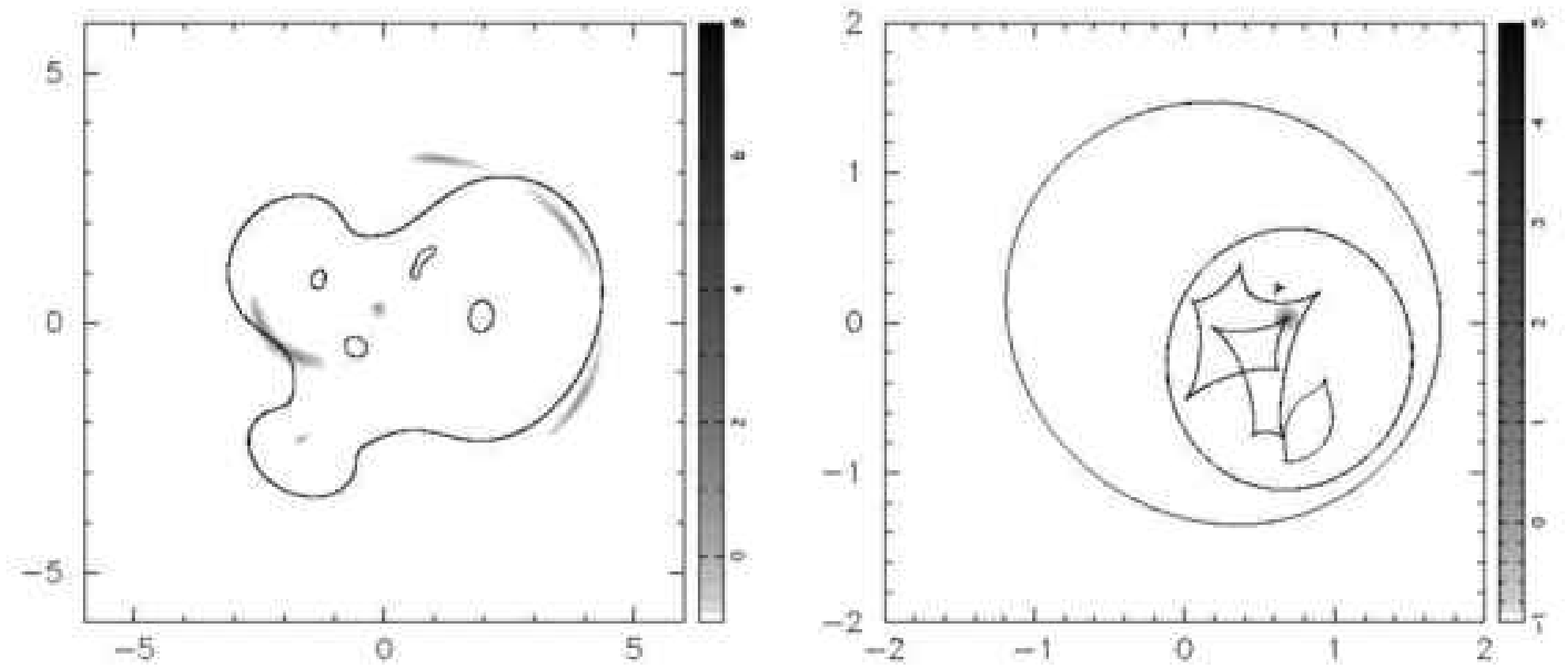}\hfill
\includegraphics[width=8cm]{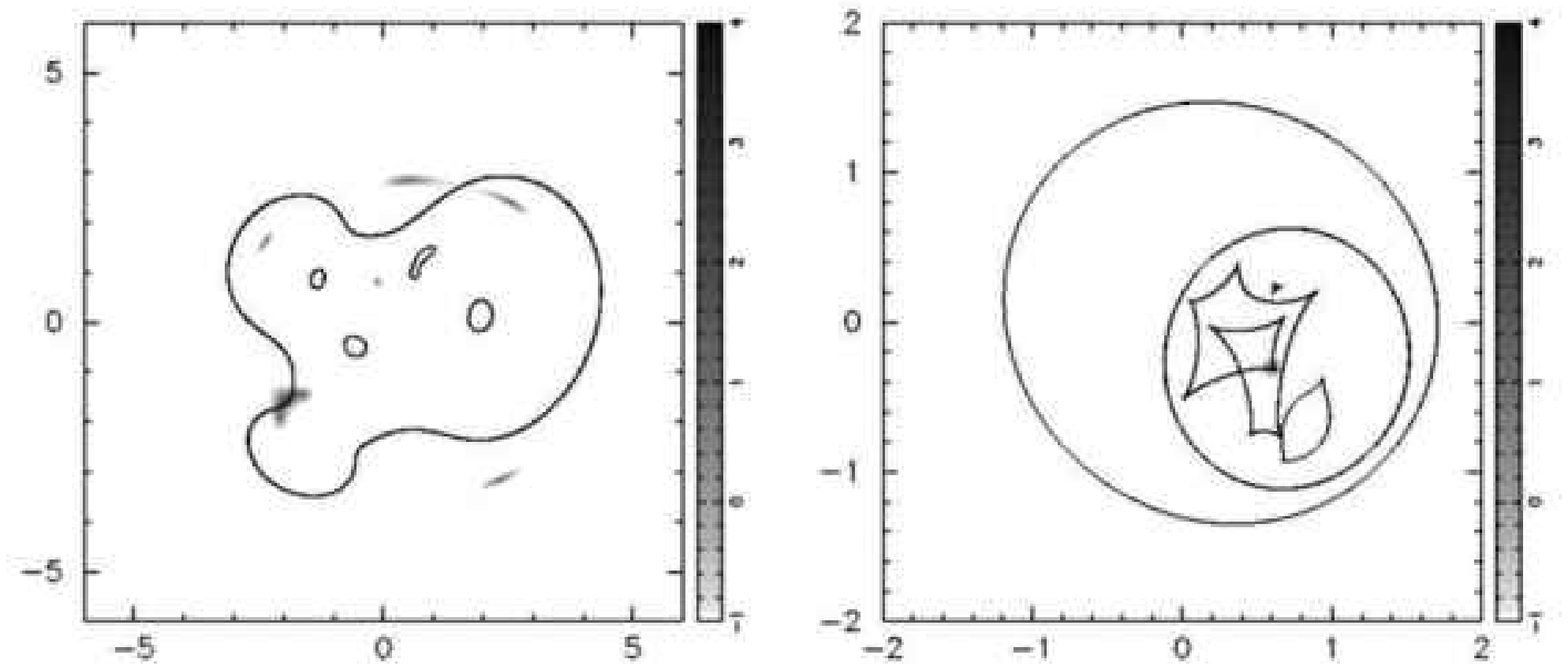}\hfill
\end{minipage}
\caption{Alternative histories for the SL2SJ0859$-$0345 system: 
the image configurations ({\it left}) and multiplicities resulting from
various source positions ({\it right}). Left column, from top to bottom: 
1. Double image, of a source in the first outer caustic. 
2. Quad: second outer caustic. 
3. Sextuplet:  outer astroid caustic. 
4. Quad: lips caustic.
5. Octuplet: inner-astroid caustic.
Right column, from top to bottom:
1. Sextuplet, including 4 merging images of an elliptic umbilic configuration:
  source in a deltoid caustic. 
2. Octuplet:  source at the astroid fold crossing point. 
3, 4, 5. Octuplets: sources on each of the three inner-astroid cusps.}
\label{fig:085914m}
\end{figure*}

\begin{figure*}
\centering
\includegraphics[width=17cm]{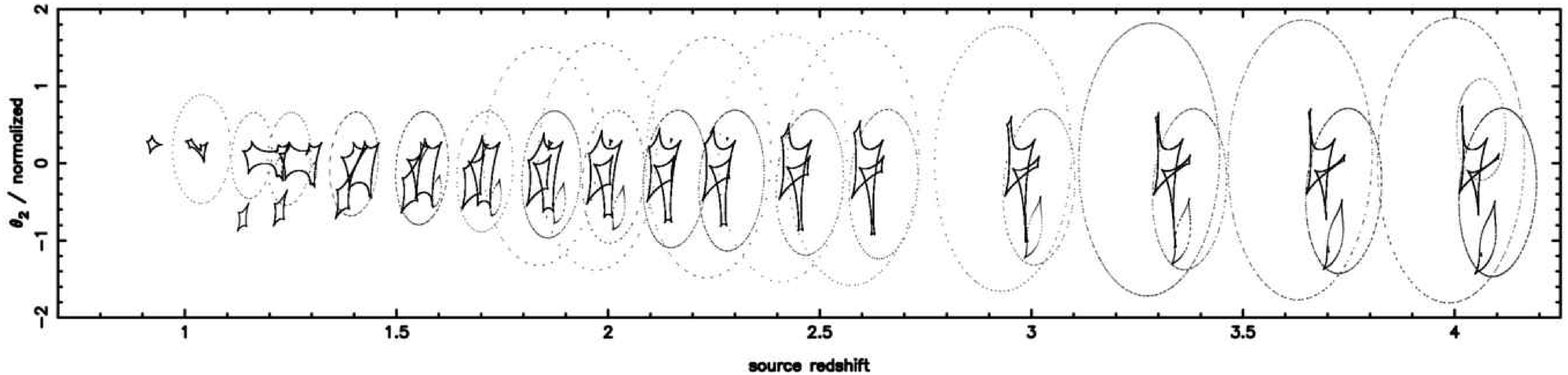}
\caption{Evolution of the caustics with source redshift
for our qualitative model of SL2SJ0859$-$0345. The ovoid 
outer curves are sometimes only sparsely
sampled by the adaptive mesh code used to compute the magnification maps 
(see appendix).}
\label{fig:085914red}
\end{figure*}

In \fref{fig:085914m}, we present an array of different image configurations
produced by varying the source position, illustrating what might have been.
Several different  multiplicities are possible, including an elliptic umbilic
image configuration. The caustic structure is composed of two outer ovoids, a
folded astroid caustic, a lips, and a deltoid caustic. The highest image
multiplicity obtained is nine (in practice eight images, and referred to as an
octuplet). We also plot the evolution of these caustics with source redshift
in \fref{fig:085914red}: several beak-to-beak calamities result in the
extended and folded-over central astroid caustic, while  the elliptic umbilic
catastrophe occurs at around $\zs \simeq 2.3$. 
As a caveat, this four-lens model is in one sense maximally complex:
any smooth group-scale halo of matter would act to smooth the
potential and lead to somewhat simpler critical curves and caustics. However,
it serves to illustrate the possibilities, and may not be far from a realistic
representation of such compact, forming groups.

\subsubsection{Magnification}

We choose to not discuss extensively the magnification of this system, having
already examined the available magnifications in the simpler instances of the
various caustic structures in previous sections.  Nevertheless, we would still
like to know if the deltoid caustic (when we are close to an elliptic umbilic
catastrophe) can lead to a higher magnification than other regions of the
source plane. In order to do that, we compute the total magnification for two
particular positions of the source:  in the central astroid caustic (leading
to eight observable images in a broken Einstein ring, bottom left panel of
\fref{fig:085914m}), and in the deltoid caustic (top right panel of
\fref{fig:085914m}). 

We again consider our three fiducial exponential sources  ($\re = \{0.5, 1.5,
2.5\}$~kpc.  The central position leads to magnifications of $\mu  \simeq
\{55,39,27\}$  respectively, and for the deltoid position  we obtain a total
magnification of $\mu \simeq \{49,33,24\}$. The deltoid caustic again does not
give the maximum total magnification -- it is a region of image multiplicity
seven, while the central astroid caustic of multiplicity nine.  However, the
deltoid does concentrate, as a cusp does, the magnification into a single
merging image object. In the next section we will see an example of how such
focusing by a higher order catastrophe leads to a highly locally-informative
image configuration.


\subsection{Abell 1703}\label{ssec:a1703}

\begin{figure*}
\centering
\includegraphics[height=4cm]{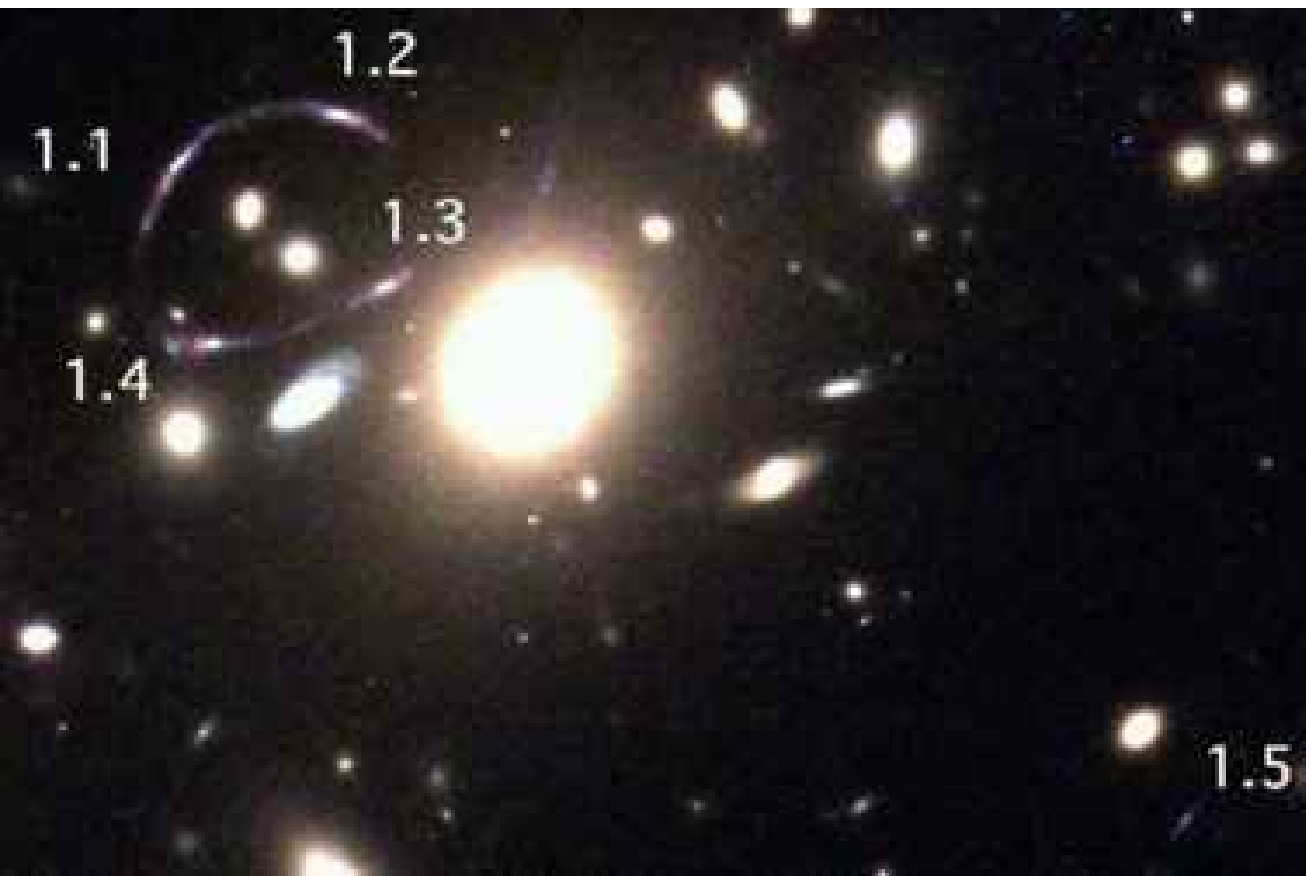}
\includegraphics[height=4cm]{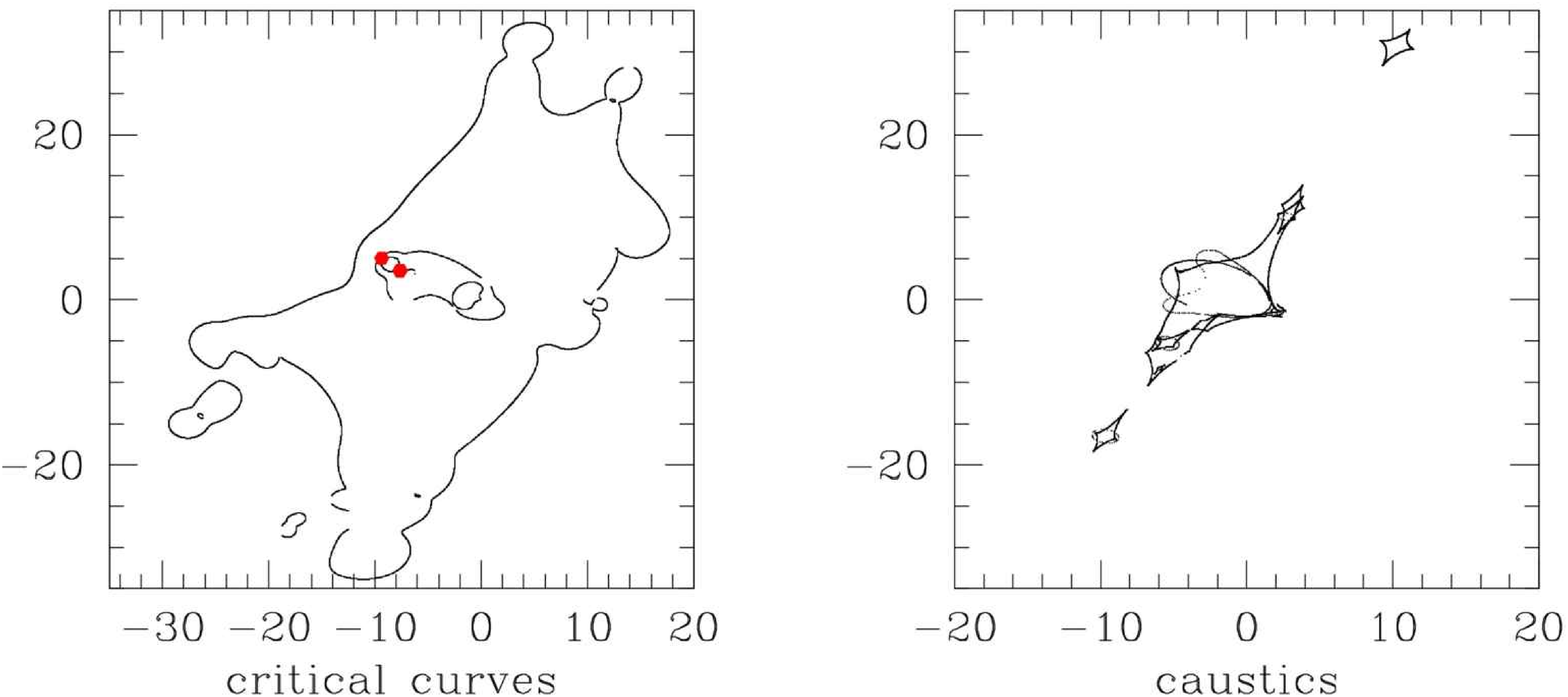}
\caption{Abell 1703 and the lens model modeled of \citet{L08}
{\it Left:} the core of the cluster observed with \hst/ACS \citep{Sto07}. The image
is approximately 40 arcsec wide.
Critical curves ({\it centre}) and caustics ({\it right}) in the lens
model of \citet{L08}. The two galaxies inside the central ring are
represented by two filled red circles.}
\label{fig:abell1703-real}
\end{figure*}

\fref{fig:abell1703-real} shows the central part of the cluster Abell 1703,
and its critical curves and caustics as modeled by \citet{L08}.  This
elliptical cluster, 
modelled most recently by \citet{Ogu++09} and \citet{Ric++09}, 
contains an interesting
wide-separation ($\simeq 8$ arcsec) 4-image system  (labelled 1.1--1.4 in the
left-hand panel of \fref{fig:abell1703-real}). This ``central ring'' quad has
a counter-image on the opposite side of the cluster (labelled  1.5). Two red
galaxies lie inside the ring; they are marked as  two red dots in the central
panel of \fref{fig:abell1703-real}.

\begin{figure*}
\raggedright
\begin{minipage}[t]{0.95\linewidth}
  \begin{minipage}[t]{0.32\linewidth}
    \vspace{-0.94\linewidth}
    \raggedleft\includegraphics[width=0.78\linewidth]{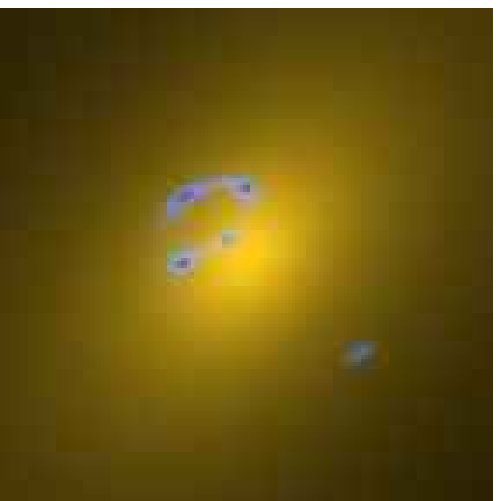}
  \end{minipage}\hfill
  \begin{minipage}[t]{0.67\linewidth}
    \centering\includegraphics[width=\linewidth]{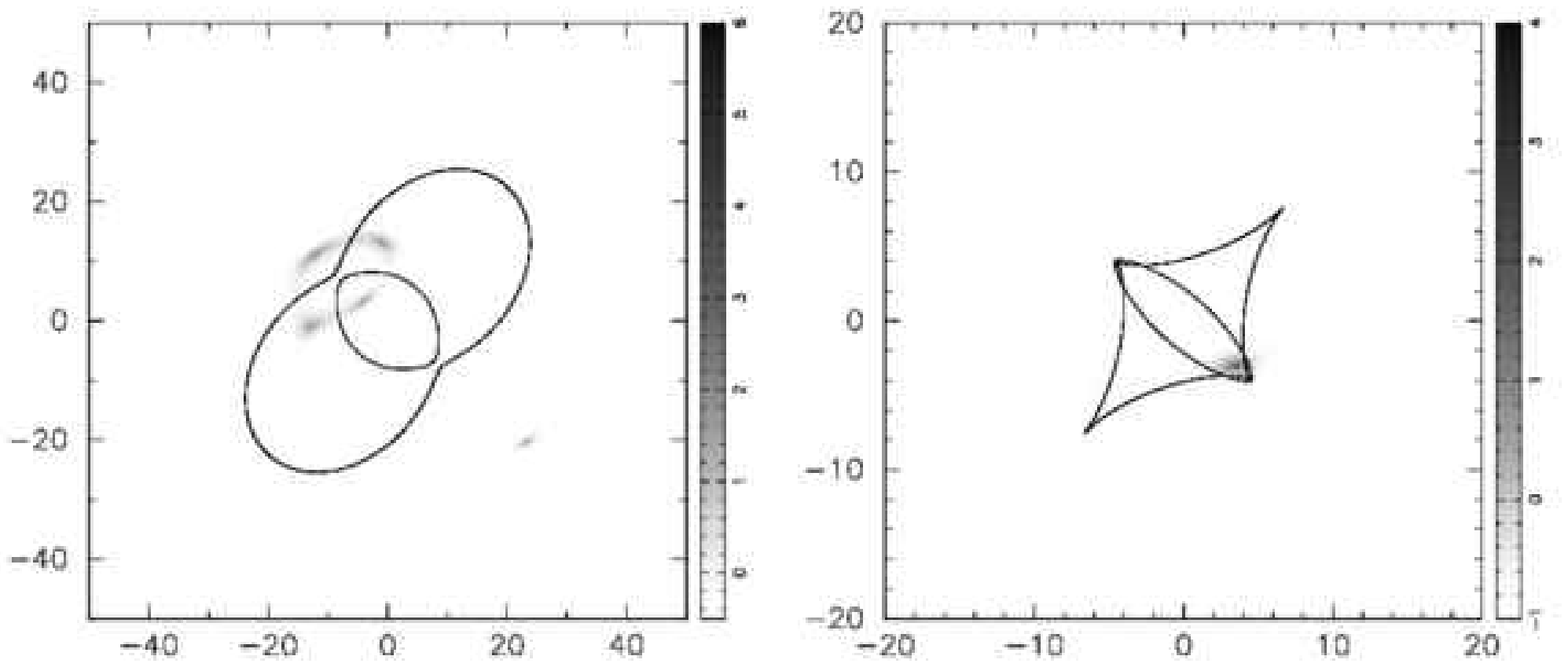}
  \end{minipage}
\end{minipage}
\raggedright
\begin{minipage}[t]{0.95\linewidth}
  \begin{minipage}[t]{0.32\linewidth}
    \vspace{-0.94\linewidth}
    \raggedleft\includegraphics[width=0.78\linewidth]{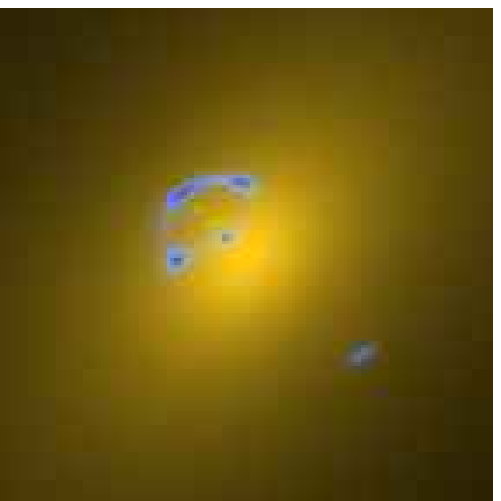}
  \end{minipage}\hfill
  \begin{minipage}[t]{0.67\linewidth}
    \centering\includegraphics[width=\linewidth]{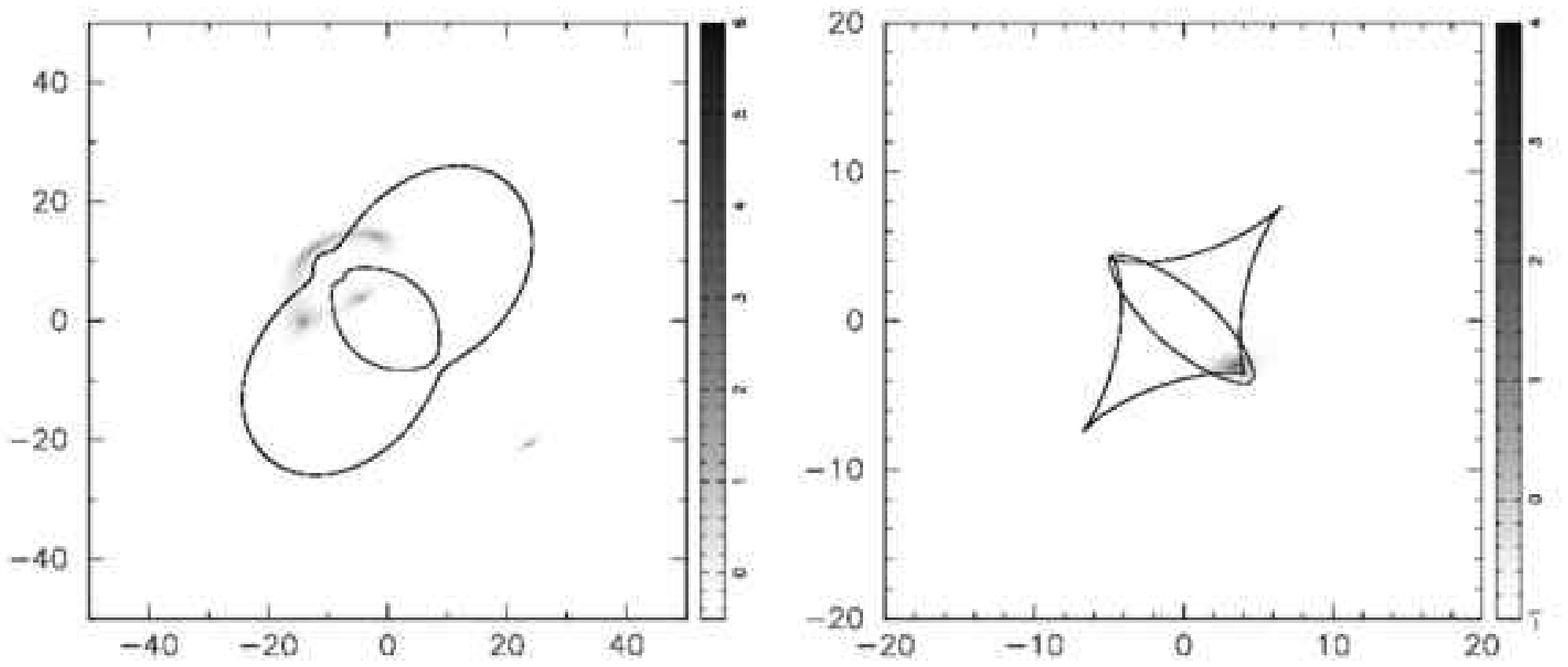}
  \end{minipage}
\end{minipage}
\caption{Simple models for understanding Abell 1703, with predicted images 
(left), critical curves (centre) and caustics (right). The effect of the 
two galaxies inside the central ring is shown in the lower panels; they are
absent from the model in the upper panels.}
\label{fig:abell1703-model}
\end{figure*}

Explaining the central ring as one image of a double system being split by the
two red galaxies is difficult; in the \citet{L08} mass model they are
represented as isothermal spheres with velocity dispersion $\simeq
100$~km~s$^{-1}$ each. The wide separation of the ring is a feature of the
cluster-scale mass distribution, as we now show. We made a rough modeling of
the cluster using a NIE profile with velocity dispersion of $\sim$
1200~km~s$^{-1}$, a core radius of $\sim$~30~kpc, and an ellipticity of the
potential of $\epsilon_{\rm p}' = 0.2$. This single halo model reproduces the
central ring image configuration very well. We show the predicted images,
critical curves and caustics in the upper panels of
\fref{fig:abell1703-model}.  This image configuration is associated with a
hyperbolic umbilic catastrophe;  one can see in the right-most panel of
\fref{fig:abell1703-real} and more clearly in he corresponding panel of
\fref{fig:abell1703-model} that the source lies very close to the point where
the ovoid and astroid caustics almost touch. Were the source actually at the
catastrophe point, the four images of the central ring would be merging; as it
is, we see the characteristic pattern of a hyperbolic umbilic image
configuration. As far as we know this is the first to have ever been
observed. 

When we add the two $\sigma = 100$km s$^{-1}$ SIE galaxies inside the central
ring (lower panels of \fref{fig:abell1703-model}), the central ring system is
perturbed a little. This perturbation allowed \citet{L08} to measure them
despite their small mass, illustrating the notion that sources near higher
order catastrophes provide opportunities to map the local lens mass
distribution in some detail. Indeed, this image system provides quite a strong
constraint on the mass profile of the cluster \citep{L08}.

\subsubsection{An even simpler model}

To try and estimate the abundances of such hyperbolic umbilic image 
configurations, we consider a slightly more representative (but still very
simple) model of a cluster: cluster and its brightest cluster galaxy (BCG),
both modelled by a NIE profile. The parameters used are taken after considering
the sample of \citet{Smi++05} who modelled a sample of clusters using NIE
profiles. The model is as follows:
\begin{center}
	\begin{tabular}{lrrr}
	\hline
	type &    $\sigma$ & $\epsilon_{\rm p}'$ & $ \rc$  \\
	\hline\hline
	NIE - Cluster & 900 & 0.2& 80 \\
	NIE - BCG     & 300  & 0.2& 1 \\	
	\hline
	\end{tabular}
\end{center}
where $\sigma$ is in km~s$^{-1}$ and $\rc$ is in kpc.

\subsubsection{Critical curves, caustics and image configurations}\label{sssec:hucriterion}

Using a simple NIE profile, we can obtain an hyperbolic umbilic catastrophe,
provided that the central convergence is sufficiently shallow ({\ie a
significant core radius is required}), and sufficiently elliptical. 
\citet{KSB94}
give a criterion for a NIE lens to be capable of generating 
a hyperbolic umbilic catastrophe: 
\be
b_{\rm c} = \dfrac{q^{2/3}}{2},
\ee
where $b_{\rm c}$ is the core radius in units of the Einstein radius. and the
axis ratio~$q$ refers to the mass distribution and not the lens potential.
\citet{K+K93} also discuss elliptical mass distributions and elliptical
potentials, and describe the different metamorphoses when varying the core
radius.

\begin{figure*}
\includegraphics[width=12 cm]{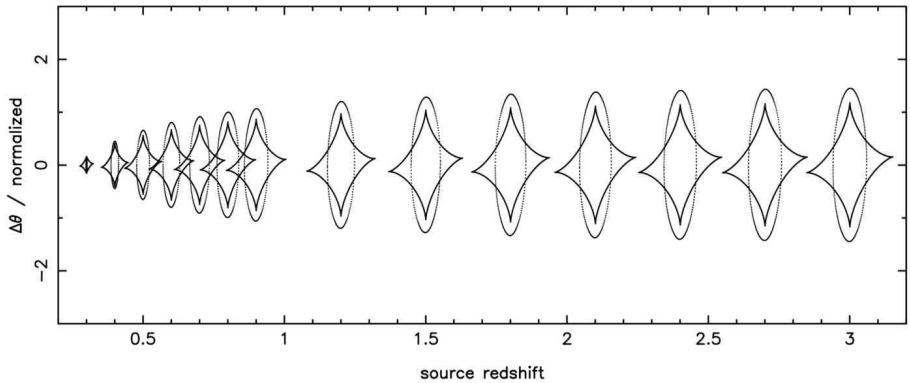}
\caption{Evolution of a simple elliptical cluster lens caustics with
source redshift. The closeness of the cusps to the outer ovoid caustic
persists over a wide range of redshifts.}
\label{fig:abell-redshift-evolution}
\end{figure*}

We show the evolution of this model's caustics with source redshift in
\fref{fig:abell-redshift-evolution}. At low redshift, the caustic structure
consists of two lips (with relative orientation of 90 degrees); as $\zs$
increases, the inner lips becomes bigger and meets the outer lips in a
hyperbolic umbilic catastrophe. The hyperbolic umbilic metamorphosis conserves
the number of cusps: at higher source redshift, we see the familiar (from
elliptical galaxies) outer ovoid caustic with an inner astroid caustic. 

In the light of these considerations, we don't expect hyperbolic umbilic
catastrophes to be generic in elliptical galaxy lenses, as the central
convergence is not sufficiently shallow. However, this catastrophe point should
be present behind many clusters, as the combination of a BCG and a cluster
with appropriate core radii and ellipticities can lead to the central
convergence and ellipticity required. Of course the hyperbolic umbilic image
configuration is more likely to happen than the catastrophe itself. Indeed, as
\fref{fig:abell-redshift-evolution} shows, the hyperbolic umbilic image
configuration can occur over a wide range of source redshift. 

\subsubsection{Magnification}

\begin{figure}
\centering
\includegraphics[width=8cm]{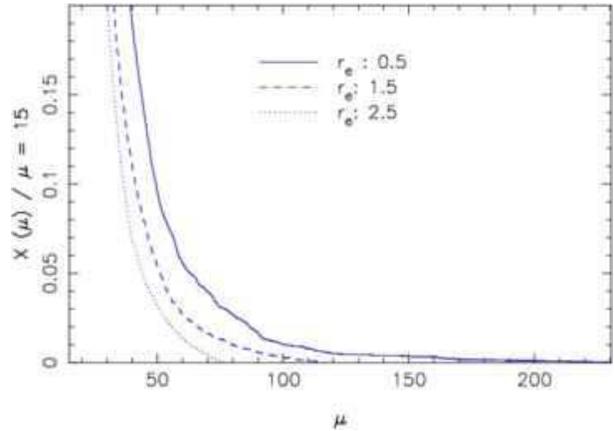}
\caption{Fractional cross-section (relative to the area with 
magnification $\mu \geq 15$) as a function of the threshold
magnification  for our simple NIE cluster model.}
\label{fig:amag}
\end{figure}

\fref{fig:amag} shows the  fractional cross-section area of our NIE cluster
model (at $\zd=0.28$)  as a function of magnification threshold. Here, the
area of the source plane providing total magnification greater than $\mu$ is
compared to the area providing total magnification greater than $\mu \geq 15$,
which is approximately the area of the astroid caustic.  We consider again our
fiducial exponential sources, and observe that the system produces some very
high magnifications.  Analyzing the different convolved maps of source plane 
magnification, we find that it is indeed the regions close to the hyperbolic
umbilic catastrophes that provide the highest magnifications. 
For $\re=0.5$~kpc, the cross-section for $\mu \geq 100$ is entirely due to 
the two regions
close to the catastrophe. We obtain the same result for the two other sources
(respectively $\mu \simeq 75$ and $\mu \simeq 25 $ for $\re = 1.5$ and
$2.5$ kpc): it is likely that A1703 is magnifying its source by something like
these factors.
\fref{fig:amag} shows that the region with a characteristic magnification of a
hyperbolic umbilic catastrophe is around 1--2\% of the  multiple-imaging area.

\section{The abundance of higher-order catastrophes}\label{sec:abundance}

In the previous sections, we have illustrated a range of ``exotic'' lenses,
systems presenting higher-order catastrophes such as swallowtails,
butterflies, elliptic umbilics and hyperbolic umbilics, and their related
metamorphoses. In this section we make the first very rough approximations of
the abundances of lenses showing such exotic image configurations.  

{\it Procedure.}
For each given type of lens, we first estimate the basic abundances of the
more common, lower magnification configurations, based on the number densities
observed so far in \slacs and \sls.  We then estimate, for a given source
redshift, the fractional cross-section area for production of the desired
image configuration. To do this we consider both a) the information provided
by the magnification maps and their associated fractional cross-sectional
areas, and b) our rough estimates of the ``area of influence'' of the relevant
caustic feature. Finally, we estimate the typical window in source redshift
for which the image configuration is obtainable, and hence the fraction of
sources lying at a suitable position along the line of sight. The product of
these two fractions with the basic abundance gives us an order-of-magnitude
estimate of the number density of exotic lenses detectable in a future imaging
survey (which we take to cover $\sim10^4$ square degrees).

{\it Galaxy scales.}
On galaxy scales, we expect some 10 strong gravitational lenses observable per
square degree, provided the image resolution is high enough: this is
approximately what is found in the COSMOS \citep{Fau++08} and EGS
\citep{Mou++07} surveys. The majority of these lens galaxies are
massive ellipticals, as expected \citep{TOG84}. In the larger 
\slacs sample of similar galaxy-galaxy lenses, the fraction of systems
showing   strong evidence of a disk component is approximately 10\%, while
only one or two (a few percent) have massive satellites within the Einstein
radius \citep{Bol++08}. Making the link between galaxy-scale lenses
expected in imaging surveys and the lenses found by the \slacs survey should be
done with care, but it is not unreasonable to start from this extrapolation --
especially in the light of the \slacs studies demonstrating the normality of
the lens galaxies relative to the parent massive galaxy population
\citep{Tre++06,Tre++09}. This gives us basic abundances of disky lenses
and binary/merging lenses of $\sim 1$ and $\sim 0.1$ per square degree
respectively. This is what we might expect a space-based survey to recover;
the number visible from the ground will be much reduced by the image
resolution.

Therefore, our estimate of the abundance of butterflies is  $1 \times (10 /
180) \times 0.001 \sim 5 \times 10^{-5}$ per square degree, where $10/180$ is
the angular range where we can expect butterflies, 0.001 is our estimation of
the cross-section of the butterflies and nine-image multiplicity regions from
\sref{ssec:d+b}. Note that we have optimistically considered a uniform
distribution of disk-bulge misalignments: this abundance might be taken as an
upper limit for these types of lens mass distributions.  We expect to have
roughly 10 times more swallowtails, since the range of misalignment angles
leading to swallowtails is $\sim 10$ times greater than that for butterflies. 
We therefore obtain an upper limit of $\sim 1$  observable disk/bulge
butterfly on the sky, and $\sim$10 disk/bulge swallowtails.

{\it Group scales.} On group scales, we might expect the ground-based surveys
to be as complete as those from space. Moreover, the depth of the CFHT Legacy
Survey images is comparable to that expected from future surveys such as that
planned with LSST (which will cover 2 orders of magnitude more sky area).
Extrapolating the number of group-scale gravitational lenses from the SL2S
survey \citep{Cab++07}  is therefore a reasonable starting point. The observed
abundance of group-scale lenses is $\sim 0.1$ per square degree, and perhaps a
third show multiple mass components within the Einstein radius. To order of
magnitude we therefore estimate the basic abundances of compact group lenses
as $\sim 0.01$ per square degree. 

The two systems studied here show a variety of complex caustic structure. In
the case of the binary systems in \sref{ssec:binary}, we estimated
geometrically the fractional cross-section of a deltoid caustic to be $\sim$
10$^{-3}$ over a wide redshift range. We therefore  obtain an order of
magnitude estimate of  $0.01 \times 10^{-3} \sim$10$^{-5}$ per square degree
for the abundance of elliptic umbilic image configurations, or approximately
one observable Y-shaped image feature in our future survey. This cross-section
area is unlikely to be very different (depending on the lens)  to that for the
production of swallowtails and butterflies by group lenses. A Monte Carlo
simulation would perhaps be the best way to estimate these abundances more
accurately.

{\it Clusters.} For clusters, as shown in section~\ref{sssec:hucriterion}, the
properties needed to obtain an hyperbolic umbilic catastrophe are well
understood: the problem lies in estimating the ranges of ellipticities,
convergences and core radii of the cluster and BCG mass distributions that
combine to provide a sufficiently shallow inner density slope for the system
to generate a hyperbolic umbilic.  As a first attempt,  we consider the
distribution of cluster lens potential ellipticities estimated from numerical
simulations: this is approximately a Gaussian with mean 0.125 and width 0.05
(\citeauthor{Mar06}~\citeyear{Mar06}, following
\citeauthor{J+S02}~\citeyear{J+S02}). In general the BCG increases the central
mass slope (decreasing the core radius of an effective NIE profiles), such
that we need high ellipticity in order to be close to an hyperbolic umbilic
catastrophe.  We estimate that only the 1--10\% highest ellipticity clusters
will  meet this criterion. Our qualitative analysis in \sref{ssec:a1703}
suggests that once the ellipticity and density profile condition is met then
the  hyperbolic umbilic metamorphosis is slow with source redshift.  The
fractional cross-section area of this ``hyperbolic umbilic region'' is
approximately 2\%. As the number of cluster lenses is expected to be $\sim
0.1$ per square degree, we estimate the abundance to be 
$\sim 10^{-4}$ per square degree, 
again predicting  $\sim 1$
hyperbolic umbilic catastrophe image configuration  per sky survey. A question
comes to mind: is Abell 1703 the only hyperbolic umbilic gravitational lens we
will ever see? 

{\it Discussion.} Throughout this work we have focused on the most
abundant sources, the faint blue galaxies, and tried to account for
their finite size when estimating magnifications and now abundances. To
first order this should take care of any magnification bias. However,
for the less common point-like sources (such as quasars and AGN) it is
possible that magnification bias may play a much more important role.
Investigation of this effect would be a good topic for further work. 
%
The sensitivity of galaxy-scale  gravitational lensing of point sources
to small-scale CDM substructure \citep[\eg][]{M+S98,D+K02,Bra++04}  is
such that  more investigation of the exotically-lensed quasar abundance
is certainly warranted. There are 
similar, but perhaps more constraining,
relations between the signed fluxes of
images of sources lying close to 
higher order catastrophes \citep{A+P08} as there are for 
folds and cusps \citep{KGP03,KGP05}. 

Another possibility that might be included in a more complete analysis
of exotic lens abundance is that of generating higher order catastrophes
with multiple lens planes, a topic studied by \citep{K+A88}. If a
substantial fraction of the observed group-scale lenses are in fact due
to superpositions of massive galaxies rather than compact groups this
could prove a more efficient mechanism for the production of higher
order catastrophes.


\section{Conclusions}\label{sec:conclude}

Inspired by various samples of complex lenses observed, and with future
all-sky imaging surveys in mind, we have compiled an atlas of realistic
physical  gravitational lens models capable of producing exotic image
configurations associated with higher-order catastrophes in the lens map. For
each type of lens considered, we have investigated the caustic structure, 
magnification map, and example image configurations, and estimated approximate
relative cross-sections and abundances. We draw the following conclusions:

\begin{itemize}

\item Misalignment of the disk and bulge components of elliptical galaxies
gives rise to swallowtail (for a wide range of misalignment angles) 
and butterfly catastrophes (if the misalignment is close to 90~degrees). The
image configuration produced by the butterfly caustic would be observed
as a broken (8-image) Einstein ring.

\item The central nine-imaging region has cross-section $\sim10^{-3}$ relative
to the total cross-section for multiple imaging; combining this with rough
estimates of the abundance of such disky lens galaxies in the \slacs survey,
we estimate an approximate abundance of $\lesssim1$ such butterfly lens  per
all-sky survey. We estimate that the swallowtail lenses could be $\sim10$
times more numerous.

\item Binary and merging galaxies produce elliptic umbilic catastrophes when
the separations of the mass components are comparable to their Einstein
radii. The configuration formed when the source lies within the deltoid
caustic is a Y-shaped pattern of 4 merging images between, but offset from,
the two lenses. 

\item In this case the deltoid caustic does not provide the
maximum total magnification available to the lens; the relevant cross-section
must be estimated from the area within the deltoid. We find the relative
cross-section to also be $\sim10^{-3}$, leading to an  approximate abundance
of $\sim1$ binary galaxy elliptic umbilic lens per all-sky survey. 

\item More complex group-scale lenses, of the kind being discovered by the
\sls survey, offer a wide range of critical points in their caustics, and so
are a promising source of exotic lenses. For example, our model of 
SL2SJ0859$-$0345 shows an elliptic umbilic catastrophe at around~$z=2.4$.
At present, our understanding of the distribution of group-scale lens
parameters, and the extreme variety of caustic structure prevents us making
accurate estimates of the exotic lens abundance for these systems.

\item As noted by previous authors, elliptical clusters with appropriate
inner density profile slopes produce hyperbolic umbilic catastrophes. We
find that just such a simple model is capable of reproducing the
central ring image configuration of Abell~1703, demonstrating the source
to lie close to a hyperbolic umbilic point. The total magnification of
this source could be $\sim100$, depending on the source size.

\item Despite the rather general properties of clusters needed to produce such
image configurations, we estimate that the all-sky abundance of such 
hyperbolic umbilic cluster lenses may still only be $\sim1$.

\end{itemize}
 
In some cases, proximity to a catastrophe point does not guarantee
maximal magnification. Further studies of the abundances of exotic
lenses should explore more advanced diagnostics of catastrophic
behavior. While high magnification is certainly a desirable feature of
gravitational lenses for some applications (\eg cosmic telescope
astronomy), high local image  multiplicity is perhaps the more relevant
property for studies of small scale structure in the lens potential, 
using the information on the gradient and curvature of the mass
distribution that is present. 
The constraining power of the hyperbolic
umbilic image configuration in Abell 1703 is a good example of this on
cluster scales. 
The exploitation of the plausible lenses we have predicted here in this way
would be an interesting  topic for further research.


\section*{Acknowledgments}

We thank Tommaso Treu, Maru\v{s}a Brada\v{c}, Masamune Oguri, Jean-Paul Kneib,
Marceau Limousin and Roger Blandford for useful discussions, and Raphael
Gavazzi and the \sls and \slacs teams for providing some of the illustrative
data shown here.  We are grateful to Ted Baltz for supporting our use of the
\glamroc software.  
We thank the anonymous referee for a very encouraging report, and for
correcting several technical details of lensing singularity theory.
GOX thanks the members of the UCSB astrophysics group for
their hospitality during his undergraduate research internship during which
this work was carried out. The work of PJM was supported by the TABASGO
foundation in the form of a research fellowship.





\label{lastpage}
\bsp


\clearpage
\appendix

\section{Lens and source models}\label{sec:models}

In this appendix, we describe in a little more detail the different models
used throughout the  paper: the isothermal profiles to model the lens
potentials  and the Sersic profiles for the background source galaxies' light
distributions (or for the stellar mass component of a lens galaxy). 

For all image simulations in this paper we uses the publically-available
\glamroc code (Gravitational Lens Adaptive Mesh Ray-tracing of Catastrophes)
written by
E.~A.~Baltz.\footnote{\texttt{http://kipac.stanford.edu/collab/research/lensing/glamroc/}}
Here we give a very brief introduction to this code and its
capabilities.

Each \glamroc lens model is made of an arbitrary number of lensing objects on
an arbitrary number of lens planes.  The individual lensing object all have
analytic lens potentials, such that all their derivatives are also analytic. 
In this way the deflection angles, magnification matrices, and combinations of
higher derivatives can be calculated as sums of terms coming from each lens in
turn.   Higher derivatives are required to identify catastrophes of the lens
map, which include all the catastrophes define above. An adaptive mesh is used
to improve the resolution either near the critical curves of the lens system
or in regions of high surface brightness. Lens types currently implemented
include point masses, isothermal spheres with and without core and truncation
radii, NFW profiles with and without truncation, and Sersic profiles.  For
each type, elliptical isopotentials can be used and either boxiness, diskiness
and skew added. Sources are modelled with  (superpositions of)
elliptically-symmetric Sersic profiles.


\subsection{Isothermal lenses}

Throughout the paper, we used the non-singular isothermal ellipsoid (NIE)
model for our lens components. 
The NIE (three-dimensional) mass density diverges as
$\rho \varpropto r^{-2}$; 
the surface (projected) mass density profile is 
\be
\Sigma(r) = \dfrac{\sigma}{2 G (r^2 + \rc^2)^{1/2}},
\ee
where $r$ is now the projected radius, 
$\rc$ is the core radius, and $\sigma$ is the velocity dispersion of the
lens. The core radius is that at which the rising density profile turns over into a
uniform distribution in the central region.
This type of profile has been shown to be a very good approximation of the
lens potential on both galaxy scales~\citep[\eg][]{Koo++06}, provided the core
radius is very small ($\lesssim 0.1$~kpc),  and also on cluster scales, with
much larger core radii \citep[$\sim 50.0$~kpc, \eg][]{Smi++05}. 
In practice a small core allows more convenient
plotting of caustics and critical curves. However, we are careful not to allow
cores larger than typically permitted by  the data: the non-singular  core
prevents the infinite de-magnification of the central image, which would be
observable.

The NIE model isopotential lines are ellipses with constant
\begin{equation}
r^2=(1-\epsilon_{\rm p}')x^2+(1+\epsilon_{\rm p}')y^2.
\end{equation}
where $x$ and $y$ denote the Cartesian coordinates in the lens plane.
The potential is, therefore, a function of $r$ only, and the 
ellipticity~$\epsilon_{\rm p}'$ is the ellipticity of this potential.
The ellipticity must be carefully chosen in order to keep the potential
physically meaningful. 
Indeed, for $\epsilon_{\rm p}' > 0.2$ the isodensity contours 
become dumbbell-shaped \citep[see \eg][]{K+K93}. 
\citet{BMO09} provide  a simple practical solution to this problem:
add several (suitably chosen and weighted) elliptical
potentials at the same location. 
Using their algorithm, we are able to model nearly elliptical isodensity
contours with ellipticities  as large as $\epsilon'=0.8$. This allows us to
model, for example, the combination of a bulge and a disk in a galaxy. (Note
that the NIE model described here is different from that defined by
\citet{KSB94}, where the ellipticity pertains to the mass distribution, not to
the lens potential.)

We follow the \glamroc notation and  use as our main ellipticity definition
$\epsilon' = \frac{a^2 - b^2}{a^2 + b^2}$, where $a$ and $b$ are the major and
minor axis lengths of the ellipse in question. This is different from the
definition often used in  weak lensing,  $\epsilon = \frac{a - b}{a + b}$. The
relation between these two definitions of the ellipticity is $\epsilon' =
\frac{2 \epsilon}{1 + \epsilon^2}$.


\subsection{Sersic sources}

The Sersic power law profile is one of the most frequently used in the study
of galaxy morphology.  It has the following functional
form~\citep[\eg][]{Pen++02},
\be
\label{eq:sersic} I(r) = \Ie \exp{\left[ - \kappa \left( \left( \dfrac{r}{\re} \right)^{1/n} - 1 \right)\right]}, 
\ee
where $\Ie$ is the surface brightness at the effective radius $\re$. The
parameter $\re$ is known as the effective, or half-light, radius,  defined
such that half of the total flux lies within $\re$. The parameter $n$ is the
Sersic index: with $n=1/2$ the profile is a  Gaussian, $n=1$ gives an
exponential profile typical of elliptical galaxies,  and $n=4$ is the de
Vaucouleurs profile found to well-represent galaxy bulges.
We compared these three standard profiles when exploring the observability of
exotic lenses in the text. 

The half-light radius is an important parameter 
when considering the observable magnification of extended sources. 
For the sizes of the expected faint blue source galaxies, we adopted the
size-redshift relation measured by \citet{Fer++04}. For reference, this gave
effective radii of 4.0~kpc (0.5") at $\zs=1$, decreasing to
2.9~kpc (0.35") at $\zs=2$ and 1.7~kpc (0.21") at $\zs=4$.
We used sources having equal half-light radii, instead of the 
same total flux (the relative intensity
is, as we explain in the paper, not important as the 
magnification calculation based on \eref{eq:realmag} 
is independent of the total flux of the source).

\end{document}